\documentclass[journal]{IEEEtran}
\usepackage[utf8]{inputenc}
\usepackage{siunitx}
\usepackage{hyperref} 
\usepackage{amssymb}
\usepackage{amsmath}
\usepackage{amsthm}
\usepackage{bm}
\usepackage{float}
\usepackage{xcolor}
\usepackage{colortbl}
\usepackage{scalerel,stackengine}
\usepackage{tabularx}
\usepackage{multicol}
\usepackage{multirow}
\usepackage{multirow}
\usepackage{graphicx}
\usepackage{multirow}
\usepackage{tabularx}
\usepackage{threeparttable}
\usepackage{array}
\usepackage{balance}
\newtheoremstyle{remarkstyle}
  {2pt} 
  {2pt} 
  {}    
  {}    
  {\bfseries} 
  {.}   
  {.5em} 
  {}    

\theoremstyle{remarkstyle}
\newtheorem{remark}{Remark}
\usepackage[ruled,vlined,noresetcount]{algorithm2e} 

\makeatletter
\newenvironment{Procedure}[1][!t]{%
    \begin{algorithm}[#1]%
}{%
    \end{algorithm}
}
\makeatother

\title{A Bayesian Adaptive Spectral Surrogate Model for Efficient Probabilistic Optimal Power Flow Evaluation}
\author{Xiaoting Wang,~\IEEEmembership{Member,~IEEE},  Xiaozhe Wang,~\IEEEmembership{Senior Member,~IEEE},
Gregory Kish,~\IEEEmembership{Senior Member,~IEEE}, and Yunwei (Ryan) Li,~\IEEEmembership{Fellow,~IEEE} 
\thanks{This work was supported in part by the Natural Sciences and Engineering Research Council of Canada (NSERC) Discovery Grant RGPIN-2022-03236, in part by the Canada Research Chairs Program under Grant CRC-2023-00006, in part by the Major Innovation Fund (MIF) from the Government of Alberta, and in part by Future Energy Systems at the University of Alberta, funded by the Canada First Research Excellence Fund (CFREF).}}
\date{January 2024}

\begin{document}

\maketitle

\begin{abstract}
This paper presents an adaptive stochastic spectral embedding (ASSE) method to solve the probabilistic AC optimal power flow (AC-OPF), a critical aspect of power system operation. The proposed method can efficiently and accurately estimate the probabilistic characteristics (e.g., mean, variance, median, and 
 quantile-based metrics) of AC-OPF solutions while minimizing power losses. Based on estimated AC-OPF decisions (i.e., generator outputs), the 
 confidence interval (CI)-based production cost index can be determined. Specially, an adaptive domain partition strategy is adopted to guide refinement domain selection and partition. The Bayesian compressive sensing-based coefficient calculation algorithm is integrated to enhance its performance. Numerical studies on modified IEEE 9-bus and IEEE 118-bus systems demonstrate that the proposed ASSE method offers accurate and fast evaluations compared to Monte Carlo simulations. Comparisons with a sparse polynomial chaos expansion, Gaussian process regression,  and deep neural networks,    
further illustrate its efficacy in accurately assessing the responses with strongly localized behavior and non-symmetric distributions, providing practical
decision-making bounds for generator outputs and operating
costs under uncertainty.
\end{abstract}



\section*{Nomenclature}

\subsection*{Power System Variables}
\begin{tabular}{l p{0.68\columnwidth}}
$P_{\mathrm{loss}}$ &
Total active power transmission loss. \\

$P_{G_i},\,Q_{G_i}$ &
Active and reactive power outputs of generator $i$. \\

$\bm{x}$ &
State vector of bus voltage magnitudes $\bm{V}$ and angles $\bm{\theta}$. \\

$\bm{u}$ &
Decision variable vector of the optimal power flow problem. \\

$C_{i,g},g=0,1,2$ &
Generation cost coefficients of generator $i$.\\

$C^{\mathrm{cost}}_{\mathrm{low}}$ &
Lower cost bound calculated based on the 5\% quantile of $P_{G_i}$.  \\

$C^{\mathrm{cost}}_{\mathrm{up}}$ &
Upper cost bound calculated based on the 95\% quantile of $P_{G_i}$. \\
\end{tabular}

\vspace{1ex}

\subsection*{Uncertainty Modeling}
\begin{tabular}{l p{0.70\columnwidth}}
$\bm{\zeta}$ &
Vector of random inputs (e.g., wind speed, solar irradiation, load). \\

$\bm{Z}$ &
Stochastic response of the power system. \\

$F(\cdot)$ &
Stochastic input--output mapping, $\bm{Z}=F(\bm{\zeta})$. \\

$\mathcal{D}_{\bm{\zeta}}$ &
Entire random input domain. \\

$\mathcal{D}_{\bm{\zeta}}^{\bm{\alpha}}$ &
Subdomain of $\mathcal{D}_{\bm{\zeta}}$ indexed by $\bm{\alpha}$. \\
\end{tabular}

\vspace{1ex}

\subsection*{Polynomial Chaos and Domain Decomposition}
\begin{tabular}{l p{0.70\columnwidth}}
${\mathcal{R}}_{\mathrm{S}}^{\bm{\alpha}}(\bm{\zeta})$ &
True residual on subdomain $\mathcal{D}_{\bm{\zeta}}^{\bm{\alpha}}$. \\

$ {\mathcal{R}}_{\mathrm{PCE}}^{\bm{\alpha}}(\bm{\zeta})$ &
Local polynomial chaos expansion approximation of ${\mathcal{R}}_{\mathrm{S}}^{\bm{\alpha}}(\bm{\zeta})$. \\
$\bm{\alpha}=(k,d)$ &Index pair identifying the expansion level $k$ and subdomain $d$. \\

$k$ &
Expansion level index. \\

$K$ &
Total expansion level. \\

$d$ &
Subdomain index at level $k$. \\

$D_k$ &
Number of subdomains at expansion level $k$. \\

$\mathcal{A} $ &
Set of all subdomain indices. \\

$\mathcal{J}$ &
Set of terminal (unsplit) subdomain indices. \\
$M$ & Total number of local polynomial chaos expansion terms on $\mathcal{D}_{\bm{\zeta}}^{\bm{\alpha}}$. \\
$c_{m}^{{\bm{\alpha}}}$ & Local polynomial chaos expansion coefficients  on
$\mathcal{D}_{\bm{\zeta}}^{\bm{\alpha}}$, $m=\{1,\cdots,M\}$. \\
$\widehat{\bm{C}}^{\bm{\alpha}}$ &
Local polynomial chaos expansion coefficient matrix on
$\mathcal{D}_{\bm{\zeta}}^{\bm{\alpha}}$. \\
$\Psi_{m}^{{\bm{\alpha}}}(\bm{\zeta})$ & Local multivariate orthonormal polynomial bases on $\mathcal{D}_{\bm{\zeta}}^{\bm{\alpha}}$. \\
$\bm{\Phi}^{\bm{\alpha}}$ &
Local polynomial basis matrix. \\
$N_{\mathrm{ref}}$ &
Minimum number of training samples required in each subdomain for constructing the local residual approximation. \\
$N_{\bm{\alpha}}$  & Number of samples in subdomain $\mathcal{D}_{\bm{\zeta}}$ \\
$\mathcal{K}$ &
Number of folds in cross-validation. \\
$s $ & Current terminal/unsplit subdomain index. $\mathcal{D}^{(k,s)}$ is the corresponding current terminal subdomain.\\
\end{tabular}
\section{Introduction}
AC Optimal power flow (AC-OPF) is critical for power system operations, supporting 
security assessment \cite{Wang2024}, economic dispatch \cite{Hu2021}, and resilience enhancement \cite{Liu2020}. Yet today's grid faces mounting uncertainty: the deep penetration of renewable energy and the increase in electric vehicle adoption create highly variable generation and demand profiles.
Traditional deterministic AC-OPF methods 
often struggle to adequately manage these uncertainties. To address this issue, %
variants of the OPF formulation have been developed, generally falling into two categories: (i)  formulations aiming to determine a single (point) schedule that remain feasible under uncertainty, 
such as robust OPF \cite{Lorca2015} and stochastic OPF \cite{Ruiz2009}; and (ii) probabilistic OPF, which 
explicitly focused on predicting the full probability distributions of optimal decision variables themselves \cite{Zou2014}. Knowledge of the complete distribution of optimal solutions enables grid operators and market participants to quantify risks (e.g., tail costs and constraint violation probabilities) and perform robust budget planning, offering advantages that single-point solutions cannot provide. Therefore, this paper focuses on probabilistic OPF. 

Nevertheless, conventional methods for estimating the full OPF solution distributions typically rely on 
extensive Monte Carlo (MC) simulations \cite{Zou2014}, limiting their practical applications. To alleviate this computational burden, linearized OPF models \cite{Yang2018} and heuristic search strategies \cite{Onate2008, Niu2014} have been proposed. Though these methods speed up solving a single OPF instance, repeated runs for probabilistic analyses remain computationally costly. 

Recently, learning-based methods 
have emerged to accelerate OPF solutions, 
primarily predicting single-point schedules or active constraints under uncertainty.  
Several studies \cite{Liu2023OPF,Hasan2021,guha2019machine,Dong2020} employed deep neural network (DNN) methods to predict warm start points or identify active constraint sets, thereby speeding up conventional OPF solvers. For example, Liu et al. \cite{Liu2023OPF} integrated a DNN with a knowledge graph to learn active constraints, reducing computational time and enhancing adaptability to evolving conditions.  Dong et al. \cite{Dong2020} developed an NN framework generating high-quality initial solutions to accelerate numerical OPF simulations.  Other studies directly map input variables (e.g., load profiles, renewable outputs) to OPF solutions \cite{Pan2023,Gao2023,Falconer2022,Park2024}.  Pan et al. \cite{Pan2023} proposed a specialized loss function incorporating accuracy and power flow-based regularization for physical feasibility.
Gao et al. \cite{Gao2023} used a graph convolutional NN with physics-embedded kernels and iterative feature construction to capture network topology and operational constraints.
While these methods significantly improve computational efficiency, most of them focus on point-to-point estimations aimed at rapidly identifying feasible solutions or active constraints. They often require extensive training datasets and lack closed-form representations of the solutions’ probabilistic statistics.  %
In contrast, the primary objective of this paper is the probabilistic OPF, which explicitly estimates the full probability distributions of the optimal decision variables and is different from pointwise prediction. 

Analytical methods for probabilistic OPF %
(e.g., Cumulant method \cite{Schellenberg2005,Tamtum2009}) were also proposed, which 
%
improve efficiency by directly computing first- and second-order statistical moments 
of OPF solutions. However, these approaches often struggle with accurately capturing statistical moments for complex, non-Gaussian distributions.  
Alternatively, surrogate model techniques, such as Gaussian process regression (GPR) \cite{Pareek2021} and polynomial chaos expansion (PCE) \cite{Mühlpfordt2019,Ly2023}, offer accurate OPF solutions with limited training samples, providing closed-form statistics (means and variances) 
and probability distributions.  
For example, Pareek et al. \cite{Pareek2021} presented a GPR-based method reducing the required sample size, though GPR can perform poorly with noisy, multimodal distributions or complex global behavior \cite{Rajabi2019review}.
Mühlpfordt et al. \cite{Mühlpfordt2019} proposed a PCE-based approach that derives statistical moments directly from PCE coefficients, but efficiency degrades as the input dimensionality grows. 
To mitigate this challenge, adaptive sparse PCE (SPCE) \cite{Sheng2018} and Lite-PCE \cite{Ly2023} methods were developed.
Nonetheless, accurately capturing strongly localized, multimodal, or noisy distributions remains challenging without computationally costly high-order expansions.
 Indeed, decision-variable distributions in probabilistic OPF are not necessarily Gaussian; they can be multimodal, skewed, or heavy-tailed, complicating their statistical characterization.

More recently, stochastic spectral embedding (SSE) \cite{Marelli2021stochastic} emerged as an alternative by partitioning the stochastic space into local spectral surrogates, effectively modeling highly nonlinear behaviors. Chi et al. \cite{chi2024hybrid} applied the SSE to electromechanical transient studies. However, the original SSE method recursively refines all terminal regions, increasing computational costs unnecessarily in already-accurate domains. Wagner et al. \cite{wagner2021bayesian}  proposed SSLE as an extension of SSE  for Bayesian model inversion, where localized spectral expansions are used to approximate likelihood functions and recover posterior-related quantities such as the evidence, posterior moments, and posterior marginals. While SSLE provides an effective framework for Bayesian inversion, its objective is different from the forward uncertainty propagation setting considered in probabilistic OPF. 
To date, the SSE/SSLE's ability to estimate the full probability distributions  of probabilistic OPF solutions, including quantiles and tail behavior, critical for quantifying risk and informing robust budget planning, has not been systematically evaluated. 

To address the limitations of existing methods, this paper proposes an efficient adaptive stochastic spectral embedding (ASSE) framework specifically designed 
 
for accurate distributional estimation of probabilistic AC-OPF solutions. Different from point forecasting surrogates that primarily target sample-wise regression accuracy or mean prediction, ASSE is designed to approximate the full probabilistic behavior of OPF responses, including output distributions, quantiles, tail behavior, and decision-relevant confidence bounds. In addition,
ASSE integrates Bayesian compressive sensing (BCS), a modified $\mathcal{K}$-fold CV error, and Sobol'-guided adaptive partitioning to selectively refine locally complex regimes.
 
Leveraging advanced surrogate modeling techniques, the proposed ASSE simultaneously achieves computational efficiency and precise probabilistic characterization, providing practical decision-making bounds for generator outputs and operating costs under uncertainty. 
Simulation results on the IEEE 9-bus and 118-bus systems demonstrate the proposed ASSE’s effectiveness in 
accurately handling complex, non-symmetric, or tailed distributions with lower-order spectral representations. 
Its overall performance is compared to  representative surrogate models, 
SPCE \cite{Sheng2018} and GPR \cite{Pareek2021} (commonly used in small-sample settings) as well as a widely adopted deep-learning method (DNN) \cite{guha2019machine}. 
 
The contributions of the paper are summarized below. 
\begin{enumerate}
\item  For accurate distributional estimation of probabilistic AC-OPF solutions, we 
integrate the BCS algorithm and a modified $\mathcal{K}$-fold CV scheme into fitting each local polynomial chaos expansion, allowing adaptive parameter selection (e.g., polynomial order and truncation norm) and yielding stable fits even with a small training dataset.
\item We propose a modified refinement indicator that couples CV error with subdomain size and then combine this indicator with a Sobol’ index–guided partition strategy to refine only the most error-contributing terminal subdomain. This avoids indiscriminate partitioning 
and  unnecessary global refinement. 
\item  The proposed ASSE framework retains the analytical advantages of spectral surrogates by providing closed-form expressions for the mean and variance. 
By improving surrogate accuracy in localized and strongly non-Gaussian regions via adaptive  subdomain refinement and sparse local expansions, it enhances the accuracy of these statistics and enables efficient estimation of quantiles and full distributions, 
thus offering clear decision bounds for generator outputs and operating costs.

\end{enumerate}


The remainder of this paper is organized as follows. Section \ref{sec:AC-OPF} introduces the mathematical formulation of the probabilistic AC-OPF problem. Section \ref{sec:ASSE} elaborates on the ASSE representation to solve the probabilistic AC-OPF. Section \ref{sec:ASSE_ACOPF} presents a detailed ASSE-based approach for the probabilistic AC-OPF. Section \ref{sec: Simulation} shows the numerical studies. Section \ref{sec:cons} gives the conclusions. Appendix \ref{sec:append_bcs} provides details of the Bayesian compressive sensing-based coefficients calculation.  Appendix~\ref{appen:evaluation} summarizes the evaluation metrics used in this paper. Appendix~\ref{app:training_time} provides supplementary results for the case studies. Appendix~\ref{re:101_dimension} discusses the high-dimensional test case.

\section{The Probabilistic AC-OPF  Problem} \label{sec:AC-OPF}



 Consider a power system with $N_b$ buses and $N_g$ conventional generators. Although the general AC-OPF problem aims to optimize generator power outputs and bus voltages while minimizing power generation costs without violating physical and operational constraints \cite{Gao2023physics}, alternative objectives can sometimes be selected. In this paper, minimizing the power loss 
is set as a priority to enhance the voltage profile and reduce thermal loading, thereby improving the efficiency and reliability of the transmission network.
  We model uncertainty through the random input vector $\bm{\zeta}$, which captures variability in renewable generation and load (e.g., wind speed, solar irradiation, and demand). 
 Then, the AC-OPF with the objective of minimizing the power loss while considering the uncertainties is defined as \cite{Leeton2010}: 
\begin{subequations}
\label{eq:DOPF}
\setlength{\abovedisplayskip}{2pt}
\setlength{\belowdisplayskip}{2pt}
\begin{align}
    P_{\mathrm{loss}} & =   \min_{\bm{u}}  \sum_{l_b\in \mathbb{B}} P_{l_b}(\bm{\zeta})  \label{eq:objectives} \\
    \mathrm{s.t.} &  \quad  {\bm{f}}\left( \bm{x}(\bm{\zeta}), \bm{u}(\bm{\zeta})\right) =0 
    \label{eq:powerflow}\\
    &  \quad  \bm{h}(\bm{x}(\bm{\zeta}), \bm{u}(\bm{\zeta})) \leq 0\label{eq:detinequalitys} 
 \end{align} 
 \end{subequations}
where the objective \eqref{eq:objectives} denotes minimum total power transmission loss $P_{\mathrm{loss}}$, with $P_{l_b}(\bm{\zeta}) = g_{ij}[(V_i(\bm{\zeta}))^2+(V_j(\bm{\zeta})^2-2V_{i}(\bm{\zeta})V_{j}(\bm{\zeta})\cos\theta_{ij}(\bm{\zeta})]$ being the power loss at each branch $l_b$ and $\mathbb{B}$ being the branch index set. $g_{ij}$ denotes the conductance of line $l_{b}$ connecting bus $i$ and bus $j$, $V_i(\bm{\zeta})$ denotes the voltage magnitude, $\theta_{ij}(\bm{\zeta})$ denotes the voltage angle difference between bus $i$ and $j$. Constraints \eqref{eq:powerflow} denote the power flow equations 
with the  state variable vector $\bm{x}(\bm\zeta) = [\bm{V}(\bm{\zeta}),\bm{\theta}(\bm{\zeta})]$ and the decision variable vector $ \bm{u}(\bm{\zeta}) = \{{P_{G_i}}(\bm{\zeta}),{Q_{G_i}}(\bm{\zeta}),V_i(\bm{\zeta})| i \in \mathbb{G}\}$; 
$\mathbb{G}$ is the generator bus index set.  Constraints \eqref{eq:detinequalitys} indicate the inequality 
constraints under uncertainties, including the generation capacity limits, voltage limits at each bus, and branch flow limits. 

Growing uncertainties $\bm{\zeta}$ introduce variability into the objectives and solutions of AC-OPF, making their probabilistic characteristics (mean, variance, median, and distributions) crucial for power system operations. Once these statistics for variables such as $P_{\mathrm{loss}}$, $\bm{u}(\bm{\zeta})$, and power flows are determined, the corresponding production cost can be computed. 
Let $\mathrm{Index}[P_{G_i}(\bm{\zeta})]$ denote the statistics of $P_{G_i}(\bm{\zeta})$ distribution (e.g., mean or certain confidence interval ( CI)). The production cost based on certain statistics can be calculated as:
\begin{equation}
\setlength{\abovedisplayskip}{3pt}
\setlength{\belowdisplayskip}{3pt}
\label{eq:Q_index}
     {C}^{\mathrm{cost}}_{\mathrm{Index}}(\bm{\zeta}) 
    =  C_{i,2}\mathrm{Index}[{P_{G_i}}]^2 + C_{i,1}\mathrm{Index}[P_{G_i}] + C_{i,0}
\end{equation}
where $  {C}^{\mathrm{cost}}_{\mathrm{Index}}(\bm{\zeta}) $ 
represents the production cost associated with $ P_{G_i} $ under a specific index. The parameters $ C_{i,g}, g \in \{0,1,2\} $ denote the generation cost coefficients for generator $ i $. 
For example, in this paper $  {C}^{\mathrm{cost}}_{\mathrm{low}}$ 
is defined as the  value calculated based on  5\%  quantile  
of $ P_{G_i} $, while $  {C}^{\mathrm{cost}}_{\mathrm{up}}$ 
corresponds to the  95\%  quantile  
of $ P_{G_i} $, representing the lower and upper bounds of the production cost, respectively, while accounting for uncertainties.  By tracking the generator production costs \eqref{eq:Q_index} that result from the loss-optimized formulation \eqref{eq:DOPF}, one can determine a budget plan aligned with minimal power loss. This integrated perspective highlights how minimal‐loss strategies can be balanced against operational budgets and regulatory requirements (e.g., allowable levels of system losses), providing important insights for both system planners and market regulators.
\section{The Adaptive Stochastic Spectral Embedding Method} \label{sec:ASSE}
\begin{figure}[ht!]
\centering
\setlength{\abovecaptionskip}{-4pt}
\setlength{\belowcaptionskip}{-4pt}
\includegraphics[width=0.43\textwidth]{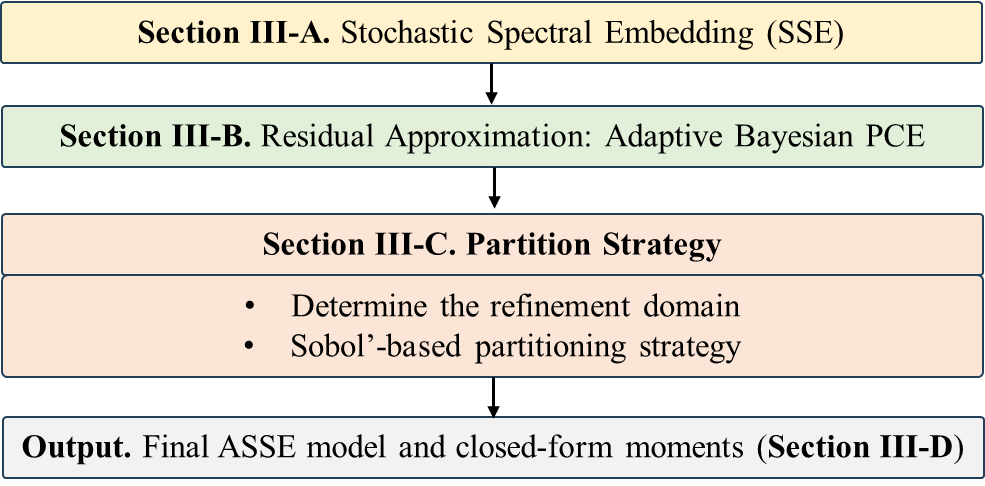}
\caption{ High-level structure of the proposed ASSE algorithm, including the general representation of SSE (Section~\ref{sec:SSE_representation}), residual approximation via adaptive Bayesian PCE (Section~\ref{sec:ABPCE}), sensitivity-guided partition strategy (Section~\ref{sec:partition}), and the final ASSE surrogate with closed-form moment evaluation (Section~\ref{sec:closed-form_statistics}).
}
\label{fig:SSE_algorithm}
\vspace{-10pt}
\end{figure}

%
This section introduces 
the ASSE method, %
which leverages the SSE technique to sequentially expand residuals through adaptive Bayesian PCE, creating an efficient surrogate model for probabilistic AC-OPF problems.
Fig.~\ref{fig:SSE_algorithm} presents a  high-level structure 
of the ASSE algorithm and its key components. %
The ASSE is constructed from a general SSE representation introduced in Section~\ref{sec:SSE_representation}. It 
 mainly involves two key components: residual approximation (Section \ref{sec:ABPCE}) and partition strategy (Section \ref{sec:partition}). The residual approximation employs adaptive Bayesian PCE models to achieve accurate expansions within subdomains, ensuring global convergence and preserving closed-form statistical properties. The partition strategy adaptively selects the subdomain to refine based on residual errors and employs Sobol' sensitivity indices to determine optimal splitting directions. 

\subsection{The SSE Representation} \label{sec:SSE_representation}
Denote the probabilistic AC-OPF formulation \eqref{eq:DOPF} by $\bm{Z}=F(\bm{\zeta})$, where $\bm{\zeta} \in \mathbb{R}^{\mathcal{M}_{\mathrm{in}}}$ is the system random inputs (e.g., wind speed, load power), and $\bm{Z}$ could be any individual or group of stochastic responses, including the solutions of \eqref{eq:DOPF} (e.g., the decision variables $\bm{u}$, state variables  $\bm{x}$, 
the objective function $P_{\mathrm{loss}}$, and the power flow). Assume all random variables (responses $\bm{Z}$ and inputs $\bm{\zeta}$) have finite second-order moments (e.g., $\mathbb{E}[{\bm{Z}}^2] < + \infty$), 
the SSE 
method aims to build a sequence of spectral expansions, each with a manageable level of complexity, across successively smaller subdomains ${\mathcal{D}_{\bm{\zeta}}^{\bm{\alpha}}}$ within the entire input domain ${\mathcal{D}}_{\bm{\zeta}}$. Following this approach, %
we approximate the AC‑OPF mapping $\bm{Z}=F(\boldsymbol{\zeta})$ \eqref{eq:DOPF} with a SSE representation constructed through the sequential expansion of residuals $\widehat{\mathcal{R}}_{\mathrm{S}}^{\bm{\alpha}}(\bm{\zeta})$ over the subdomains $\mathcal{D}^{\bm{\alpha}}_{\bm{\zeta}}$ \cite{Marelli2021stochastic}:
%
\begin{equation}
\setlength{\abovedisplayskip}{2pt}
\setlength{\belowdisplayskip}{3pt}
\label{eq:SSE}
\bm{Z} = {F}(\bm{\zeta}) \approx 
{F}_{\mathrm{SSE}}(\bm{\zeta})=\sum_{\bm{\alpha} \in \mathcal{A}} \bm{1}_{\mathcal{D}_{\bm{\zeta}}^{\alpha}}(\bm{\zeta}) \widehat{\mathcal{R}}_{\mathrm{S}}^{\alpha}(\bm{\zeta}) 
\end{equation}
%
 where the index pairs $\bm{\alpha} = (k,d) \in \mathcal{A} \subseteq \mathbb{N}^{2}$, i.e.,  $\mathcal{D}_{\bm{\zeta}}^{\bm{\alpha}} = \mathcal{D}_{\bm{\zeta}}^{k,d}$. For the expansion level $k \in \{0,\dots,K\}$, the subdomain index is $d \in \{1,\dots,D_k\}$. For classical SSE the number of subdomains at level $k$ is $D_k = 2^{k}$.
$\bm{1}_{\mathcal{D}_{\bm{\zeta}}^{\alpha}}$ is a subdomain index function with $\bm{1}_{\mathcal{D}_{\bm{\zeta}}^{\alpha}} = 1 $ when evaluated within the $\alpha$th subdomain and takes the value $0$ in all other regions.  

\begin{figure}[ht!]
\centering
\vspace{-8pt}
\setlength{\abovecaptionskip}{-8pt}
\setlength{\belowcaptionskip}{-4pt}
\includegraphics[width=0.45\textwidth]{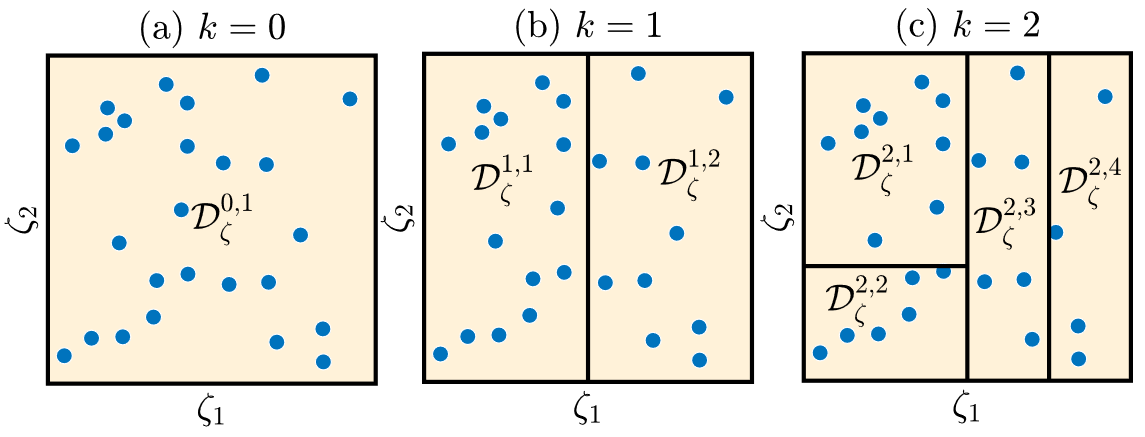}
\caption{Example of classical SSE domain partition for different  expansion level $k$ and subdomain index $d$ with maximum level $K = 2$ and $D_k =4$ subdomains. 
} 
\label{fig:SSE_Domain}
\end{figure}

Fig. \ref{fig:SSE_Domain} presents examples of subdomains for different $\bm{\alpha} = (k,d)$. Specially, when $k=0$, $\mathcal{D}^{0,1}_{\bm{\zeta}}$ is the entire input domain $\mathcal{D}_{\bm{\zeta}}$ and the residual ${R^{0}_{\mathrm{S}}}(\bm{\zeta}) = F(\bm{\zeta})$. As $k$ increases, the total expansion level grows, and the number of subdomains $D_k$ expands at an exponential rate. %
To construct the SSE-based model \eqref{eq:SSE}, the first important aspect is to approximate the residuals ${R^{k}_{\mathrm{S}}}(\bm{\zeta})$ over subdomains. Fundamentally, any spectral expansion technique could be used to approximate the residuals ${R^{k}_{\mathrm{S}}}(\bm{\zeta})$. Specially, this paper adopts the adaptive Bayesian PCE method to approximate the residuals, i.e., $\widehat{\mathcal{R}}_{\mathrm{S}}^{\alpha}(\bm{\zeta}) = \mathcal{R}_{\mathrm{PCE}}^{\bm{\alpha}}(\bm{\zeta})$, where $\mathcal{R}_{\mathrm{PCE}}^{\bm{\alpha}}(\bm{\zeta})$ denotes the PCE-based model constructed in each subdomain. Details will be discussed in Section~\ref{sec:ABPCE}.  %
Incorporating the adaptive model enables the original SSE representation with the ability 
to adaptively adjust its key parameters. This enhanced method is hereafter referred to as adaptive SSE (ASSE).     
The other critical element to construct 
the ASSE-based model is 
to determine a refinement domain for partitioning. 
This involves distinguishing between domains that have already been split and terminal domains, which remain intact without any splits. Let $\mathcal{J}$ denote the set of indices for terminal domains (i.e., unsplit domain index set). With this, the ASSE as expressed in \eqref{eq:SSE} can be restructured as follows:
 \begin{equation}
 \setlength{\abovedisplayskip}{2pt}
\setlength{\belowdisplayskip}{2pt}
\label{eq:SSE_sep}
{F}_{\mathrm{SSE}}(\bm{\zeta})=\sum_{\bm{\alpha} \in \mathcal{A}\backslash{\mathcal{J}}} \bm{1}_{\mathcal{D}_{\bm{\zeta}}^{\bm{\alpha}}}(\bm{\zeta}) \widehat{\mathcal{R}}_{\mathrm{S}}^{\alpha}(\bm{\zeta}) + \sum_{\bm{\alpha} \in {\mathcal{J}}} \bm{1}_{\mathcal{D}_{\bm{\zeta}}^{\bm{\alpha}}}(\bm{\zeta}) \widehat{\mathcal{R}}_{\mathrm{S}}^{\bm{\alpha}}(\bm{\zeta})
\end{equation}
where $\mathcal{A}\backslash{\mathcal{J}}$ denotes the index set for domains that have already been split. The entire input domains can be constructed as the union of all terminal domains:  $\mathcal{D}_{\bm{\zeta}} = \cup_{\bm{\alpha}\in \mathcal{J}} \mathcal{D}_{\bm{\zeta}}^{\bm{\alpha}}$. For $k=0$, the terminal domain $\mathcal{J}$ will be the entire domain $\mathcal{D}_{\bm{\zeta}}$. 
Detailed partition strategy to select the refinement domain will be elaborated in Section~\ref{sec:partition}.

\subsection{The Adaptive Bayesian  PCE-based Residual Expansions} \label{sec:ABPCE}
Various spectral surrogate modeling methods (e.g., Fourier series, polynomial chaos expansion (PCE)) \cite{Xiu2002wiener} can be applied to approximate the residual expansions $\widehat{\mathcal{R}}_{\mathrm{S}}^{\bm{\alpha}}(\bm{\zeta})$. Among these, we select the PCE method because it requires fewer sample evaluations, provides analytical solutions for statistical properties, and efficiently quantifies uncertainties %
\cite{blatman2009adaptive}. 
%
Then, the PCE  to approximate the residuals 
${\mathcal{R}}_{\mathrm{S}}^{\bm{\alpha}}(\bm{\zeta})$ over 
each subdomain 
\cite{Xiu2002wiener} can be represented as:  
\begin{equation}
\setlength{\abovedisplayskip}{2pt}
\setlength{\belowdisplayskip}{2pt}
\label{eq:PCE}
\widehat{\mathcal{R}}_{\mathrm{S}}^{\bm{\alpha}}(\bm{\zeta}) = {\mathcal{R}}_{\mathrm{PCE}}^{\bm{\alpha}}(\bm{\zeta}) = \sum_{m=1}^{M} c_{m}^{{\bm{\alpha}}}\Psi_{m}^{\bm{\alpha}}(\bm{\zeta})  
\end{equation}
%
with $M$ number of unknown PCE coefficients $c_{m}^{{\bm{\alpha}}}$ and  multivariate orthonormal polynomial bases $\Psi_{m}^{{\bm{\alpha}}}(\bm{\zeta})$ at subdomain $\mathcal{D}_{\bm{\zeta}}^{\bm{\alpha}}$.

%
The construction of residual expansion (i.e., the PCE model \eqref{eq:PCE}) includes two key steps: building the multivariate polynomial bases $\Psi_{m}^{\bm{\alpha}}$ and calculation of the residual expansion coefficients $c_{m}^{\bm{\alpha}}$. Typically, $\Psi_{m}^{\bm{\alpha}}$ can be constructed via the tensor product of the univariate orthonormal polynomial bases integrated with standard truncation scheme ((10) in \cite{wang2022}) or sparse truncation scheme (i.e., $q$-norm truncation (22) \cite{wang2022}), which is applied to improve the calculation efficiency. While the univariate orthonormal polynomials can be selected based on known probability distributions (see Table 4.1 \cite{Xiu2002wiener}), or moment-based methods (see Section III-B \cite{Wang2021} ) and Stieltjes procedure \cite{Xu2019PPF}). Specifically, this paper adopts the discretized Stieltjes procedure (three-term recurrence relation \cite{Xu2019PPF,Sheng2018}) to construct univariate orthonormal polynomial bases for input variables with arbitrary marginal distributions \cite{Sheng2018}.  Importantly, the univariate polynomials generated by the discretized Stieltjes procedure are orthonormal with respect to the restricted probability measure on each subdomain $\mathcal{D}_{\bm{\zeta}}^{\bm{\alpha}}$ rather than global marginals. The associated local inner product is defined by integration over $\mathcal{D}_{\bm{\zeta}}^{\bm{\alpha}}$ with the restricted density. In practice, this inner product is evaluated using a local sample-based quadrature over the samples falling within $\mathcal{D}_{\bm{\zeta}}^{\bm{\alpha}}$. This enables the ASSE framework to handle various uncertain sources (e.g., wind speed, solar irradiance, and load), which follow different distribution families, by building a dedicated orthonormal basis for each input and combining these into the multivariate basis
functions $\Psi_m^{\bm{\alpha}}$ in~\eqref{eq:PCE}.

Next, we employ the Bayesian compressive sensing algorithm with a modified $\mathcal{K}$-Fold cross-validation scheme 
to calculate the coefficients $c_{m}^{\bm{\alpha}}$, 
allowing for adaptively choosing the PCE model parameters (e.g., PCE order, truncation norm) %
to promote sparsity while maintaining model accuracy.  

\noindent \textbf{Bayesian Compressive Sensing-based Coefficients Calculation:}
After constructing polynomial basis functions $\Psi_{m}^{\bm{\alpha}}(\bm{\zeta})$, the corresponding coefficients $c_{m}^{\bm{\alpha}}$ are calculated using regression methods %
such as ordinary least squares (OLS), least angle regression (LAR), and Bayesian compressive sensing (BCS). 
 
Specifically, BCS, a sparse regression technique, is employed due to its effectiveness in selecting optimal polynomial bases and determining coefficients for high-dimensional problems with limited data %
\cite{luthen2021sparse}. 

Given a training input dataset $\bm{\zeta}_{\mathrm{ED}}= \{\bm{\zeta}^{(1)},\dots,\bm{\zeta}^{(N_{\mathrm{ED}})}\}\in \mathbb{R}^{N_{\mathrm{ED}}\times \mathcal{M}_{\mathrm{in}}}$ and corresponding responses 
$\bm{Z}_{\mathrm{ED}} = \{\bm{Z}^{(1)},\dots,\bm{Z}^{(N_{\mathrm{ED}})}\}\in \mathbb{R}^{N_{\mathrm{ED}}\times \mathcal{M}_z}$, where $N_{\mathrm{ED}}$ is the initial domain sample size (level $k=0$), $\mathcal{M}_{\mathrm{in}}$ is the input dimension, and $\mathcal{M}_z$ is the output dimension (e.g., scalar or vector outputs), the coefficients on each subdomain $\mathcal{D}_{\bm{\zeta}}^{\bm{\alpha}}$ are computed via a compressive sensing optimization problem:  
%
\begin{equation}
\setlength{\abovedisplayskip}{2pt}
\setlength{\belowdisplayskip}{2pt}
    \label{eq:coe_optmization}
    \widehat{\bm{C}}^{\bm{\alpha}} = \arg \min_{\bm{C}^{\bm{\alpha}}} \left\{\|\bm{Z}^{\bm{\alpha}} - \bm{\Phi}^{\bm{\alpha}}\bm{C}^{\bm{\alpha}}\|^{2}_2 + \tau \|\bm{C}^{\bm{\alpha}}\|_{1}\right\},
\end{equation}
where $\bm{C}^{\bm{\alpha}} \in \mathbb{R}^{M\times \mathcal{M}_z}$ is the coefficient matrix, $\bm{\Phi}^{\bm{\alpha}}\in\mathbb{R}^{N_{\alpha}\times M}$ is the local polynomial basis matrix with $\bm{\Phi}^{\bm{\alpha}}_{nm} = \Psi_{m}^{\bm{\alpha}}(\bm{\zeta}^{(n)})$, $n=1,\dots,N_{\alpha}$, $m=1,\dots,M$, and $\tau$ is a sparsity-controlling hyperparameter. %
Particularly, $N_{\alpha}$ is the number of training samples in each subdomain $\mathcal{D}_{\bm{\zeta}}^{\bm{\alpha}}$.   %
We then solve the optimization problem \eqref{eq:coe_optmization} in a Bayesian framework (Appendix \ref{sec:append_bcs}) to obtain coefficients $\bm{C}^{\bm{\alpha}}$,  from which the residual expansion in each subdomain can be readily computed. 

\noindent\textbf{Modified $\mathcal{K}$-Fold Cross-Validation:} 
%
%
The accuracy of the BCS algorithm for PCE coefficients calculation is based on $\mathcal{K}$-fold cross-validation (CV).  However, conventional CV errors may become overly optimistic when the number of retained basis functions is comparable to the number of available samples, which is a common situation in local PCE construction on small subdomains. In such cases, the raw CV error tends to favor over-parameterized models and provides an unreliable stopping criterion. To account for model complexity and sample-size effects, %
we introduce a multiplicative correction factor $F_{\mathrm{mod}}$ into the CV error $e_{\mathrm{cv}}$.  The resulting modified CV error, denoted as $e_{\mathrm{mcv}}$,  penalizes overly complex local models and  is used as a stopping criterion. 

%
\begin{equation}
\setlength{\abovedisplayskip}{2pt}
\setlength{\belowdisplayskip}{2pt}
\label{eq:emcv}
 e_{\mathrm{mcv}} = \underbrace{\frac{N_{\alpha}}{N_{\alpha} -M}\left(1+{\mbox{tr}\left[\left([\bm{\Phi^{\alpha}}]^\top{\bm{\Phi^{\alpha}}}\right)^{-1}\right]}\right)}_{F_{\mathrm{mod}}} e_{\mathrm{cv}} 
\end{equation}
\normalsize
%
where $e_{\mathrm{cv}} $ denotes the $\mathcal{K}$-fold CV error:
\small
\begin{eqnarray}
\setlength{\abovedisplayskip}{0pt}
\setlength{\belowdisplayskip}{2pt}
\label{eq:ecv}
 e_{\mathrm{cv}} &=& \frac{1}{\mathcal{K}} \sum_{k_f =1}^{\mathcal{K}}e_{\mathrm{cv},k_f} \\ \label{eq:e_cv}
 e_{\mathrm{cv},k_f} &=& \frac{\frac{1}{N_{\alpha k}}\sum_{n =1}^{ N_{\alpha k}}\left[ Z^{\bm{\alpha},{(n)}} - {\mathcal{R}}_{\mathrm{PCE}}^{\bm{\alpha}}(\bm{\zeta}^{\bm{\alpha},(n)})\right]^2}{\frac{N_{\alpha}}{N_{\alpha}-1}\sum_{n=1}^{N_{\alpha}}[Z^{\bm{\alpha},(n)}-\hat{\mu}_Z]^2} \label{eq:cv_k}
\end{eqnarray}
\normalsize
$e_{\mathrm{cv},k_f}$ denotes the error evaluated for the points in the $k_f$-th fold, 
$k_f\in\{1,2,...\mathcal{K}\}$,  
and $N_{\alpha k}$ is the corresponding  fold size ($N_{\alpha k} =N_{\alpha }/\mathcal{K}$), with $N_{\alpha}$ being the total number of training samples in subdomain $\mathcal{D}_{\zeta}^{\alpha}$.
$\hat{\mu}_Z = \frac{1}{N_{\alpha}}\sum_{n=1}^{N_{\alpha}}Z^{\bm{\alpha},(n)}$ is the sample average; %
 The modification factor $F_{\mathrm{mod}}$ incorporated in \eqref{eq:emcv}  increases with the number of retained basis functions $M$ relative to
the available samples $N_{\alpha}$, thereby discouraging overfitting and improving
robustness in small-sample regimes.  Particularly, $F_{\mathrm{mod}}$ approaches $1$ as $N_{\alpha} \rightarrow {\infty}$ \cite{blatman2009adaptive}.  In practice, $\operatorname{tr}\!\big((\Phi^{\alpha\top}\Phi^{\alpha})^{-1}\big)$ is evaluated
using a pseudoinverse for numerical stability; when $M \ge N_{\alpha}$, we set
$F_{\mathrm{mod}}=\infty$ to exclude underdetermined local models.
Note that here $\mathcal{K}$ denotes the number of cross-validation folds, whereas the previously defined notation $K$ in Subsection \ref{sec:SSE_representation}  referred to the maximum expansion level.

%

\subsection{Partition Strategy} \label{sec:partition}
%
%
The other crucial component in building the ASSE model \eqref{eq:SSE} is selecting the appropriate domain to refine and deciding how to partition it. At each step of the algorithm, we perform two primary tasks: 1) refinement domain selection, i.e., choosing which current terminal (unsplit) domain $ \mathcal{J}$  to split; 2) partitioning strategy, i.e., deciding along which direction to split the selected subdomain.  By repeatedly refining subdomains where the response exhibits significant nonlinearity or discontinuity, we efficiently achieve more accurate local expansions via adaptive Bayesian PCE approximations ($\widehat{\mathcal{R}}_{\mathrm{S}}^{\bm{\alpha}}(\bm{\zeta}) = {\mathcal{R}}_{\mathrm{PCE}}^{\bm{\alpha}}(\bm{\zeta})$ \eqref{eq:PCE}). The detailed adaptive domain selection and partitioning strategy is described below.

\noindent  \textbf{Refinement Domain Selection:} At each iteration, 
one subdomain is chosen from the unsplit domain set $\mathcal{D}_{\bm{\zeta}}^\mathcal{J}$ (i.e., the terminal domains) to be further split.  
%
Denote $N_{\mathrm{ref}}$ as the minimum number of training points required in each subdomain to construct a local residual approximation ${\mathcal{R}}_{\mathrm{PCE}}^{k, d}$ using 
adaptive Bayesian PCE. 
This model can only be built when $N_{\bm{\alpha}} \geq N_{\mathrm{ref}}$, i.e., $\exists \widehat{\mathcal{R}}_{\mathrm{S}}^{k, d}$ when $ N_{\bm{\alpha}} \geq N_{\mathrm{ref}}$.  
Let $ \mathcal{D}_{\bm{\zeta}}^{k,s}$ with $s \in \{1,2...D_k\}$ be the current terminal subdomains.  To adaptively select the refinement domain,  we first design a refinement score $\chi^{\bm{\alpha}}, \bm{\alpha} =(k,s)$, based on the modified CV error $e_{\mathrm{mcv}}$:
%
%
%
%
 \begin{equation}
\setlength{\abovedisplayskip}{2pt}
\setlength{\belowdisplayskip}{2pt}
\label{eq:refine_score}
\chi^{k,s} = \begin{cases} 
e_{\mathrm{mcv}}^{k, s} {\mathbb{P}}^{k, s}, & \mathrm{ if } \quad  \exists \widehat{{R}}_S^{k, s}, \\ e_{\mathrm{mcv}}^{k-1, s} {\mathbb{P}}^{k, s},
&\mathrm{ otherwise}\end{cases}
\end{equation}
%
where 
${\mathbb{P}}^{k, s}$ denotes the empirical probability mass of subdomain $\mathcal{D}_{\bm{\zeta}}^{k,s}$, calculated as ${\mathbb{P}}^{k, s} = \frac{N_{\bm{\alpha}}}{N_{\mathrm{ED}}}$. 

Using \eqref{eq:refine_score}, %
the updated score $\chi^{k,s} $ incorporates the subdomain size %
$N_{\bm{\alpha}}$  (via the number of training samples in the subdomain) 
and the estimation accuracy 
(via the $\mathcal{K}$-fold CV error). 
The intuition is that denser subdomains with higher prediction error should be prioritized for refinement. 
%

A threshold $\tau_{\mathrm{thr}}$ is introduced to define the set of refinement candidates: 
\begin{equation}
\setlength{\abovedisplayskip}{2pt}
\setlength{\belowdisplayskip}{2pt}
\label{eq:threshold} 
\widetilde{\mathcal{J}}_k = \{(k,s)\in\mathcal{J}:\chi^{k,s}>\tau_{\mathrm{thr}}\}
\end{equation}
Only subdomains with refinement score exceeding $\tau_{\mathrm{thr}}$ remain candidates for splitting. Examples of the unsplit domain index set $\mathcal{J}$ and refinement candidate index set $\widetilde{\mathcal{J}}_k$ are provided in \textbf{Remark}~\ref{re:split_domains}.

 If $\widetilde{\mathcal{J}}_k = \emptyset$, or if no $\mathcal{D}_{\bm{\zeta}}^{k,s}$ has $N_{\bm{\alpha}}\geq N_{\mathrm{ref}}$, or if we have reached a maximum number of splits $K_{\mathrm{max}}$, the algorithm terminates. Otherwise, we select 
the refinement domain 
with the largest score ${\chi^{\bm{\alpha}}}$, i.e.,
\begin{equation}
\setlength{\abovedisplayskip}{2pt}
\setlength{\belowdisplayskip}{2pt}
\label{eq:refine}
 %
 {\bm{\alpha}}_\mathrm{refine}= \arg \max_{(k,s)\in\widetilde{\mathcal{J}}_k}\chi^{k,s}
\end{equation} 
where ${\bm{\alpha}}_\mathrm{refine}$ denotes the chosen refinement domain index. 

This refinement criterion helps mitigate heterogeneous local residual behavior, since highly nonlinear OPF responses, generator-limit activations, or poorly fitted tail regions tend to yield larger modified CV errors. By weighting this error with the empirical probability mass, ASSE prioritizes subdomains that contribute more to the overall distributional approximation error.

\begin{remark} \label{re:split_domains}
Illustrative example of $\mathcal{J}$ and  $\widetilde{\mathcal{J}}_k$ evolves during the algorithm. At $k=0$, the entire domain is terminal, $\mathcal{D}_{\bm{\zeta}}^\mathcal{J}=\mathcal{D}_{\zeta}^{0,1}$; after the first split $k=1$, the two resulting subdomains are $\{\mathcal{D}_{\bm{\zeta}}^{1,1},\mathcal{D}_{\bm{\zeta}}^{1,2}\}$, so
the unsplit set 
$\mathcal{D}_{\bm{\zeta}}^\mathcal{J}=\{\mathcal{D}_{\bm{\zeta}}^{1,1},\mathcal{D}_{\bm{\zeta}}^{1,2}\}$. Suppose in the next iteration ($k=2$), we split $\mathcal{D}_{\bm{\zeta}}^{1,1}$ into subdomains $\{\mathcal{D}_{\bm{\zeta}}^{2,1},\mathcal{D}_{\bm{\zeta}}^{2,2}\}$, while $\mathcal{D}_{\bm{\zeta}}^{2,1}$ remains unsplit. Then, the updated unsplit set $ \mathcal{D}_{\bm{\zeta}}^\mathcal{J} = \{\mathcal{D}_{\bm{\zeta}}^{1,2}, \mathcal{D}_{\bm{\zeta}}^{2,1}, \mathcal{D}_{\bm{\zeta}}^{2,2}\}$. However, if refinement score $\chi^{2,1}\leq \tau_{\mathrm{thr}}$, then the candidate refinement set for the next iteration becomes $\widetilde{\mathcal{J}}_{2}= \{\mathcal{D}_{\bm{\zeta}}^{1,2}, \mathcal{D}_{\bm{\zeta}}^{2,2}\}$.
\end{remark}
%

%
\noindent  \textbf{Sobol’ Index-Based Partitioning Strategy:} 
After the refinement domain is selected, we 
split it into two subdomains ($N_{\mathrm{split}}=2$) of equal probability mass along the input direction that maximizes the variability of $\widehat{\mathcal{R}}_{\mathrm{S}}^{\alpha}(\bm{\zeta})$. This involves two steps: 1) identifying the dominant splitting direction; 2) performing an equal-probability-mass division.  Leveraging the salient property of PCE, we effectively compute the first-order Sobol' index $\mathcal{S}_{r}^{\bm{\alpha}}$, a variance-based global sensitivity analysis measure, quantifying the contribution of each uncertain input $\zeta_r, r= \{1,\cdots,\mathcal{M}_{\mathrm{in}}\}$ and its associated variation $\mathrm{Var}[{\mathcal{R}}_{\mathrm{PCE}}^{\bm{\alpha}}({\zeta_r})]$ to the total residual variance $\mathrm{Var}[{\mathcal{R}}_{\mathrm{PCE}}^{\bm{\alpha}}(\bm{\zeta}) ]$.
Consequently, the dominant  splitting direction is chosen as the input  dimension  
with the largest first-order Sobol' index $\mathcal{S}_{r}^{k}$:
\begin{equation}
\setlength{\abovedisplayskip}{2pt}
\setlength{\belowdisplayskip}{2pt}
\label{eq:Sobol_based}
\mathbb{D}_{\mathrm{split}}^{\bm{\alpha}} =\underset{r \in\{1, \cdots, \mathcal{M}_{\mathrm{in}}\}}{\arg \max } \mathcal{S}_{r}^{\bm{\alpha}}  
\end{equation}
where $\mathcal{S}_{r}$ can be directly derived from the coefficients of the adaptive Bayesian PCE model \eqref{eq:PCE} when random inputs $\bm{\zeta}$ are mutually independent \cite{Sudret2008global}: 
\small
\begin{equation}
\setlength{\abovedisplayskip}{3pt}
\setlength{\belowdisplayskip}{3pt}
\label{eq:FirstOrderSI}
\mathcal{S}_{r}^{\bm{\alpha}}= \frac{\mathrm{Var}[\mathbb{E}[\mathcal{R}^{\bm{\alpha}}_{\mathrm{S}}|\zeta_r]]}{\mathrm{Var}[\mathcal{R}^{\bm{\alpha}}_{\mathrm{S}}]}\approx
\frac{\mathrm{Var}[{\mathcal{R}}_{\mathrm{PCE}}^{\bm{\alpha}}({\zeta_r}) ]}{\mathrm{Var}[{\mathcal{R}}_{\mathrm{PCE}}^{\bm{\alpha}}(\bm{\zeta}) ]}
=
\frac{\sum_{m\in \mathbb{I}_{r}} [{c^{\bm{\alpha}}_m}]^2}{\sum_{m=2}^M [{c^{\bm{\alpha}}_m}]^2}
\end{equation}
\normalsize
where $ \mathbb{I}_{r} = \{ m\in \{ 1,...,M\}: \beta_{r}^{m}\not= 0\}$ with $\beta_{r}^{m}$ being the PCE degree index. When random inputs are correlated, $\mathcal{S}_{r}^{\bm{\alpha}}$ can be evaluated through the covariance-based approach using the PCE model \cite{Wang2024gsa}.   Note that if multiple candidate directions yield identical (tied) first-order Sobol' indices, i.e., several $\mathcal{S}_{r}^{\bm{\alpha}}$ take the same value, we resolve the tie by selecting the smallest dimension index. For example, if $\mathcal{S}_{1}^{\bm{\alpha}}=\mathcal{S}_{2}^{\bm{\alpha}}$, then direction $r=1$ is selected. 

\begin{figure}[ht]
\vspace{-10pt}
\centering
\setlength{\abovecaptionskip}{-0.3cm}
\setlength{\belowcaptionskip}{-0.4cm}
\includegraphics[width=0.41\textwidth]{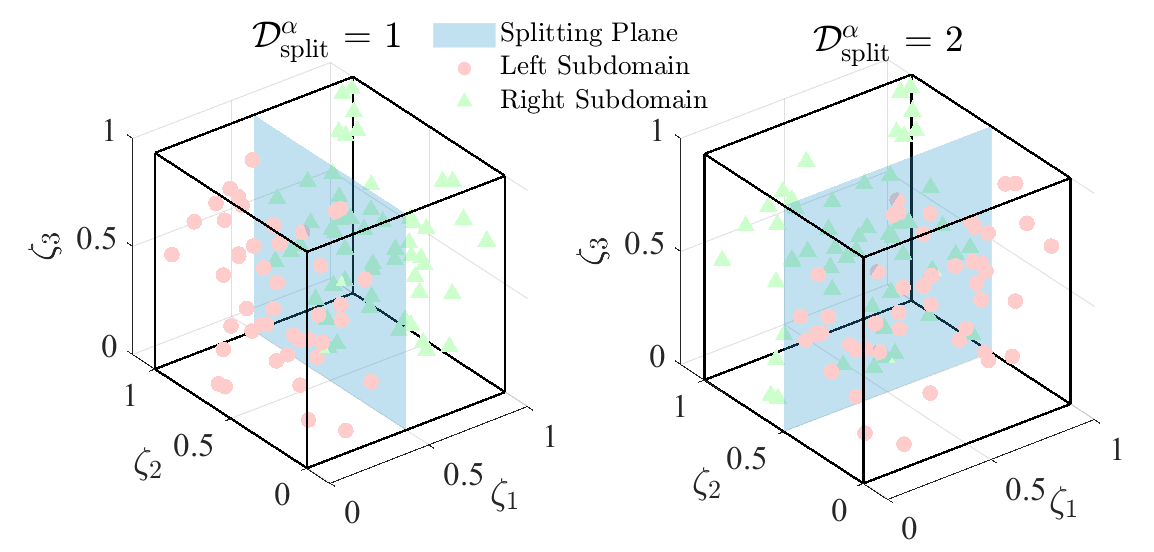}
\caption{Illustrative examples of Sobol’-index-based partitioning for a three-dimensional domain with independent, standard uniform inputs.  {Left:} Input $\zeta_1$ has the largest first-order Sobol’ index, thus the domain is split at $\zeta_1 = 0.5$. 
{Right:} Input $\zeta_2$ is dominant, and the split is performed at $\zeta_2 = 0.5$. 
} 
\label{fig:Domain_split_Dir}
\vspace{-12pt}
\end{figure}

To divide the subdomain with an equal probability mass along the selected dominant direction, we bisect the selected input dimension $\mathbb{D}_{\mathrm{split}}^{\bm{\alpha}}$ at its median (50\% quantile). The theoretical median is used if the input distribution is analytically known; otherwise, the empirical (sample) median is adopted. Only the chosen input dimension is split, while other dimensions remain unchanged.
Fig.~\ref{fig:Domain_split_Dir} illustrates the partitioning for a domain with three random inputs.

\begin{remark}

\label{re:asse_ssle}
 The proposed ASSE framework and SSLE~\cite{wagner2021bayesian} both rely on localized spectral expansions based on SSE, but they are designed for different objectives. SSLE targets Bayesian model inversion by approximating the likelihood function and recovering posterior quantities from the likelihood expansion coefficients. 
In contrast, ASSE targets forward uncertainty propagation in probabilistic AC-OPF by approximating the OPF response map $\bm Z=F(\bm\zeta)$, with the goal of estimating output probability distributions, quantiles, tail behavior, and decision-relevant bounds. Accordingly, their refinement mechanisms are different. 
SSLE focuses on localized likelihood-support regions, whereas ASSE refines subdomains with large modified CV error and non-negligible empirical probability mass. 
Moreover, ASSE selects the splitting direction using Sobol' indices derived from local adaptive Bayesian PCE residuals, enabling refinement along the most influential uncertain input direction. 
Together with BCS-based sparse local PCEs, this design allows ASSE to handle heterogeneous and non-exchangeable local approximation errors across OPF regimes such as  nonlinear power flow effects and tail operating conditions.  
\end{remark}

\subsection{Closed-Form Representation of Moments} \label{sec:closed-form_statistics}
%
 
Once an  accurate  ASSE model \eqref{eq:SSE} is constructed, the estimated mean values and variance of $\bm{Z}$ can be represented as:
\small
\begin{subequations}
\setlength{\abovedisplayskip}{2pt}
\setlength{\belowdisplayskip}{2pt}
\label{eq:moments}
\begin{align}
  & \mathrm{E}[\hat{\bm{Z}}] = \mathrm{E}[F_{\mathrm{SSE}}(\bm{\zeta})] = \sum_{\bm{\alpha}\in\mathcal{J}}c_{1}^{\bm{\alpha}}\mathbb{P}^{\alpha}  \label{eq:mean} \\
  & \mathrm{Var}[\hat{\bm{Z}}] =  \mathrm{Var}[F_{\mathrm{SSE}}(\bm{\zeta})] = \left( \sum_{\bm{\alpha}\in\mathcal{J}}\mathbb{P}^{\bm{\alpha}}\sum_{m=2}^{M}[c_{m}^{\bm{\alpha}}]^2\right) -  \mathrm{E}[\hat{\bm{Z}}]^2 \label{eq:variance}
\end{align}
\end{subequations}
\normalsize  
where $c_{1}^{\bm{\alpha}}$ denotes the constant term of PCE expansion in the terminal domain with $\mathcal{D}_{\bm{\zeta}}^{\bm{\alpha}}$.  
%

\section{The Proposed ASSE Framework for Probabilistic  AC-OPF} \label{sec:ASSE_ACOPF}
\begin{figure*}
\setlength{\abovecaptionskip}{-0.21cm}
\setlength{\belowcaptionskip}{-0.2cm}
\centering
\includegraphics[width=1.0\textwidth]{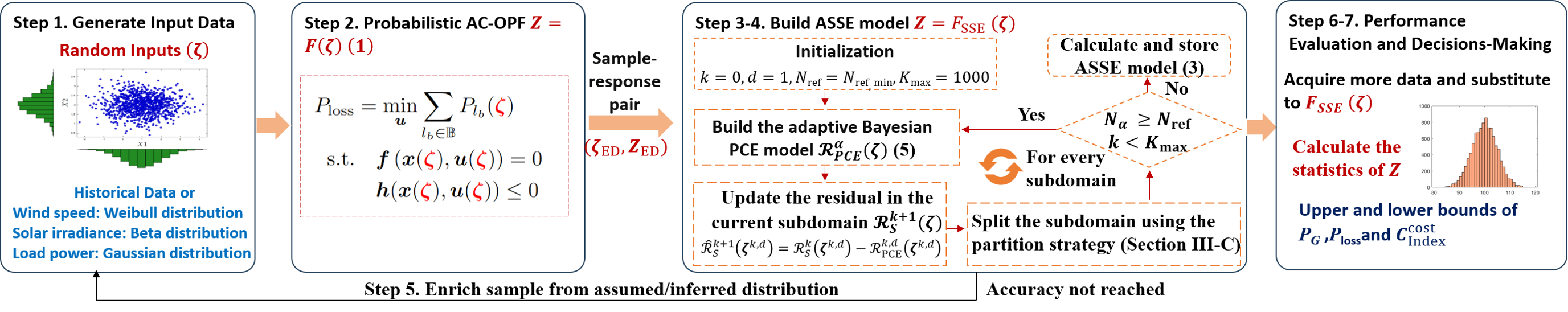}
\caption{Overview of the adaptive stochastic spectral embedding (ASSE) workflow for probabilistic AC–OPF.
\textbf{Step 1:} Acquire the input samples $\bm{\zeta}_{\mathrm{ED}}$ (see Remark~\ref{rem:remark1}). 
\textbf{Step 2:} Solve the probabilistic AC–OPF \eqref{eq:DOPF} for these samples to obtain responses $\bm{Z}_{\mathrm{ED}}$ and pass the pairs $(\bm{\zeta}_{\mathrm{ED}},\bm{Z}_{\mathrm{ED}})$ to ASSE. 
\textbf{Step 3-4:} Start from the full domain $\mathcal{D}_{\zeta}$, build local Bayesian-PCE residuals, compute refinement scores, and iteratively select and split refinement subdomains until the stopping criterion is met. 
\textbf{Step 5:} If the target accuracy is unmet, enrich the training set and repeat Steps 3-4. 
\textbf{Step 6-7:} Use the final surrogate to evaluate additional samples, yielding the desired probability distributions and the performance index $C_{\mathrm{Index}}^{\mathrm{cost}}$ via \eqref{eq:Q_index}.}
\label{fig:ASSE_overview}
\vspace{-12pt}
\end{figure*}
This section outlines the complete mechanism of the proposed ASSE method for solving the probabilistic AC-OPF problem \eqref{eq:DOPF}. Fig. \ref{fig:ASSE_overview} provides a general overview and \textbf{Procedure \ref{Procedure:ASSE}} details each step of the algorithm.
\begin{Procedure}
\small
\noindent\textbf{Step 1.} 
Input network data. Acquire $N_{\mathrm{ED}}$ realizations of random inputs $\bm{\zeta}_{p} = (\zeta^{(1)},\cdots,\zeta^{(N_{\mathrm{ED}})}) \in \mathbb{R}^{N_{\mathrm{ED}} \times \mathcal{M}_{\mathrm{in}}}$ ( e.g., wind speeds, solar radiation, and real load power) according to assumed probability distribution or historical data. \\
\noindent\textbf{Step 2.} 
Solve the probabilistic AC-OPF problem in \eqref{eq:DOPF} to obtain solutions $\bm{Z}_{\mathrm{ED}} = (\bm{Z}^{(1)},\cdots,\bm{Z}^{(N_{\mathrm{ED}})}$), e.g., through  Matpower \cite{Zimmerman2011}. Pass the training data set $[\bm{\zeta}_{\mathrm{ED}},\bm{Z}_{\mathrm{ED}}]$ to \textbf{Step 3.} \\
\noindent\textbf{Step 3.} Algorithm Initialization: \\
\begin{enumerate}
    \item[3a)] $k=0,d=1$, $K_{\mathrm{max}} = 1000$, $\tau_{\mathrm{thr}} = 10^{-5}$, and let $N_{\mathrm{ref}}$ be a user-defined constant, positive integer.  $\mathcal{D}^{0,1}_{\bm{\zeta}} = \mathcal{D}_{\bm{\zeta}}$, \\ \noindent ${\mathcal{R}}^{0,1}_{\mathrm{S}}(\bm{\zeta})$ = $F(\bm{\zeta})$. 
    \item[3b)] Construct the full adaptive Bayesian PCE model ${\mathcal{R}}_{\mathrm{PCE}}^{0,1}(\bm{\zeta})$ \eqref{eq:PCE}  for residual  ${\mathcal{R}}_{\mathrm{S}}^{0,1}(\bm{\zeta})$ in the entire domain, compute its refinement score $\chi^{0,1}$ via \eqref{eq:refine_score} and initialize  $\widetilde{\mathcal{J}}_0 = \{(0,1)\}$. 
    \item[3c)] Compute the new residual $\mathcal{R}^{1}(\bm{\zeta}) = {\mathcal{R}}_{\mathrm{S}}^{0,1}(\bm{\zeta}) - {\mathcal{R}}_{\mathrm{PCE}}^{0,1}(\bm{\zeta})$. 
\end{enumerate}    
\noindent\textbf{Step 4.} Adaptive refinement loop: \\
Repeated while $N_{\bm{\alpha}} \geq N_{\mathrm{ref}}$ and $k< K_{\mathrm{max}}$
\begin{enumerate}
    \item[4a)] Select the refinement domain via $(k,d) =\bm{\alpha}_{\mathrm{refine}}$ via \eqref{eq:refine}.
    \item[4b)] Split the current subdomain $\mathcal{D}^{k,d}_{\bm{\zeta}}$ into two 
    sub-parts 
    $\mathcal{D}_{\bm{\zeta}}^{k+1,\{s_1,s_2\}}$ ($s_1 $ and $s_2$ are indices of two resulting subdomains) by the Sobol' index-based partitioning \eqref{eq:Sobol_based}-\eqref{eq:FirstOrderSI}; update $\widetilde{\mathcal{J}}_{k+1} = (\widetilde{\mathcal{J}}_{k}\setminus\{(k,d)\}\cup \{k+1,s_1\}\cup\{k+1,s_2\})$.
     \item[4b)] For each split index $s=\{s_1,s_2\}$:
    \begin{enumerate}
    \item[i)]
     
    Build a local adaptive Bayesian PCE $\mathcal{R}_{\mathrm{PCE}}^{k+1,s}(\bm{\zeta})$ \eqref{eq:PCE} to approximate the current residual ${\mathcal{R}}_{\mathrm{S}}^{k+1}(\bm{\zeta})$ using only the samples $\bm{\zeta}$ inside the refined subdomain $\mathcal{D}_{\bm{\zeta}}^{k+1,s}$ (see Section~\ref{sec:ABPCE}).
    \item[ii)]Update the residual  only on the refined subdomain $\mathcal{D}_{\bm{\zeta}}^{k+1,s}$ via
    ${\mathcal{R}}_{\mathrm{S}}^{k+2}(\bm{\zeta})
\leftarrow {\mathcal{R}}_{\mathrm{S}}^{k+1}(\bm{\zeta})
-\mathbf{1}_{\mathcal{D}_{\bm{\zeta}}^{k+1,s}}(\bm{\zeta})\,
\mathcal{R}_{\mathrm{PCE}}^{k+1,s}(\bm{\zeta})$, 
where $\mathbf{1}_{\mathcal{D}_{\bm{\zeta}}^{k+1,s}}(\bm{\zeta})=1$ if $\bm{\zeta}\in\mathcal{D}_{\bm{\zeta}}^{k+1,s}$ and $0$ otherwise.
    \item[iii)] Recompute the refinement score $\chi^{k+1,s}$ using \eqref{eq:refine_score}
    \end{enumerate}
    \item [4c)] Increment $k \leftarrow k+1$.
\end{enumerate}
\noindent\textbf{Step 5.} Compute the ASSE model in \eqref{eq:SSE} with the full sequence of $\mathcal{D}_{\bm{\zeta}}^{k,d}$ and $\mathcal{R}_{\mathrm{PCE}}^{k,d}(\bm{\zeta})$ with $\bm{\zeta} \in \mathcal{D}_{\bm{\zeta}}^{k,d}$. If the ASSE model has reached the prescribed accuracy, (e.g.,  $e_{\mathrm{asse}} <e_{\mathrm{pre}}$, where $e_{\mathrm{asse}} =[ \sum_{d=1}^{D_k} e_{\mathrm{cv}}^{K,d}\mathbb{P}^{K,d}]/\mathrm{Var}[Z]$), or $e_{\mathrm{asse}}$ has a slightly decrement, go to \textbf{Step 6}, otherwise, enrich the sample and go to \textbf{Step 2}. 
\\
\noindent\textbf{Step 6.}  Generate/Acquire a large number of $N_{\mathrm{Val}}$ random input samples ($N_{\mathrm{Val}} \gg N_{\mathrm{ED}}$) and evaluate the probabilistic AC-OPF problem solutions using the built ASSE model \eqref{eq:SSE}. \\ 
%
%
\noindent\textbf{Step 7.} Generate the results report: assess the probabilistic information (e.g., mean and variance using \eqref{eq:moments}, median, PDF, CDF, quantiles, and  distributional metrics \eqref{eq:pbl}-\eqref{eq:WD_1}) of the AC-OPF solutions; calculate the validation error according to \eqref{eq:val_error}; compute the production cost  ${C}^{\mathrm{cost}}$  
and its associated index $  {C}^{\mathrm{cost}}_{\mathrm{Index}} $  
at a specific quantile using \eqref{eq:Q_index}. 
\caption{\small The ASSE Model for Probabilistic AC-OPF}
\label{Procedure:ASSE}
\end{Procedure}
\normalsize
\begin{remark} \label{rem:remark1} 
This paper considers random inputs $\bm{\zeta}$ consisting of wind speed, solar irradiance, and load power. The corresponding random input samples can be obtained from historical data directly or generated from prescribed distributions. Specially,  in the parametric cases,  wind speed follows a Weibull distribution \cite{Karki2006} and is converted to wind power via the turbine power curve \cite{Wang2021}; solar irradiance is drawn from a Beta distribution \cite{Salameh1995} and converted to solar power through the irradiance–power curve \cite{Wang2021}; load power is modeled as Gaussian with specified mean and variance \cite{Billinton2008}. When additional samples are required to reach the target accuracy in these parametric cases, we resample from the  same distributions.
  Specially, our simulation studies consider both independent and correlated inputs. For correlated inputs, we model dependence using a vine copula (see Section III, (3)–(11) in \cite{Xu2020coula}  for details) and generate correlated samples accordingly. The ASSE framework is then applied by treating domain-wise marginals as independent when building the polynomial basis $\Psi_{m}^{\bm{\alpha}}$ \cite{wagner2021bayesian,Wang2024gsa}.

\end{remark}

\begin{remark} \label{rem:remark2}
In addition to the probabilistic information (mean, variance, PDF and cumulative distribution functions (CDF)) of the response $\bm{Z}$, we evaluate the performance of the proposed ASSE method \eqref{eq:SSE}  using the validation error $e_{\mathrm{Val}}$ and three distributional metrics: pinball loss (PBL) \cite{wang2019probabilistic}, Kolmogorov–Smirnov distance (KSD) \cite{massey1951kolmogorov}, and the Wasserstein-1 distance (WD) \cite{panaretos2019}. Formal definitions are provided in Appendix~\ref{appen:evaluation}. 
\end{remark}

\begin{remark} \label{rem:remark 4} 
A key tuning parameter in ASSE is $N_{\mathrm{ref}}$ (i.e., the minimum required sample size in $\mathcal{D}_{\bm{\zeta}}^{\bm{\alpha}}$). A smaller $N_{\mathrm{ref}}$ allows each local PCE to adapt finely to asymmetric or localized behavior, but increases variance and runtime due to more refinements. 
 Conversely, a larger $N_{\mathrm{ref}}$ stabilizes estimates and shortens training, yet may oversmooth important tail behavior and introduce 
 bias. In Section~\ref{sec:case9}, we vary $N_{\mathrm{ref}}$ and assess its impact using  $e_{\mathrm{val}}$ as the performance metric.
\end{remark}

\begin{remark}
 The maximum expansion level $K_{\max}$ provides a hard upper bound on the refinement depth that guarantees termination and bounds the worst-case computation. We set $K_{\max}$ sufficiently large so that refinement typically stops by the data-driven criteria rather than by reaching this cap (e.g., $K_{\max}=1000$). The threshold $\tau_{\mathrm{thr}}$ controls whether a candidate subdomain is split: refinement is performed only when the refinement score exceeds $\tau_{\mathrm{thr}}$; we use a small value (e.g., $\tau_{\mathrm{thr}}=10^{-5}$) to avoid premature stopping. Together, $N_{\mathrm{ref}}$, $K_{\max}$, and $\tau_{\mathrm{thr}}$ act as stopping criteria; for earlier stopping, one may reduce $K_{\max}$ and/or increase $\tau_{\mathrm{thr}}$.  
\end{remark}
\begin{remark} \label{rem:remark 5}
In all examples, when constructing the adaptive Bayesian PCE model \eqref{eq:PCE} to approximate the residual expansion in each subdomain, we use the BCS algorithm to automatically select the PCE order and truncation $q$-norm within a given range rather than fixing them a priori. The procedure starts from a low polynomial order and small $q$-norm, and gradually increases these values, each time recomputing the surrogate and monitoring the modified CV error $e_{\mathrm{mcv}}$. The approximation of residual expansion stops once $e_{\mathrm{mcv}}$ is sufficiently small or no longer decreases significantly.
\end{remark}



\section{Numerical Studies} \label{sec: Simulation}
In this section, we validate the proposed ASSE method for solving the probabilistic AC-OPF problem \eqref{eq:DOPF} on the modified 9-bus and 
IEEE 118-bus systems.  
 Particularly, we present five case studies that vary in the system size, dimensionality, dependence structure, and data source of the random inputs. 
Case 1 (9-bus, $\mathcal{M}_{\mathrm{in}}=5$) and Case 2 (118-bus, $\mathcal{M}_{\mathrm{in}}=12$) focus on independent random inputs, whereas Case 3 ($\mathcal{M}_{\mathrm{in}}=30$) and Case 4 ($\mathcal{M}_{\mathrm{in}}=51$) consider dependent inputs on the 118-bus system.  Case 5 (118-bus) presents a real-world data test to further assess the robustness of ASSE under realistic uncertainty patterns.



For comparison, results from MC simulations conducted with MATPOWER  \cite{Zimmerman2011} serve as our benchmark. For Case~1 (IEEE 9-bus system), all network parameters and cost coefficients in
\eqref{eq:DOPF}–\eqref{eq:Q_index} are taken directly from the default
MATPOWER settings. 
For Cases~2–5 (IEEE 118-bus system), the detailed network data and
modified cost coefficients are available at
\href{https://github.com/TxiaoWang/ASSE.git}{this link}.
 
Across all five cases, we compare ASSE with two established surrogate approaches: adaptive sparse polynomial chaos expansion (SPCE)~\cite{Sheng2018} and Gaussian process regression (GPR)~\cite{Pareek2021}.  
In addition, a deep neural network (DNN) \cite{guha2019machine} is included as a representative deep learning surrogate in Cases 3 and 5. Case 3 involves higher input dimensionality and stronger nonlinearity,  while Case 5 evaluates the surrogate performance in the real-data setting. 

Specially, the SPCE surrogate is built using the discretized Stieltjes procedure to construct the polynomial basis and uses BCS to adaptively select the polynomial order and truncation ($q$-norm) from prescribed ranges and estimate the coefficients. The GPR surrogate adopts an ordinary-trend model with an anisotropic Mat\'ern-$5/2$ kernel; hyperparameters are determined by cross-validation. The DNN model is implemented as a multilayer perceptron with ReLU activations and the Adam optimizer; its depth and width are selected via cross-validation. All surrogates are trained and evaluated on the same datasets to ensure a fair comparison.

All simulations were performed using MATLAB R2024b on a  desktop equipped with an AMD Ryzen 9 9950X  (4.30 GHz)  
with 64GB RAM. Toolbox UQLab is adopted to build ASSE, SPCE, and GPR  models \cite{UQdoc_20_118,UQdoc_20_104}.  

\subsection{Case 1: The Modified 9-Bus System with 5 Independent Inputs} \label{sec:case9}
In this case, one wind farm and one solar PV power plant both with 100 MW are connected to Bus \{2,3\}, respectively. Three loads at Bus \{5,7,9\} are considered random. 
As such, there are 5 random inputs in total ($\mathcal{M}_{\mathrm{in}}=5$). 
Table \ref{tab:rv_param} presents the parameters of the random inputs (wind speed, solar irradiation). 
Particularly, random loads follow the Gaussian distributions with 5\% variations of their base power. 
%
\begin{table}[!ht]
\setlength{\abovecaptionskip}{0pt}   
\setlength{\belowcaptionskip}{0pt}   
\renewcommand{\arraystretch}{1.0}
\caption{Random input parameters for Case 1}
\label{tab:rv_param}
\resizebox{\linewidth}{!}{
\begin{tabular}{ccccccccc}
\hline
\multicolumn{9}{c}{Wind Speed and Wind Turbine Parameters  \cite{Karki2006}}                                                                                                                                                                                                                                                \\ \hline
\multicolumn{1}{c|}{Bus}  & \multicolumn{1}{c|}{${c}$}    & \multicolumn{1}{c|}{${\gamma}$}       & \multicolumn{1}{c|}{${P_{\mathrm{r}}}$}      & \multicolumn{1}{c|}{${V_{\mathrm{r}}}$} & \multicolumn{1}{c|}{${V_{\mathrm{in}}}$} & \multicolumn{1}{c|}{${V_{\mathrm{out}}}$} & \multicolumn{2}{c}{PF}                                    \\ \hline
\multicolumn{1}{c|}{2}    & \multicolumn{1}{c|}{11.153}     & \multicolumn{1}{c|}{3.289}            & \multicolumn{1}{c|}{100}            & \multicolumn{1}{c|}{12.00}     & \multicolumn{1}{c|}{3.50}       & \multicolumn{1}{c|}{24.00}       & \multicolumn{2}{c}{0.95}                                  \\ \hline
\multicolumn{9}{c}{Solar Irradiation and Solar Generator Parameters \cite{Salameh1995}}                                                                                                                                                                                                                                     \\ \hline
\multicolumn{1}{c|}{Bus}  & \multicolumn{1}{c|}{$\alpha$} & \multicolumn{1}{c|}{$\beta$}        & \multicolumn{1}{c|}{$P_{\mathrm{r}}$}        & \multicolumn{1}{c|}{$r_{\mathrm{min}}$} & \multicolumn{1}{c|}{$r_{\mathrm{max}}$}  & \multicolumn{1}{c|}{$R_{\mathrm{c}}$}       & \multicolumn{1}{c|}{$R_{\mathrm{std}}$} & \multicolumn{1}{c}{PF}  \\ \hline
\multicolumn{1}{c|}{3}   & \multicolumn{1}{c|}{1.700}      & \multicolumn{1}{c|}{0.74}             & \multicolumn{1}{c|}{100}            & \multicolumn{1}{c|}{0.00}      & \multicolumn{1}{c|}{1000.00}    & \multicolumn{1}{c|}{150.00}      & \multicolumn{1}{c|}{1000.00}   & \multicolumn{1}{c}{1.0} \\ \hline
\end{tabular}}
\vspace{-10pt}
\end{table}

\noindent{\textbf{ASSE Model Construction:}} 
First, we start from $N_{\mathrm{ED}} = 50$ to generate random input samples 
and apply the generated samples  to solve the probabilistic AC-OPF problem \eqref{eq:DOPF} to obtain the sample response pairs $[\bm{\zeta}_{\mathrm{ED}},\bm{Z}_{\mathrm{ED}}]$ (\textbf{Steps 1--2}).
Pass  $[\bm{\zeta}_{\mathrm{ED}},\bm{Z}_{\mathrm{ED}}]$ to construct the proposed ASSE, SPCE, and GPR models (\textbf{Steps 3--5}) for estimating all decision variables $\bm{u}$ (i.e., generator power output $P_{G_i}$,$ i = \{1,2,3\}$) 
and the objective function $P_{\mathrm{loss}}$. 
Particularly,  the BCS algorithm adaptively determines parameters (PCE max order varied from 2 to 4  and truncation norms from 0.5 to 0.8 in 0.05 increments).

\begin{figure}[htbp!]
\centering
\includegraphics[width=0.5\textwidth]{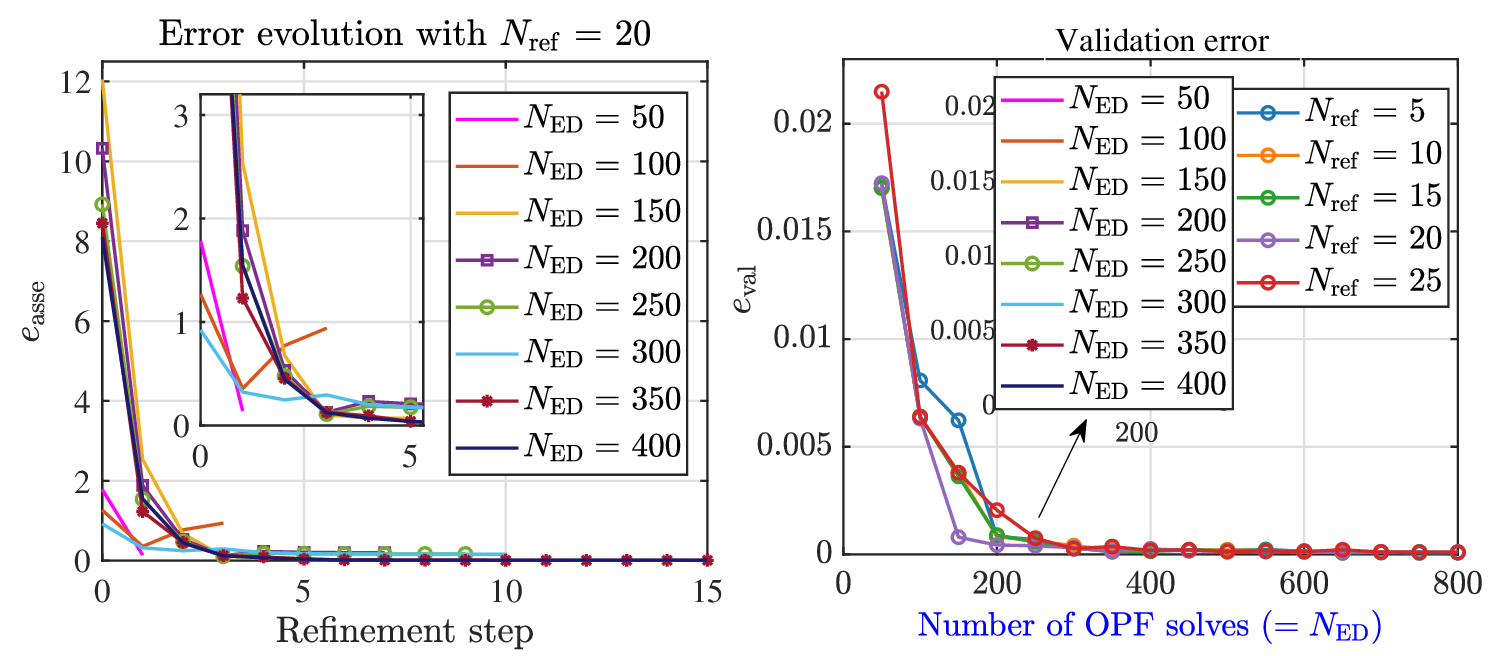}
\caption{(Left) The evolution of $e_{\mathrm{asse}}$ during training with $N_{\mathrm{ED}} = (50:50:400)$ at $ N_{\mathrm{ref}} = 20$  and (Right) the validation errors $e_{\mathrm{val}}$ versus $N_{\mathrm{ED}} = (50:50:800)$ and $N_{\mathrm{ref}}= (5:5:25)$  for estimating $P_{G_1}$ using the proposed ASSE method.  $e_{\mathrm{val}}$ is calculated with $N_{\mathrm{val}}=10,000$.}
\label{fig:ASSE_emp_error}
\vspace{-10pt}
\end{figure}

\noindent We enrich samples from $N_{\mathrm{ED}} = 50$ to $800$ with increments of 50 and repeat \textbf{Steps 1--5}. 
Particularly, we vary $N_{\mathrm{ref}}$ from $(5:5:25)$ 
 to test how the minimum number of points $N_{\mathrm{ref}}$ per subdomain affect estimation accuracy. 
Fig.~\ref{fig:ASSE_emp_error} shows the accuracy trends for estimating \(P_{G_1}\) with the proposed ASSE.  
{Left:} The evolution of training error $e_{\mathrm{asse}}$ with varying $N_{\mathrm{ED}}$ at $N_{\mathrm{ref}}=20$. Improvement levels off after $N_{\mathrm{ED}} \geq 150$ and is marginal beyond $350$.  
{Right:} Validation errors $e_{\mathrm{val}}$ with various $N_{\mathrm{ED}}$ and $N_{\mathrm{ref}}$. Accuracy degrades for $N_{\mathrm{ED}}\leq 100$, especially at larger $N_{\mathrm{ref}}$ (e.g., at $N_{\mathrm{ED}} = 100$, $N_{\mathrm{ref}} = 5$ outperforms $N_{\mathrm{ref}} = 25$). For $N_{\mathrm{ED}} \geq 350$, $N_{\mathrm{ref}}$ has little effect on error;  smaller $N_{\mathrm{ref}}$  captures finer local behavior but require extra refinements and longer training  (see Fig.~\ref{fig:case_1_ASSE_N_ref_pmax_metrics}). 
Therefore, we set $N_{\mathrm{ED}} = 350$ and $N_{\mathrm{ref}} = 20$ for ASSE models in this case. 
%
%

\begin{figure}
\setlength{\abovecaptionskip}{-0.51cm}
\setlength{\belowcaptionskip}{-0.5cm}
\centering
\includegraphics[width=0.5\textwidth]{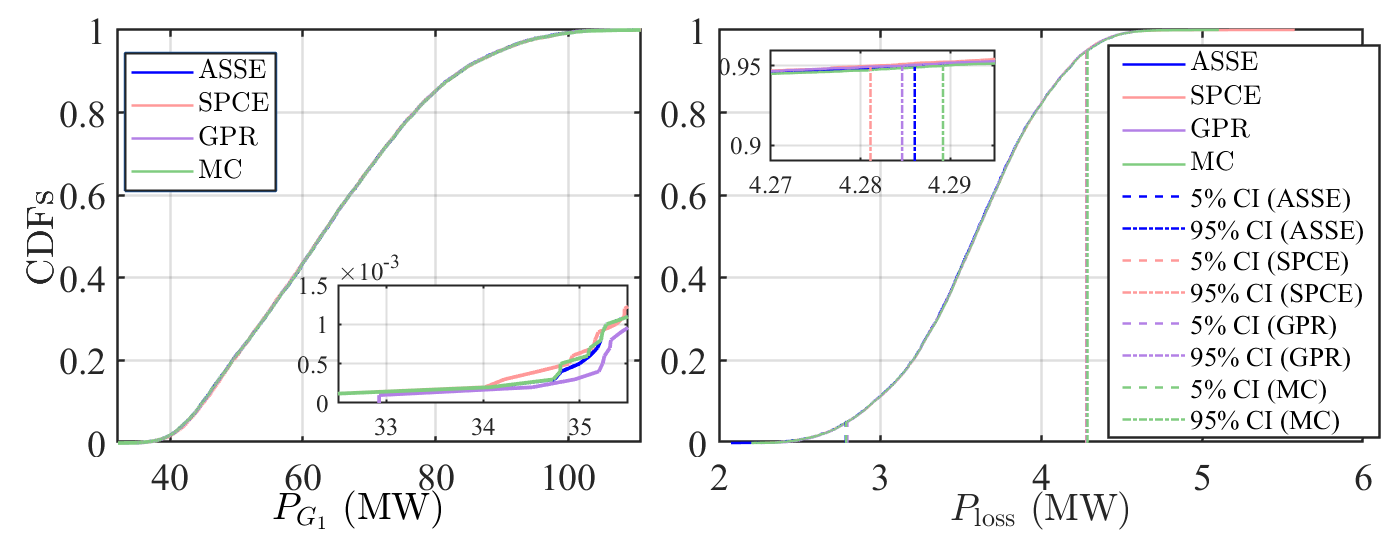}
\caption{Case 1: The estimated CDFs of $P_{G_1}$ (left) and the objective function (minimum power loss lower $P_{\mathrm{loss}}$) (right) from the ASSE (blue), SPCE (red), GPR (purple), and MC simulations (green). The corresponding dashed-dot indicate the $95\%$  quantiles  
setting as upper bounds of $P_{\mathrm{loss}}$ and dashed lines lines indicate $5\%$  quantiles,  
setting as lower bounds of $P_{\mathrm{loss}}$.  }
\label{fig:ASSE_P_loss_cdfs}
\vspace{-8pt}
\end{figure}

\noindent \textbf{Performance Evaluation:} 
We validate the proposed ASSE method using $N_{\mathrm{Val}} = 10,000$ MC samples of $\bm{\zeta}$. Table \ref{tab:statistics_comp_case9} summarizes the statistical estimates of $P_{G_1}$ for ASSE, SPCE, and GPR methods. The normalized error comparisons indicate that ASSE outperforms SPCE and GPR. Although all methods show similar accuracy against the MC benchmark, ASSE aligns most closely in the tail regions, as shown in Fig.~\ref{fig:ASSE_P_loss_cdfs} (left).

Fig. \ref{fig:ASSE_P_loss_cdfs} (right) presents the comparison of CDFs of $P_{\mathrm{loss}}$ from the proposed ASSE (blue), the SPCE (red), the GPR (purple) and the MC benchmark simulations (green). The ASSE results closely align with the MC simulations. Moreover, as highlighted by the blue dashed and dashed-dot lines, the ASSE provides more accurate estimations in the distribution tails.   Tables~\ref{tab:tail_pg_comp_5_case9} summarize the PBL~\eqref{eq:pbl}, KSD~\eqref{eq:KSD}, and WD~\eqref{eq:WD_1} metrics for the estimated $P_{G_1}$ and $P_{\mathrm{loss}}$ across all models. For Gaussian-like distributions, ASSE, SPCE, and GPR exhibit comparable performance. Notably, ASSE yields the lowest WD values for both $P_{G_1}$ and $P_{\mathrm{loss}}$, while maintaining competitive PBL and KSD results relative to the other surrogates.

 For the $95\%$ quantile of $P_{\mathrm{loss}}$, the proposed ASSE stays within 0.07\% of the MC benchmark, whereas SPCE can deviate by as much as $0.20\%$.  Scaled to a 40~GW system, such a $0.20\%$ error implies an unnecessary 80~MW reserve about a mid-sized gas turbine, while ASSE’s smaller error equates to roughly one-third of that amount. 
While Fig.~\ref{fig:ASSE_P_loss_cdfs} illustrates ASSE's better accuracy in tails, the relatively Gaussian-like distributions limit the visible accuracy differences among the methods. The following case studies  further highlight ASSE’s distinct advantage for asymmetric or otherwise non-Gaussian distributions, especially in high-dimensional settings.

\begin{table}[!t]
\setlength{\abovecaptionskip}{0pt}
\setlength{\belowcaptionskip}{0pt}
\renewcommand{\arraystretch}{1.3}
\caption{Case 1: Comparisons of metrics to evaluate the estimated distribution and tail behaviour for $P_{G_1}$ and $P_{\mathrm{loss}}$ from ASSE, SPCE, GPR, and MC simulations.}
\label{tab:tail_pg_comp_5_case9}
\centering
\resizebox{.98\columnwidth}{!}{
\begin{tabular}{c|c|c|c|c|c|c|c|c}
\hline
\multirow{2}{*}{Methods}
& \multicolumn{4}{c|}{$P_{G_1}$}
& \multicolumn{4}{c}{$P_{\mathrm{loss}}$} \\
\cline{2-9}
& $\operatorname{PBL}_{0.05}$ & $\operatorname{PBL}_{0.95}$ & $\operatorname{KSD}$ & $\operatorname{WD}$
& $\operatorname{PBL}_{0.05}$ & $\operatorname{PBL}_{0.95}$ & $\operatorname{KSD}$ & $\operatorname{WD}$ \\
\hline
ASSE & $\bm{1.1806}$ & 1.5729 & $\bm{0.0018}$ & $\bm{0.0334}$ & 0.0471 & 0.0422 & 0.0042 & $\bm{0.0020}$ \\
\hline
SPCE & 1.1810 & 1.5729 & 0.0073 & 0.0987 & 0.0471 & 0.0423 & 0.0037 & 0.0029 \\
\hline
GPR  & 1.1811 & 1.5729 & 0.0026 & 0.04416 & 0.0471 & 0.0422 & 0.0034 & 0.0024 \\
\hline
\end{tabular}}
\begin{tablenotes}
\item *$\operatorname{PBL}_{0.05}$ and $\operatorname{PBL}_{0.95}$ denote the pinball loss 
at the quantile levels $\tau_{\mathrm{qtl}} = 0.05$ and $\tau_{\mathrm{qtl}} = 0.95$, 
respectively. Note that all surrogate models are trained with the same sample size, $N_{\mathrm{ED}}=350$, to ensure a fair comparison.  For context, the pinball loss at   $\tau_{\mathrm{qtl}}=0.05$ normalized by the mean of $P_{G_1}$ yields a relative error of $1.85\%$ ( $\frac{\operatorname{PBL}_{0.05}}{\mathbb{E}[P_{G_1}]}\times 100 \%$).
\end{tablenotes}
\end{table}

\begin{table}[ht]
\vspace{-6pt}
\centering 
\setlength{\abovecaptionskip}{0pt}   
\setlength{\belowcaptionskip}{0pt}   
\caption{Statistics of generator output power $P_{G_1}$ from the ASSE, SPCE, GPR and MCS methods. 
}
\label{tab:statistics_comp_case9}
\resizebox{.98\columnwidth}{!}{
\begin{tabular}{c|c|c|c|c|c|c|c}
\hline
Index & \textbf{ASSE} & ${\Delta}P_{G_1}\% $ & SPCE & ${\Delta}P_{G_1}\% $  & GPR & ${\Delta}P_{G_1}\% $  & MCS \\ \hline
$\mathbb{E}[\cdot]$ & 63.6985 & $\bm{-0.017}$ & 63.653 & -0.088 & 63.688 & -0.033 & 63.709 \\ \hline
$\mathrm{Var}[\cdot]$ & 213.992 & $\bm{-0.499}$ & 213.400 & -0.775 & 213.875 & -0.554 & 215.066 \\ \hline
$\mathrm{Med}[\cdot]$& 62.724 & $\bm{0.012}$ & 62.632 & -0.135 & 62.697 & -0.031 & 62.716 \\
\hline 
\end{tabular}}
\begin{tablenotes}
\item* $\mathbb{E}[\cdot]$, $\mathrm{Var}[\cdot]$, and $\mathrm{Med}[\cdot]$ denote the mean value, variance, and median of $P_{G_{1}}$.   ${\Delta}P_{G_1}\% = \frac{\mathrm{Index}[\widehat{P}_{G_1}]-\mathrm{Index}[P_{G_1,\mathrm{mc}}]}{\mathrm{Index}[P_{G_1,\mathrm{mc}}]}\% $ denotes the normalized error of each index in \%, with $\widehat{P}_{G_1}$ estimated from the ASSE, SPCE and GPR models. 
\end{tablenotes}
\vspace{-12pt}
\end{table}
\begin{table}[ht!]
\vspace{-6pt}
\setlength{\abovecaptionskip}{0pt}   
\setlength{\belowcaptionskip}{0pt}   
\centering
\caption{Lower and upper bounds of generator output power $P_{G_i}$ from the ASSE, SPCE, GPR, and MCS methods. 
}
\label{tab:bounds_comp_case9}
\resizebox{.98\columnwidth}{!}{
\begin{tabular}{c|c|c|c|c|c|c|c}
\hline
Index & ASSE & ${\Delta}_\mathrm{asse}\% $ & SPCE &${\Delta}_\mathrm{spce}\% $& GPR & ${\Delta}_\mathrm{gpr}\% $ & MC \\
\hline
$\mathrm{CI_{1,5}}[\cdot]$  & 42.346& $\bm{0.076}$ & 42.603 & 0.683 & 42.401 & 0.207 & 42.314 \\ \hline
$\mathrm{CI}_{1,95}[\cdot]$ & 89.770 & $\bm{-0.077}$ & 89.686 & -0.172 & 89.711 & -0.144 & 89.84 \\ \hline
$\mathrm{CI}_{2,5}[\cdot]$ & 40.610 & $\bm{0.046}$ & 40.715 & 0.306 & 40.636 & 0.110 & 40.591 \\ \hline
$\mathrm{CI}_{2,95}[\cdot]$ & 88.095 & $\bm{-0.051}$ & 88.031 & -0.123 & 88.051 & -0.101 & 88.140 \\ \hline
$\mathrm{CI}_{3,5}[\cdot]$ & 40.460 & $\bm{0.120}$ & 40.531 & 0.298 & 40.471 & 0.148 & 40.411 \\ \hline
$\mathrm{CI}_{3,95}[\cdot]$ & 87.379 & $\bm{-0.060}$ & 87.343 & -0.100 & 87.335 & -0.110 & 87.431 \\
\hline
\end{tabular}}
\begin{tablenotes}
\item* 
$\mathrm{CI}_{i,5}[\cdot]$ and $\mathrm{CI}_{i,95}[\cdot]$ denote the $5\%$ and $95\%$  quantiles 
of the estimated $P_{G_i}$ distribution. ${\Delta}_\mathrm{asse}\% $,  ${\Delta}_\mathrm{spce}\% $, and ${\Delta}_\mathrm{gpr}\% $ are the corresponding normalized error in \% compared to MC results. 
\end{tablenotes}
\vspace{-6pt}
\end{table}

\noindent \textbf{Decisions under Uncertainties:} Once the generator outputs $P_{G_i}$ are estimated, their values at specific confidence intervals are obtained from the probability distributions by extracting the 5\% and 95\% quantiles. Table~\ref{tab:bounds_comp_case9} lists these lower and upper bounds of $P_{G_i}$, showing that the ASSE-based estimates closely match the benchmark MC results. Then, we use these bounds to compute the production cost, as presented in Table~\ref{tab:bounds_cost_comp_Dif}, providing a lower and upper budget for planning and operation. Particularly,  $  {C}^{\mathrm{cost}}_{\mathrm{low}}$ 
can be viewed as the minimum cost to support the generator outputs at the 5\% quantile level considering uncertainties, while $  {C}^{\mathrm{cost}}_{\mathrm{up}}$ 
corresponds to the upper cost limit associated with the 95\% quantile bounds.

\begin{table}[ht!]
\centering
\setlength{\abovecaptionskip}{0pt}   
\setlength{\belowcaptionskip}{0pt}   
\renewcommand{\arraystretch}{1.0}
\caption{Case 1: Comparison of the estimated statistics of cost $ {C}^{\mathrm{cost}}$ 
and their normalized errors by the ASSE, SPCE, GPR, and MC simulations. }
\label{tab:bounds_cost_comp_Dif}
\resizebox{.98\columnwidth}{!}{
\begin{tabular}{c|c|c|c|c|c|c}
\hline
Methods & $\mathbb{E}[{C}^{\mathrm{cost}}]$ & $\Delta \mathbb{E}\%$ & $ {C}^{\mathrm{cost}}_{\mathrm{low}}$& $\Delta  {C}^{\mathrm{cost}}_{\mathrm{low}}\%$ & $  {C}^{\mathrm{cost}}_{\mathrm{up}}$ & $\Delta  {C}^{\mathrm{cost}}_{\mathrm{up}}\%$ \\
\hline
ASSE & $2.620 \times 10^3$ & $-0.016$ & $1.751 \times 10^3$ & $0.063$ & $4.143 \times 10^3$ & $-0.091$ \\  \hline
SPCE & $2.621 \times 10^3$ & $-0.002$ & $1.756 \times 10^3$ & $0.341$ & $4.139 \times 10^3$ & $-0.192$ \\ \hline 
GPR & $2.620 \times 10^3$ & $-0.034$ & $1.752 \times 10^3$ & $0.122$ & $4.140 \times 10^3$ & $-0.172$ \\  \hline
MC & $2.621 \times 10^3$ & -- & $1.750 \times 10^3$ & -- & $4.147 \times 10^3$ & -- \\ 
\hline
\end{tabular}}
\begin{tablenotes}
\item* $\mathbb{E}[{C}^{\mathrm{cost}}]$ denotes the mean value of cost ${C}^{\mathrm{cost}}$, and  $\Delta \mathbb{E}\% = \frac{{\widehat{C}^{\mathrm{cost}}}-{{C}^{\mathrm{cost}}_{\mathrm{mc}}}}{{C}^{\mathrm{cost}}_{\mathrm{mc}}}\%$ is the corresponding normalized error, where $ \widehat{C}^{\mathrm{cost}}$
is estimated from the four different methods;  A quantile-based CI for the cost is defined by ${C}^{\mathrm{cost}}_{\mathrm{low}}$ and  ${C}^{\mathrm{cost}}_{\mathrm{up}}$. Specially,  $  {C}^{\mathrm{cost}}_{\mathrm{low}}$  
denote the lower bounds of the cost, calculated based on the $5\%$  quantile  
of $\widehat{P}_{G_i}$'s probability distribution, and  $ {C}^{\mathrm{cost}}_{\mathrm{up}}$ 
denote the upper bounds of cost, 
which are calculated using the 95\%  quantile  
of $\widehat{P}_{G_i}$; $\Delta  {C}^{\mathrm{cost}}_{\mathrm{low}}\% $ 
and $\Delta  {C}^{\mathrm{cost}}_{\mathrm{up}}\% $ 
are their corresponding normalized errors in \%. 
\end{tablenotes}
\end{table}

\noindent \textbf{Computational Cost:} Regarding computational cost, Table~\ref{tab:time_breakdown_case1} provides a detailed 
time breakdown of the proposed ASSE method for $P_{\mathrm{loss}}$ estimation, together with the corresponding number of refinement splits, for different values of $N_{\mathrm{ED}}$. As shown, the overall runtime is dominated by $t_{\mathrm{ed}}$, which accounts for generating sample--response pairs via repeated OPF solves (Step~2). Under a fixed $N_{\mathrm{ref}}$, increasing $N_{\mathrm{ED}}$ generally leads to deeper refinement (more splits), which in turn increases the model construction time $t_{\mathrm{cr}}$ (Steps~3--5). However, the accompanying accuracy improvement becomes marginal when $N_{\mathrm{ED}}$ increases from 350 to 500 in this case, indicating diminishing returns relative to the additional computational cost.

 To further elucidate the computational behavior of the proposed framework, the scaling of $t_{\mathrm{cr}}$ with respect to key ASSE parameters is quantified in Appendix~\ref{app:training_time}. 

\begin{table}[htbp]

\setlength{\abovecaptionskip}{0pt}
\setlength{\belowcaptionskip}{0pt}
\centering
\setlength{\tabcolsep}{1.7pt}
\caption{Case~1. Runtime of the ASSE method and corresponding splits for different $N_{\mathrm{ED}}$ with $N_{\mathrm{ref}}=20$.}
\label{tab:time_breakdown_case1}
\resizebox{1.01\columnwidth}{!}{
\begin{tabular}{c |c|c| c| c| c| c| c| c| c}
\hline
$N_{\mathrm{ED}}$ 
& splits  
& $t_{\mathrm{ed}}$(s)  
& $t_{\mathrm{basis}}$(s)  
& $t_{\mathrm{bcs}}$(s)  
& $t_{\mathrm{CV}}$(s)  
& $t_{\mathrm{refine}}$ (s) 
& $t_{\mathrm{other}}$(s) 
& $t_{\mathrm{ev}}$(s) 
& $t_{\mathrm{total}}$(s) \\
\hline
350 & 15 & 491.12 & 4.97 & 0.46 & 0.15 & 3.33 & 5.14 & 0.06&  505.23\\\hline
500 & 21 & 701.60 & 6.32 & 0.58 & 0.18 & 3.49 & 7.11 & 0.06 & 719.34 \\
\hline
\end{tabular}}
\begin{tablenotes}
    \item* $t_{\mathrm{ed}}$: time to generate sample-response pairs, i.e., OPF solves in Step 2.
  \item* $t_{\mathrm{basis}}$: time for polynomial basis construction; $t_{\mathrm{bcs}}$: time for BCS coefficients calculation;  $t_{\mathrm{CV}}$: time for cross-validation in \eqref{eq:emcv}-\eqref{eq:ecv}; $t_{\mathrm{refine}}$: time for refinement and splitting \eqref{eq:refine_score}-\eqref{eq:FirstOrderSI}; $t_{\mathrm{other}}$: time for framework initialization, model setup, data parsing, final ASSE representation construction, and analytical output moment calculations; 
  \item* The time to construct ASSE model exclude OPF solves is $t_{\mathrm{cr}}=t_{\mathrm{basis}}+t_{\mathrm{bcs}}+t_{\mathrm{CV}}+t_{\mathrm{refine}}+t_{\mathrm{other}}$ (Step 3-Step 5).
 \item* $t_{\mathrm{ev}}$: time for inference with $N_{\mathrm{Val}} = 10^{4}$ (Step 6).
 \item $t_{\mathrm{total}}$ denotes the total runtime (Step 2-Step 6). 
\end{tablenotes}
\end{table}
\normalsize

\begin{remark}
In Case~1, the SPCE selects the PCE order from 2 to 6 and the truncation $q$-norm from (0.5:0.05:0.8). For the results reported in Fig.~\ref{fig:ASSE_P_loss_cdfs} and Tables~\ref{tab:tail_pg_comp_5_case9}--\ref{tab:time_breakdown_case1}, SPCE chooses order 6 (with $q=0.8$) for $P_{G_1}$ and order 5 (with $q=0.6$) for $P_{\mathrm{loss}}$. While GPR  uses an ordinary-trend model with an anisotropic Mat\'ern-$5/2$ kernel for all case studies in this paper.
\end{remark}

\subsection{Case 2: The Modified 118-Bus System with 12 Independent Inputs } \label{sec:case_118_12}

In this case, the proposed method is extended to the modified 118-bus system. Random inputs consist of six wind farms, each with a capacity of 60 MW, connected to buses \{10, 25, 26, 49, 65, 66\}, and six solar PV power plants, each with a capacity of 40 MW, connected to buses \{12, 59, 61, 80, 89, 100\}. The loads are assumed to remain constant at their base power levels throughout the analysis.
In total, there are 12 random inputs (\(\mathcal{M}_{\mathrm{in}} = 12\)). The parameters for the random input distributions, including wind speed and solar irradiation, are obtained from \cite{Sheng2018}.

%
\noindent{\textbf{ASSE Model Construction:}} 
Case 2 follows a similar procedure as Case 1, we start with $N_{\mathrm{ED}} = 60$ random input samples, solve the probabilistic AC-OPF problem~\eqref{eq:DOPF}. 
Then, pass the obtained sample--response pairs, \(\bigl[\boldsymbol{\zeta}_{\mathrm{ED}}, \boldsymbol{Z}_{\mathrm{ED}}\bigr]\) (\textbf{Steps~1--2}) to construct the proposed ASSE, SPCE, and GPR models (\textbf{Steps~3--5}), enabling the estimation of all decision variables $\bm{u}$ ($P_{G_i},i \in \{1,\cdots,54\}$) and the objective function $P_{\mathrm{loss}}$. 
Specially, the BCS algorithm adaptively determines key parameters (PCE max order varied from 1 to 3 and truncation norm from 0.5 to 0.8 in steps of 0.05). 
To further enrich the training process, we increase
$N_{\mathrm{ED}}$ from 60 to 960 in increments of 60 and repeat \textbf{Steps~1--5}. The minimum number of points in each subdomain is set as $N_{\mathrm{ref}} = 12:12:36$ while $N_{\mathrm{ref}}=24$ is finally selected for most $P_{G_i}$ estimates in this case. 

\noindent \textbf{Performance Evaluation:} 
To validate the proposed method, we generate \( N_{\mathrm{Val}} = 10,000 \) MC samples of \( \bm{\zeta} \). Table \ref{tab:statistics_comp_case118} compares the statistics of $ P_{G_{20}}$ using ASSE, SPCE, and GPR, along with the MC benchmark results; ASSE yields the smallest normalized errors in these metrics. Fig.~ \ref{fig:ASSE_emp_error_case118} plots validation error versus $N_{\mathrm{ED}}$. It shows that ASSE outperforms the other two methods for moderate training samples size (e.g., $N_{\mathrm{ED}} \geq 120$), while for extremely smaller training sample size ($N_{\mathrm{ED}}$ = 60 in this case), GPR shows the highest accuracy.  Fig.~\ref{fig:ASSE_WD_comp_case2} reports the WD~\eqref{eq:WD_1} metric for $P_{G_5}$ and the average value across all generators $P_{G_i}, i = 1,\cdots, 54$ obtained from the ASSE, SPCE, and GPR by varying sample size $N_{\mathrm{ED}}$, showing the ASSE's strong ability in estimating the complete probability distribution.  Fig.~\ref{fig:ASSE_PDFall_case118} presents the PDFs for $P_{G_i},i\in \{5:8,19\}$.  Table~\ref{tab:tail_pg19_comp_12_case118} summarizes the 
PBL~\eqref{eq:pbl}, KSD~\eqref{eq:KSD}, and WD metrics for $P_{G_5}$, 
$P_{G_{19}}$ (a highly skewed case), further validating the performance of the 
proposed method in tail and full distribution estimation. Table \ref{tab:tail_pgi_ave_comp_12_case118} summarizes the average values of these metrics over all generators.   For near-Gaussian distributions ($P_{G_7}$, $P_{G_8}$), all three methods exhibit comparable accuracy. For skewed or localized cases (e.g., $P_{G_{19}}$), ASSE and GPR track the MC benchmark closely, whereas SPCE must raise its order to six, yet still lags behind, even when using more training sample points. Furthermore, for such highly skewed distributions, SPCE shows only minor performance improvements as the number of training samples increases and still underperforms compared with ASSE and GPR. 
\begin{table}[ht]
\setlength{\abovecaptionskip}{0pt}   
\setlength{\belowcaptionskip}{0pt}   
\setlength{\belowcaptionskip}{-0.5cm}
\centering 
\caption{Statistics of generator output power $P_{G_{20}}$ from the ASSE, SPCE, GPR, and MCS methods. The models are built based on the training sample size $N_{\mathrm{ED}}=780$ and $N_{\mathrm{ref}}=24$ for the proposed ASSE method.}
\label{tab:statistics_comp_case118}
\resizebox{.98\columnwidth}{!}{
\begin{tabular}{c|c|c|c|c|c|c|c}
\hline
Index & \textbf{ASSE} & ${\Delta}P_{G_{20}}\% $ & SPCE & ${\Delta}P_{G_{20}}\% $  & GPR & ${\Delta}P_{G_{20}}\% $  & MCS \\ 
\hline
$\mathbb{E}[\cdot]$ & 89.2739 & $\bm{3.880\times10^{-5}}$ & 89.2757 & 0.0021 & 89.2726 & -0.0015 & 89.2739\\ \hline
$\mathrm{Var}[\cdot]$ & 4.7193 & $\bm{-0.2311}$ &  4.7171 & -0.2767 &  4.6610 & -1.4629 &  4.7302 \\ \hline
$\mathrm{Med}[\cdot]$ & 88.9279 & $\bm{-0.0108}$ & 88.9205 & -0.0191 & 88.9335 & -0.0048 & 88.9375 \\
\hline
\end{tabular}}
\begin{tablenotes}
\item* $\mathbb{E}[\cdot]$, $\mathrm{Var}[\cdot]$, and $\mathrm{Med}[\cdot]$ denote the mean value, variance, and median of $P_{G_{20}}$.   ${\Delta}P_{G_{20}}\% = \frac{\mathrm{Index}[\widehat{P}_{G_{20}}]-\mathrm{Index}[P_{G_{20},\mathrm{mc}}]}{\mathrm{Index}[P_{G_{20},\mathrm{mc}}]}\% $ denotes the normalized error of each index in \%, with $\widehat{P}_{G_{20}}$ estimated from the ASSE, SPCE and GPR models. 
\end{tablenotes}
\end{table}
\begin{table}[]

\setlength{\abovecaptionskip}{0pt}   
\setlength{\belowcaptionskip}{0pt} 
\renewcommand{\arraystretch}{1.3}
\caption{Case 2: Comparisons of metrics to evaluate the estimated distribution and tail behaviour for $P_{G_{5}}$ and $P_{G_{19}}$ from  ASSE, SPCE, GPR, and MC simulations. }
\label{tab:tail_pg19_comp_12_case118}
\centering
\resizebox{.98\columnwidth}{!}{
\begin{tabular}{c|cccc|cccc}
\hline
\multirow{2}{*}{Methods} & \multicolumn{4}{c|}{$P_{G_{5}}$}                                                                                                                                         & \multicolumn{4}{c}{$P_{G_{19}}$}                                                                                                                                     \\ \cline{2-9} 
                         & \multicolumn{1}{c|}{$\operatorname{PBL}_{0.05}$} & \multicolumn{1}{c|}{$\operatorname{PBL}_{0.95}$} & \multicolumn{1}{c|}{$\operatorname{KSD}$} & $\operatorname{WD}$ & \multicolumn{1}{c|}{$\operatorname{PBL}_{0.05}$} & \multicolumn{1}{c|}{$\operatorname{PBL}_{0.95}$} & \multicolumn{1}{c|}{$\operatorname{KSD}$} & $\operatorname{WD}$ \\ \hline
ASSE                     & \multicolumn{1}{c|}{\textbf{0.4817}}                         & \multicolumn{1}{c|}{\textbf{0.3422}}                      & \multicolumn{1}{c|}{\textbf{0.0031}}               & \textbf{0.01551}             & \multicolumn{1}{c|}{\textbf{0.0588}}                      & \multicolumn{1}{c|}{\textbf{0.0101}}                      & \multicolumn{1}{c|}{$\bm{0.2060}$}           & $\bm{0.0240}$          \\ \hline
SPCE                     & \multicolumn{1}{c|}{0.4826}                         & \multicolumn{1}{c|}{0.3422}                      & \multicolumn{1}{c|}{0.0094}               & 0.0708              & \multicolumn{1}{c|}{0.0612}                      & \multicolumn{1}{c|}{0.0133}                      & \multicolumn{1}{c|}{0.3266}               & 0.0744              \\ \hline
GPR                      & \multicolumn{1}{c|}{0.4822}                         & \multicolumn{1}{c|}{0.3422}                      & \multicolumn{1}{c|}{0.0059}               & 0.0243              & \multicolumn{1}{c|}{0.0587}                      & \multicolumn{1}{c|}{0.0108}                      & \multicolumn{1}{c|}{0.2905}               & 0.0232              \\ \hline
\end{tabular}}
\begin{tablenotes}
\item *All surrogate models are trained with the same sample size, $N_{\mathrm{ED}}=780$.
\end{tablenotes}
\end{table}

%
%
\begin{table}[!t]

\setlength{\abovecaptionskip}{0pt}   
\setlength{\belowcaptionskip}{0pt} 
\renewcommand{\arraystretch}{1.3}
\caption{Case 2: Comparisons of average metrics to evaluate the estimated distribution and tail behaviour for $P_{G_{i}}$ from  ASSE, SPCE, GPR, and MC simulations. }
\label{tab:tail_pgi_ave_comp_12_case118}
\centering
\begin{tabular}{c|c|c|c|c}
\hline
Methods & $\overline{\operatorname{PBL}}_{0.05}$ & $\overline{\operatorname{PBL}}_{0.95}$ & $\overline{\operatorname{KSD}}$ & $\overline{\operatorname{WD}}$ \\
\hline
ASSE & \textbf{0.1556} & 0.1806 & $\bm{0.0085}$ & $\bm{0.0094}$ \\ \hline
SPCE & 0.1557 & 0.1807 & 0.0116 & 0.0141  \\ \hline
GPR & 0.1557& 0.1806& 0.0107& 0.0124  \\ \hline
\end{tabular}
\begin{tablenotes}
\item *All surrogate models are trained with the same sample size, $N_{\mathrm{ED}}=780$. $\overline{[\cdot]}$ denotes the average of the corresponding metric over all generators $P_{G_i}$, $i=1,\ldots,54$; e.g.,
$\overline{\operatorname{PBL}}_{0.95}=\frac{1}{54}\sum_{i=1}^{54}\operatorname{PBL}_{0.95}[{P}_{G_{i}}]$.  
\end{tablenotes}
\end{table}

\begin{figure}[htbp!]
\setlength{\abovecaptionskip}{-0.61cm}
\setlength{\belowcaptionskip}{-0.6cm}
\centering
\includegraphics[width=0.5\textwidth]{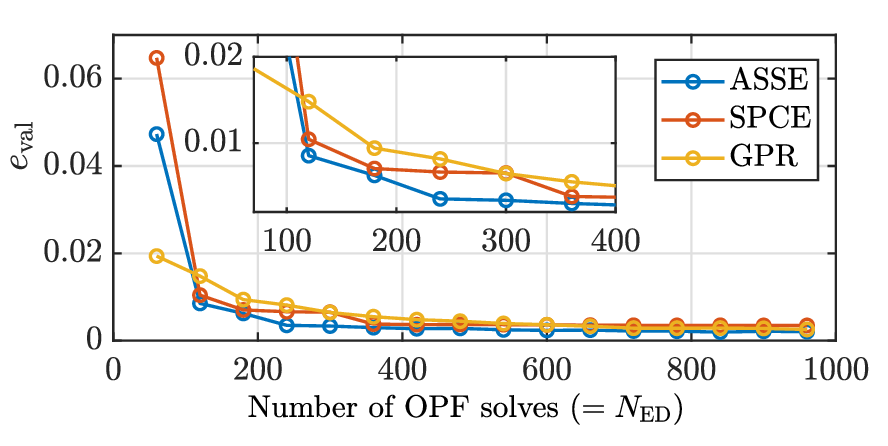}
\caption{The validation errors $e_{\mathrm{val}}$ for ASSE, SPCE, GPR versus  $N_{\mathrm{ED}} = 60:60:960$ in estimating $P_{G_{20}}$. $e_{\mathrm{val}}$ is calculated with  $N_{\mathrm{val}}=10,000$ MC samples.}
\label{fig:ASSE_emp_error_case118}
\end{figure}

\begin{figure}[htbp!]
\vspace{-6pt}
\centering
\setlength{\abovecaptionskip}{-0.61cm}
\setlength{\belowcaptionskip}{-0.6cm}
\includegraphics[width=0.5\textwidth]{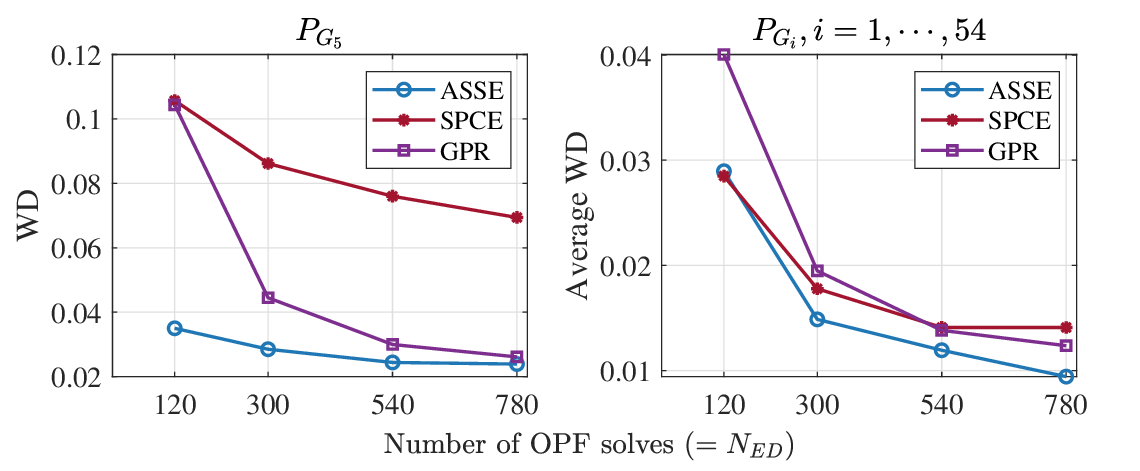}
\caption{Case 2: WD and average WD values comparison from the proposed ASSE, SPCE, and GPR under varied $N_{\mathrm{ED}}$. Left: $P_{G_{5}}$; Right: Average WD: $\overline{\operatorname{WD}}=\frac{1}{54}\sum_{i=1}^{54}\operatorname{WD}[P_{G_i}]$.}
\label{fig:ASSE_WD_comp_case2}
\end{figure}

\begin{figure*}
\setlength{\abovecaptionskip}{-0.61cm}
\setlength{\belowcaptionskip}{-0.5cm}
\centering
\includegraphics[width=1\textwidth]{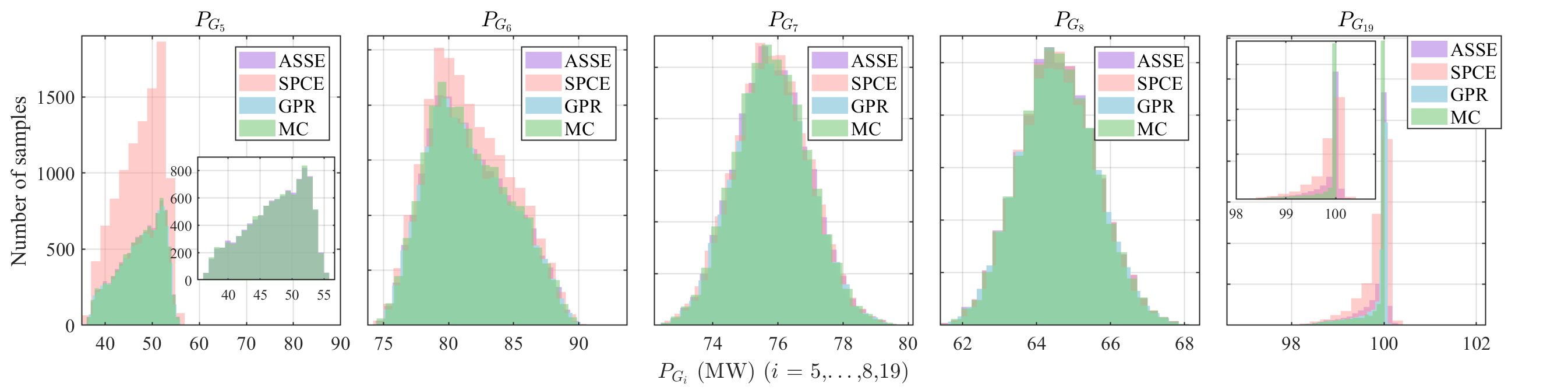}
\caption{ Estimated PDFs of the generator output power ($P_{G_{i}},i=\{5:8,19\}$) from the ASSE (purple), SPCE (red), GPR (blue) and MC simulations (green). For skewed distributions ($P_{G_5}$, $P_{G_6}$), ASSE tracks the benchmark closely, whereas SPCE's performance deteriorates significantly. For near-Gaussian distributions ($P_{G_7}$ and $P_{G_8}$), all three surrogates align well. However, for response with highly localized behavior ($P_{G_{19}}$), ASSE still outperforms SPCE. 
}
\label{fig:ASSE_PDFall_case118}
\vspace{-10pt}
\end{figure*}
%
%
%

\begin{figure}
\setlength{\abovecaptionskip}{-0.51cm}
\setlength{\belowcaptionskip}{-0.5cm}
\centering
\includegraphics[width=0.5\textwidth]{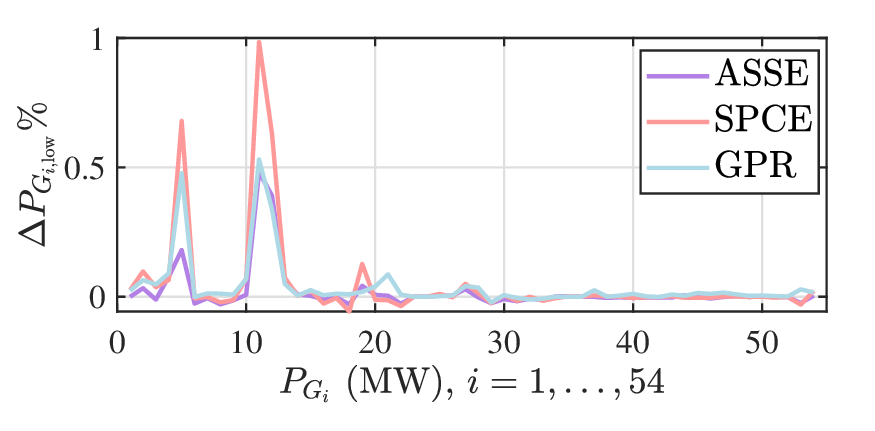}
\caption{
The normalized lower bound errors in \%: $\Delta P_{G_{i,\mathrm{low}}}\% = \frac{\hat{P}_{G_i,{\mathrm{low}}}-{P}_{G_i,{\mathrm{low}}}}{{P}_{G_i,{\mathrm{low}}}}\%$ for the estimates of $P_{G_{i,\mathrm{low}}}, (i = 1:54)$ from ASSE (purple), SPCE (red), and GPR (blue).  .
The proposed ASSE yields smaller error, especially for skewed distributions (e.g., $P_{G_5}$).}
\label{fig:bounds_comp_case118}
\vspace{-6pt}
\end{figure}

\noindent \textbf{Decisions under Uncertainties:}
After generator outputs' $ P_{G_i}$ PDFs are obtained, their $5\%$ and $95\%$  quantiles   
define the lower ($P_{G_{i,\mathrm{low}}}$) and upper ($P_{G_{i,\mathrm{up}}}$) decision bounds. These bounds feed directly into the production cost calculation. 
Fig.~\ref{fig:bounds_comp_case118} reports the normalized percentage error $\Delta P_{G_{i,\mathrm{low}}}\%$ in estimating $P_{G_{i,\mathrm{low}}}$ for ASSE, two existing methods, and the MC benchmark. ASSE consistently achieves smaller errors, especially for skewed distributions (e.g., $P_{G_5}$ and $P_{G_{19}}$), than SPCE,  consistent with the PDF comparisons in Fig.~\ref{fig:ASSE_PDFall_case118} and the results in Table~\ref{tab:tail_pg19_comp_12_case118}.
Table \ref{tab:bounds_cost_comp_Dif_case118} lists the resulting lower $  {C}^{\mathrm{cost}}_{\mathrm{low}}$ 
and upper $ {C}^{\mathrm{cost}}_{\mathrm{low}}$  
production cost estimates. Because generator output can exhibit non-Gaussian, skewed, or heavy‐tailed output distributions (e.g., $P_{G_5},P_{G_{19}}$), when these generators carry high cost coefficients, accurately capturing the full distributional shape is essential. Our proposed ASSE method delivers the smallest normalized error in \% for both bounds, approximately two to three times smaller than those of the other two existing methods, providing far more reliable bounds for operational and planning decisions.

%

\begin{table}[ht!]
\centering
\setlength{\abovecaptionskip}{0pt}   
\setlength{\belowcaptionskip}{0pt}   
\renewcommand{\arraystretch}{1.0}
\caption{Case 2: Comparison of the estimated lower and upper bounds of cost $C^{\mathrm{cost}}_{\mathrm{Index}}$  and their normalized errors in \% by ASSE, SPCE, GPR, and MC simulations. }
\label{tab:bounds_cost_comp_Dif_case118}
\resizebox{.98\columnwidth}{!}{
\begin{tabular}{c|c|c|c|c}
\hline
Methods & ${C}^{\mathrm{cost}}_{\mathrm{low}}$& $\Delta {C}^{\mathrm{cost}}_{\mathrm{low}}\%$ & ${C}^{\mathrm{cost}}_{\mathrm{up}}$ & $\Delta {C}^{{C}^{\mathrm{cost}}}_{\mathrm{up}}\%$ \\
\hline
\textbf{ASSE} &  $2.5008\times 10^5$ & $\bm{0.0625}$ & $2.9581 \times 10^5$& $\bm{0.0047}$\\ \hline
SPCE &  $2.5032\times 10^5$ & 0.1592& $2.9588 \times 10^5$& $0.0198$\\ \hline
GPR &  $2.5018 \times 10^5$ & 0.1021& $2.9558\times 10^5$ & -0.0736\\ \hline
MC & $2.4992 \times 10^5$ & -- & $2.9580\times 10^5$ & -- \\ \hline
\end{tabular}}
\end{table}

\noindent \textbf{Computational Cost and Sample Efficiency:} Regarding the computational cost to achieve comparable accuracy in the 
$P_{G_{19}}$ estimations, the time required to construct the SPCE model
($t_{\mathrm{cr}} \approx 954.04~\text{s}$) is roughly ten times that of the
ASSE model ($t_{\mathrm{cr}} \approx 33.33~\text{s}$). Moreover, ASSE and GPR require less training samples ($N_{\mathrm{ED}}$) compared to SPCE.   Detailed construction $t_{\mathrm{ed}} + t_{\mathrm{cr}}$,
evaluation $t_{\mathrm{ev}}$, and total runtimes, together with the corresponding training
sample sizes for all methods in  $P_{G_i}$ estimation, are summarized in
Table~\ref{tab:IEEE118_12_time_comp}. 

\begin{table}[htbp!]

\setlength{\abovecaptionskip}{0.cm}
\renewcommand{\arraystretch}{1.3}
\caption{Case 2: Comparison of computation time for estimating $P_{G_{19}}$ using the proposed ASSE, 
the existing methods SPCE, AND GPR, and the MC benchmark, all at comparable accuracy.}
\label{tab:IEEE118_12_time_comp}
\centering
\begin{tabular}{c|c|c|c|c|c}
\hline
Method & $N_{\mathrm{ED}}$ & ${t_{\mathrm{ed}}}$ (s) & ${t_{\mathrm{cr}}}$ (s) & ${t_{\mathrm{ev}}}$ (s) & ${t_{\mathrm{total}}}$ (s) \\
\hline
MC    &  -- &--  &  --    & 22038.97 & 22038.97 \\
\hline
\textbf{ASSE}  & 780 & $1719.04$   &   $33.33$    & 0.06&  1752.43\\
\hline
 SPCE   & 1200  & 2611.20   &   $954.04$  & 0.08  &  3565.32 \\
\hline 
 GPR & 780  & $1719.04$ & $31.83$ & 0.08 & 1750.95\\
\hline
\end{tabular}
\begin{tablenotes}
\item * $t_{\mathrm{ed}}$ denotes the time to generate the dataset to build the models, i.e., Step 2.
\item * $t_{\mathrm{cr}} = t_{\mathrm{basis}} + t_{\mathrm{bcs}} + t_{\mathrm{CV}} + t_{\mathrm{refine}} + t_{\mathrm{other}}$ denotes the time to construct the model including Step 3-Step 5, where $t_{\mathrm{basis}} = 11.03$ s, $ t_{\mathrm{bcs}} = 1.46$ s, $t_{\mathrm{CV}}=0.17$ s, $t_{\mathrm{refine}} = 11.67$ s, $t_{\mathrm{other}} = 9.00$ s. Notations are defined in Table~\ref{tab:time_breakdown_case1}.
\item * $t_{\mathrm{ev}} $ denotes the time to evaluate $N_{\mathrm{val}}$ samples (Step 6).
\item * $t_{\mathrm{total}}$ denotes the total runtime (Step 2-Step 6).
\item* SPCE requires more training samples and higher order with 6 to achieve comparable accuracy with ASSE and GPR. 
\end{tablenotes}
\end{table}

\begin{remark}
In Case 2, for the results in Tables~\ref{tab:statistics_comp_case118}--\ref{tab:IEEE118_12_time_comp} and Figs.~\ref{fig:ASSE_emp_error_case118}-\ref{fig:ASSE_PDFall_case118}, SPCE selects polynomial orders of 6 for $P_{G_i}, i = \{7,8,20\}$ ($q$-norm $=0.65$), 5 for $P_{G_5}, P_{G_6}$ ($q$-norm $=0.5$),  and 6 for $P_{G_{19}}$ ($q$-norm $=0.75$).
\end{remark}


\subsection{Case 3: The Modified 118-Bus System with 30 Dependent Inputs } \label{sec:118_bus_30}

In this case, we extend the proposed ASSE framework to a more challenging scenario on the IEEE 118-bus system, considering a high-dimensional random input space ($\mathcal{M}_{\mathrm{in}} = 30$) with dependent inputs.
Compared with Case~2, which considers 12 random inputs,  we introduce 9 additional wind generators (each rated at 80~MW) at buses $\{2, 4, 10, 13, 14, 19, 20, 22, 28\}$ and 9 additional solar PV plants (also 80~MW) at buses $\{32, 33, 39, 41, 42, 44, 51, 52, 53\}$. The system therefore contains 15 wind generators and 15 solar PV plants in total, yielding 30 stochastic inputs. To further increase the difficulty, we consider heavier-tailed marginals and stronger tail dependence. The input dependence is modeled using a vine copula with two independent blocks: a C-vine for wind speeds and a D-vine for solar irradiance, producing a flexible non-Gaussian dependence structure. Full details of the marginal and copula specifications are provided at the aforementioned \href{https://github.com/TxiaoWang/ASSE.git}{link}.

\noindent{\textbf{ASSE Model Construction:}}  
As in the previous cases, we build the ASSE framework for the probabilistic AC-OPF problem~\eqref{eq:DOPF}. Specifically,  $N_{\mathrm{ED}} $ is varied from $240$ to $1200$ in increments of $120$ 
samples of the random inputs, with $N_{\mathrm{ref}} = 40$. The BCS algorithm adaptively selects the PCE order (from $1$ to $3$) and truncation norm (from $0.5$ to $0.8$ in steps of $0.05$). In addition to comparing ASSE with SPCE, GPR, and the MC benchmark, we also compare its performance with that of a DNN-based surrogate.

\begin{figure}[htbp!]
\vspace{-6pt}
\centering
\setlength{\abovecaptionskip}{-0.51cm}
\includegraphics[width=0.5\textwidth]{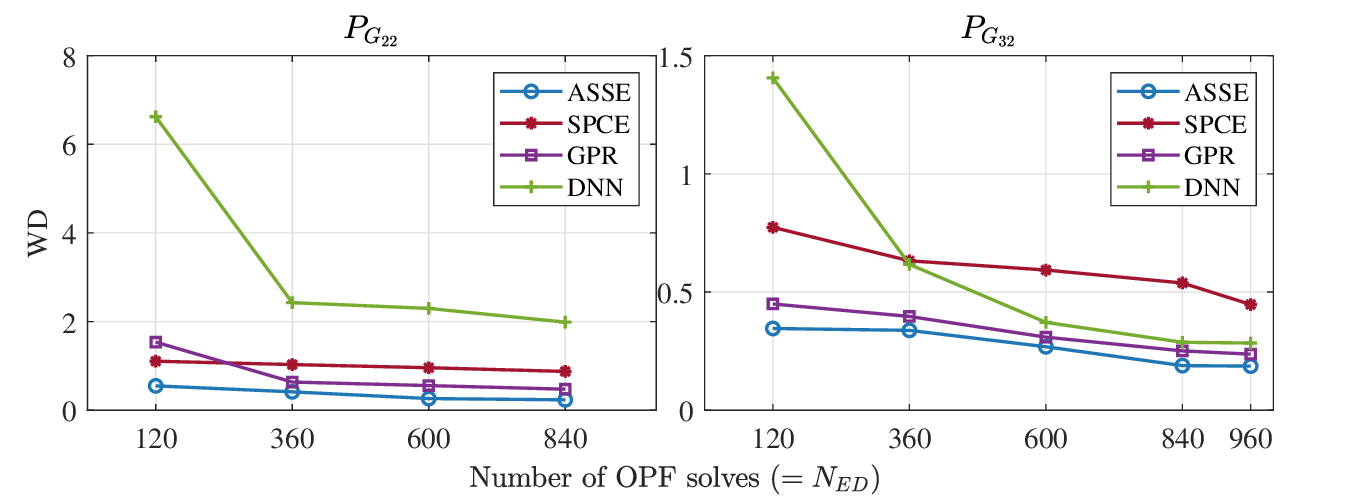}
\caption{Case 3: WD values comparison from the proposed ASSE, SPCE, GPR, and DNN under varied $N_{\mathrm{ED}}$. Left: $P_{G_{22}}$; Right: $P_{G_{32}}$.}
\label{fig:ASSE_WD_comp_case3}
\vspace{-10pt}
\end{figure}

\begin{figure}[htbp!]
\setlength{\abovecaptionskip}{-0.41cm}
\setlength{\belowcaptionskip}{-0.5cm}
\centering
\includegraphics[width=0.5\textwidth]{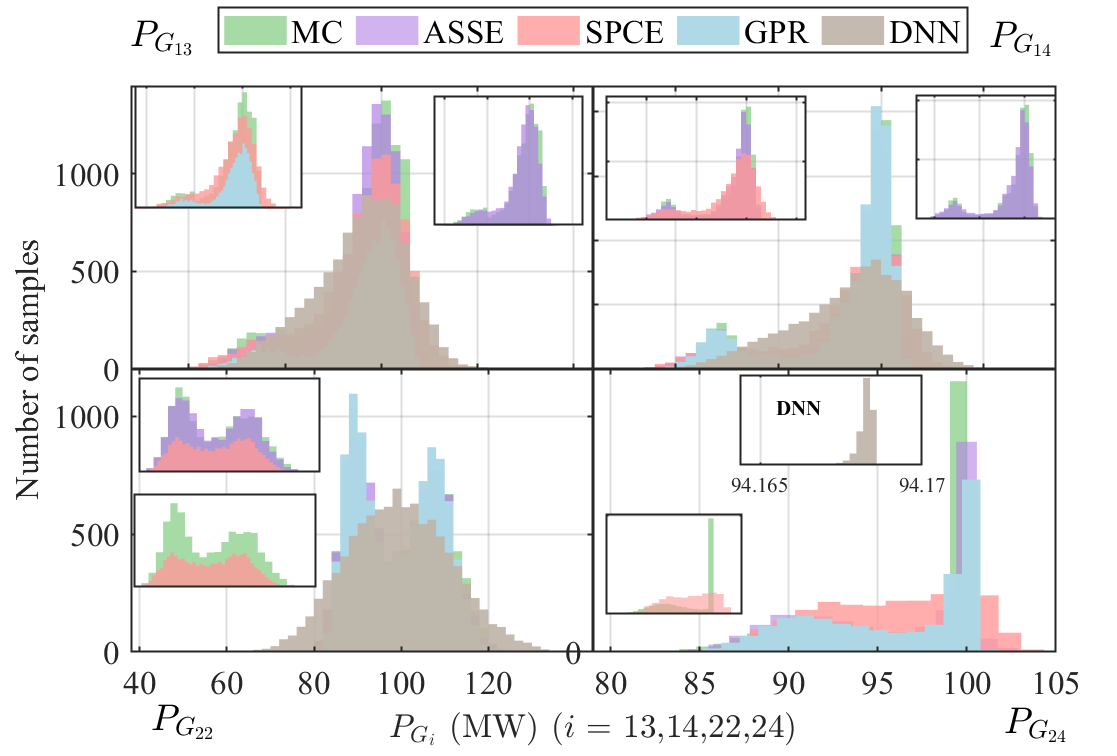}
\caption{ Case 3: Estimated PDFs of the generator output power ($P_{G_{i}}, i=\{13,14,22,24\}$) from the ASSE (purple), SPCE (red), GPR (blue), DNN (brown), and MC simulations (green). For comparison, all surrogate models are constructed using the same training sample size ($N_{\mathrm{ED}} = 840$). All generators exhibit skewed output distributions. Among the surrogate models, 
ASSE closely follows the MC benchmark, whereas the accuracy of SPCE and DNN deteriorates noticeably. While DNN with limited training samples for  $P_{G_{24}}$ (collapsing its predictions near 94) with highly skewed, failed to capture the behaviors.}
\label{fig:ASSE_PDFall_case118_30}
\vspace{-10pt}
\end{figure}

\noindent \textbf{Performance Evaluation:} 
$N_{\mathrm{val}}=10,000$ samples (Step
6) of $\bm{\zeta}$ are generated to evaluate the performance of the constructed ASSE model. Specially, Fig.~\ref{fig:ASSE_WD_comp_case3} presents the comparison of WD values from the proposed ASSE, SPCE, GPR, and DNN under different $N_{\mathrm{ED}}$, demonstrating that under a limited number of samples, ASSE provides the most accurate probability distribution estimation for skewed distributions (with the lowest WD values). Fig.~\ref{fig:ASSE_PDFall_case118_30} presents the PDFs of $P_{G_{i}} (i=\{13,14,22,24\})$ from the ASSE, SPCE, GPR, DNN, and MC simulations. It shows that when only a limited number of training samples is available (i.e., all surrogates are trained with the same sample size $N_{\mathrm{ED}}=840$), ASSE matches the MC PDFs most closely, and GPR achieves comparable accuracy. By contrast, SPCE and DNN failed to capture the behavior. 
For strongly skewed outputs (e.g., $P_{G_{22}}$ and $P_{G_{24}}$), SPCE needs to
increase the polynomial order to six and use more training samples, yet still lags behind ASSE, and only modest improvement as the sample size grows. 
With limited data, DNN performs the worst: it fails to estimate $P_{G_{24}}$, collapsing its predictions near 94, and captures the skewed behavior only with substantially larger training sets (e.g., $N_{\mathrm{ED}}\ge 4000$), at a substantial computational cost.

%


\noindent \textbf{ Decisions under Uncertainties:}   Once the PDFs of $P_{G_i}$ are accurately
estimated, their lower and upper decision bounds (i.e., the 5\% and 95\%
quantiles of $P_{G_i}$) can be determined. To further evaluate both the overall
predictive distribution and tail accuracy, Table~\ref{tab:tail_pg_comp_30_case118} reports the PBL, KSD, and WD 
metrics for the estimated $P_{G_i}$ obtained by ASSE, SPCE, GPR, and DNN. ASSE achieves the lowest PBL, KSD, and WD, indicating that its
predicted distributions are closest to the MC benchmark. The small PBL values
reflect accurate quantile (tail) estimates, while the reduced KSD and WD values
demonstrate an improved match to the CDF and overall distributional shape of $P_{G_{22}}$. In contrast, SPCE and DNN exhibit the largest errors among the
surrogate models.

Then, the obtained decision bounds of $P_{G_i}$ are directly used in the production cost calculation. 
Table~\ref{tab:IEEE118_30_cost_comp} reports the resulting lower 
$C^{\mathrm{cost}}_{\mathrm{low}}$ and upper $C^{\mathrm{cost}}_{\mathrm{up}}$ 
production cost estimates. Compared with Case~1 and Case~2, this case includes a much larger proportion of skewed, non-Gaussian, or heavy-tailed output 
distributions (about 60\% of $P_{G_i}$, versus only 11\% in Case~2), making the estimation problem more challenging. ASSE nonetheless provides noticeably more 
accurate estimates than SPCE and achieves the smallest normalized errors (in \%) 
for both bounds, offering more reliable cost bounds for risk-aware operational 
and planning decisions.

\begin{table}[htbp]
\centering

\setlength{\abovecaptionskip}{0pt}   
\setlength{\belowcaptionskip}{0pt} 
\renewcommand{\arraystretch}{1.3}
\caption{Case 3: Comparisons of metrics to evaluate the estimated distribution and tail behaviour for $P_{G_{22}}$ from  ASSE, SPCE, GPR, DNN and MC simulations. }
\label{tab:tail_pg_comp_30_case118}
\centering
\begin{tabular}{c|c|c|c|c}
\hline
Methods & $\operatorname{PBL}_{0.05}$ & $\operatorname{PBL}_{0.95}$ & $\operatorname{KSD}$ & $\operatorname{WD}$ \\
\hline
\textbf{ASSE} & $\bm{0.7616}$ & $\bm{0.8984}$ & $\bm{0.0201}$ & $\bm{0.2380}$ \\
\hline
SPCE & 0.7633 & 0.9216& 0.0545 & 0.8752 \\
\hline
GPR & 0.7679 & 0.9173 & 0.0332 & 0.4703 \\
\hline
DNN & 0.9289 & 1.0165 & 0.0767 & 1.9872 \\
\hline
\end{tabular}
\begin{tablenotes}
\item *All surrogate models are trained with the same sample size 
$N_{\mathrm{ED}} = 840$ in this table.
\end{tablenotes}
\end{table}

\begin{table}[!t]

\setlength{\abovecaptionskip}{0.cm}
\renewcommand{\arraystretch}{1.3}
\caption{Case 3: Comparison of the estimated lower and upper bounds of cost $C^{\mathrm{cost}}_{\mathrm{Index}}$  and their normalized errors in \% by ASSE, SPCE, GPR, DNN, and MC simulations.}
\label{tab:IEEE118_30_cost_comp}
\centering
\begin{tabular}{c|c|c|c|c}
\hline
Methods & $C_{\mathrm{low}}^{\mathrm{cost}}$ & $\Delta_{\mathrm{low}}^{\mathrm{cost}} \%$ & $C_{\mathrm{up}}^{\mathrm{cost}}$ & $\Delta_{\mathrm{up}}^{\mathrm{cost}} \%$  \\
\hline
ASSE & $1.8866 	\times 10^{6}$ & $\bm{0.2159}$ & $3.6638 	\times 10^{6}$ & $\bm{0.1075}$ \\
\hline
SPCE & $1.9024 	\times 10^{6}$ & $1.0540$ & $3.7747 	\times 10^{6}$ & $3.1394$ \\
\hline
GPR & $1.9127 	\times 10^{6}$ & $1.5990$ & $3.6258 	\times 10^{6}$ & $-0.9294$ \\
\hline
DNN & $2.2248 	\times 10^{6}$ & $18.1814$ & $3.3918 	\times 10^{6}$ & $-7.3244$ \\
\hline
MC & $1.8826 	\times 10^{6}$ & $--$ & $3.6598 	\times 10^{6}$ & $--$ \\
\hline
\end{tabular}
\begin{tablenotes}
\item *All surrogate models are trained with the same sample size, $N_{\mathrm{ED}}=840$.
\end{tablenotes}
\end{table}

\noindent \textbf{Computational Cost and Sample Efficiency:} Table~\ref{tab:IEEE118_30_time_comp} summarizes the runtime of ASSE and representative surrogate models (SPCE, GPR, and DNN), along with the MC benchmark. The results show that achieving comparable accuracy requires substantially more training time and samples for SPCE, and an even larger amount of data for DNN.
\begin{table}[htbp!]

\setlength{\abovecaptionskip}{0.cm}
\renewcommand{\arraystretch}{1.3}
\caption{Case 3: Comparison of computation time for estimating $P_{G_{22}}$ using the proposed ASSE, the existing methods SPCE, and GPR, DNN, and the MC benchmark, all achieving comparable accuracy.}
\label{tab:IEEE118_30_time_comp}
\centering
\begin{tabular}{c|c|c|c|c|c}
\hline
Method & $N_{\mathrm{ED}}$ & ${t_{\mathrm{ed}}}$(s) & ${t_{\mathrm{cr}}}$(s) & ${t_{\mathrm{ev}}}$(s) & ${t_{\mathrm{total}}}$(s) \\
\hline
MC    &  -- &--  &  --    & 22310.33 & 22310.33 \\
\hline
\textbf{ASSE}  & 840 &  $1874.06$   &   $58.36$    & 0.14 &  1932.56 \\
\hline
 SPCE   & 1300  &  2900.30  &   $1345.03$  & 0.19  &  4245.51 \\
\hline 
 GPR & 840  & $1874.06$  & $45.48$ & 0.17 & 1919.71\\
\hline
DNN & 4000  & $8924.00$  & $235.32$ & 0.14 & 9159.46\\
\hline
\end{tabular}
\begin{tablenotes}
\item* ASSE expansion level $K=7$, refinement splits = 33, and total expansions $D_{K}=52$. 
\item* SPCE and DNN require more training samples to achieve comparable accuracy with ASSE and GPR. 
\end{tablenotes}
\end{table}

\subsection{ Case 4: The Modified 118-Bus System with 51 Dependent Inputs } \label{sec:118_bus_51}
In this case, we further increase the input dimension to  $\mathcal{M}_{\mathrm{in}} = 51$ in the IEEE 118-bus system while retaining dependence among the stochastic inputs. Building on the 30 random inputs in
Case~3, we add 5 wind generators at buses $\{1, 6, 15, 18, 19\}$ and 16 solar PV generators at buses $\{40, 55, 56, 62, 70, 76, 77, 85, 92, 103, 104, 105,
107, 110, 116\}$, each rated at 80~MW. The system thus comprises 20 wind   generators and 31 solar PV generators, yielding 51 stochastic inputs in total.
As in Case~3, the dependence between the random inputs is modeled by a vine copula, with detailed settings given in the previously  \href{https://github.com/TxiaoWang/ASSE.git}{referenced configuration}.

\noindent{\textbf{ASSE Model Construction:}}  
Likewise, we build the ASSE framework for the probabilistic AC-OPF problem~\eqref{eq:DOPF}. Specifically,  $N_{\mathrm{ED}} $ start from $ 1000 $ and $N_{\mathrm{ref}} = 50$. The BCS algorithm adaptively selects the truncation norm (from $0.5$ to $0.8$ in steps of $0.05$) and PCE order fixed at $2$. While for the existing method SPCE, the maximum order is set as 4. 

\begin{figure}
\setlength{\abovecaptionskip}{-0.61cm}
\setlength{\belowcaptionskip}{-0.5cm}
\centering
\includegraphics[width=0.5\textwidth]{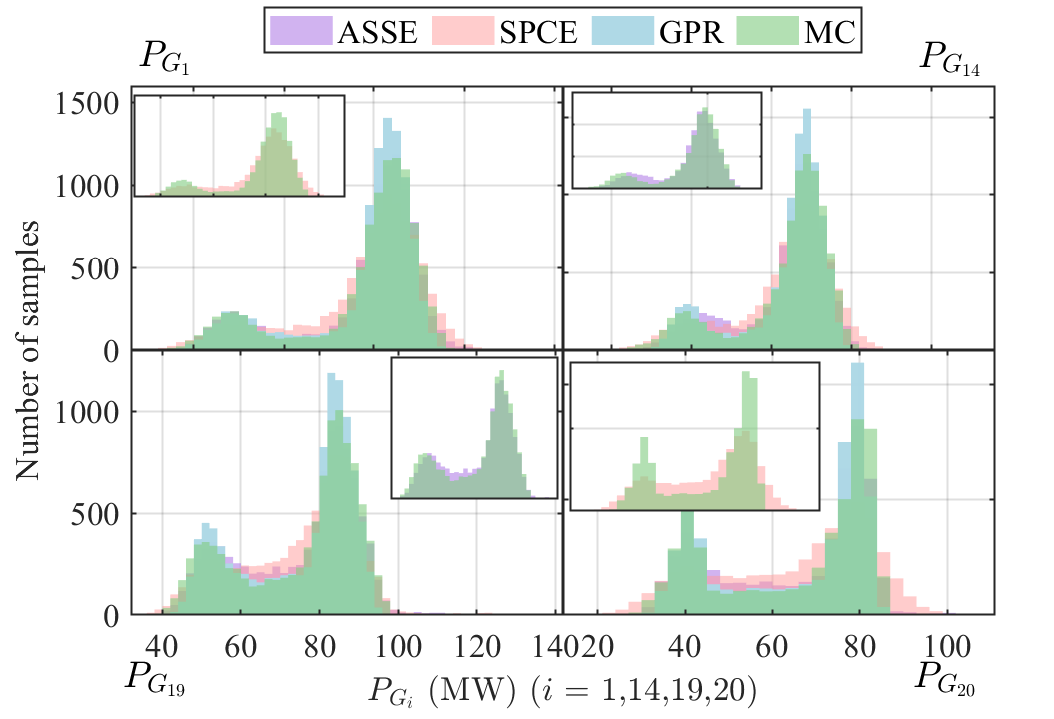}
\caption{ Case 4: Estimated PDFs of the generator output power ($P_{G_{i}},i=\{1,14,19,20\}$) from the ASSE (purple), SPCE (red), GPR (blue),  and MC simulations (green). For comparison, all surrogate models are constructed using the same training sample size ($N_{\mathrm{ED}} = 1000$). All generators exhibit skewed output distributions. Among the surrogate models, 
ASSE most closely follows the MC benchmark.}
\label{fig:ASSE_PDFall_case118_51}
\vspace{-10pt}
\end{figure}

\noindent \textbf{Performance Evaluation:} 
A total of $N_{\mathrm{val}} = 10{,}000$ validation samples of $\bm{\zeta}$
(Step~6) are generated to evaluate the performance of the constructed ASSE
model. Fig.~\ref{fig:ASSE_PDFall_case118_51} compares the PDFs of $P_{G_{i}},i=\{1, 14,19,20\}$ obtained by ASSE, SPCE, GPR,
and MC simulations. When all surrogate models are trained with the same limited sample size, ASSE consistently provides the closest match to the MC benchmark. In contrast, SPCE exhibits clearly inferior accuracy under limited data and shows only marginal improvements as the training sample size increases. 

%

\noindent \textbf{Decisions under Uncertainties:} Based on the estimated PDFs of $P_{G_i}$, we compute the 5\% and 95\% quantiles as lower and upper decision bounds. Table~\ref{tab:tail_pg_comp_51_case118} reports the PBL~\eqref{eq:pbl}, KSD~\eqref{eq:KSD}, and WD~\eqref{eq:WD_1} for the estimated $P_{G_i}$ from ASSE, SPCE, GPR, and MC. It shows that distributions estimated from ASSE remain closest to the MC benchmark even in this high-dimensional setting. Table~\ref{tab:IEEE118_51_cost_comp} summarizes the corresponding lower  $C^{\mathrm{cost}}_{\mathrm{low}}$ and upper $C^{\mathrm{cost}}_{\mathrm{up}}$ production cost bounds. In this case, ASSE continues to perform robustly,  whereas the accuracy of GPR degrades with the increased input 
dimension.

%

\begin{table}[!t]

\setlength{\abovecaptionskip}{0pt}   
\setlength{\belowcaptionskip}{0pt} 
\renewcommand{\arraystretch}{1.3}
\caption{Case 4: Comparisons of metrics to evaluate the estimated distribution and tail behaviour for $P_{G_1}$ from  ASSE, SPCE, GPR, and MC simulations. }
\label{tab:tail_pg_comp_51_case118}
\centering
\begin{tabular}{c|c|c|c|c}
\hline
Methods & $\operatorname{PBL}_{0.05}$ & $\operatorname{PBL}_{0.95}$ & $\operatorname{KSD}$ & $\operatorname{WD}$ \\
\hline
ASSE & 1.7541 & \textbf{0.7976} & \textbf{0.0177} & \textbf{0.4026} \\ \hline
SPCE & 1.7529 & 0.8507 & 0.0599 & 1.5093  \\ \hline
GPR & 1.7605 & 0.8351 & 0.0643 & 0.8843 \\ \hline
\end{tabular}
\begin{tablenotes}
\item *All surrogate models are trained with the same sample size  $N_{\mathrm{ED}}=1000$.
\end{tablenotes}
\end{table}

\begin{table}[!t]

\setlength{\abovecaptionskip}{0.cm}
\renewcommand{\arraystretch}{1.3}
\caption{Case 4: Comparison of the estimated lower and upper bounds of cost $C^{\mathrm{cost}}_{\mathrm{Index}}$  and their normalized errors in \% by ASSE, SPCE, GPR, and MC simulations.}
\label{tab:IEEE118_51_cost_comp}
\centering
\begin{tabular}{c|c|c|c|c}
\hline
Methods & $C_{\mathrm{low}}^{\mathrm{cost}}$ & $\Delta_{\mathrm{low}}^{\mathrm{cost}} \%$ & $C_{\mathrm{up}}^{\mathrm{cost}}$ & $\Delta_{\mathrm{up}}^{\mathrm{cost}} \%$ \\ 
\hline
\textbf{ASSE} & $1.6435 	\times 10^{5}$ & $\bm{2.3402}$ & $8.7929 	\times 10^{5}$ & $\bm{-1.2514}$ \\ \hline
SPCE & $1.5676 	\times 10^{5}$ & $-2.3838$ & $9.0253 	\times 10^{5}$ & $1.3585$ \\ \hline
GPR & $1.7142 	\times 10^{5}$ & $6.7442$ & $8.4651 	\times 10^{5}$ & $-4.9333$ \\ \hline
MC & $1.6059 	\times 10^{5}$ & $--$ & $8.9043 	\times 10^{5}$ & $--$ \\
\hline
\end{tabular}
\end{table}

\noindent \textbf{Computational Cost and Sample Efficiency:} 
Regarding the computational cost to achieve comparable accuracy in the
$P_{G_{1}}$ estimations, SPCE and GPR both require more training samples, and thus longer time $t_{\mathrm{ed}}+t_{\mathrm{cr}}$ required to construct the model compared with  
ASSE. Detailed construction,
evaluation, and total runtimes, together with the corresponding training sample sizes required to achieve comparable accuracy, are summarized in
Table~\ref{tab:IEEE118_51_time_comp}.

\begin{table}[htbp!]

\setlength{\abovecaptionskip}{0.cm}
\renewcommand{\arraystretch}{1.3}
\caption{Case 4: Comparison of computation time for estimating $P_{G_{1}}$ using the proposed ASSE, 
the existing methods SPCE, AND GPR, and the MC benchmark}
\label{tab:IEEE118_51_time_comp}
\centering
\begin{tabular}{c|c|c|c|c|c}
\hline
Method & $N_{\mathrm{ED}}$ & ${t_{\mathrm{ed}}}$(s) & ${t_{\mathrm{cr}}}$(s) & ${t_{\mathrm{ev}}}$(s) & ${t_{\mathrm{total}}}$(s) \\
\hline
MC    &  -- &--  &  --    & 22410.80 & 22410.80 \\
\hline
\textbf{ASSE}  & 1000 &  $2241.10$   &   $88.27$    & 0.21 &  2329.58\\
\hline
 SPCE   & 2000  & 4482.58   &   $5176.97$  & 0.21  &  9659.55 \\
\hline 
 GPR & 1200  & $2689.2$   & $80.61$ & 0.21& 2770.02\\
\hline
\end{tabular}
\begin{tablenotes}
\item * Different $N_{\mathrm{ED}}$ samples and time consumption are summarized in this table for different methods to reach similar accuracy. SPCE and GPR require larger $N_{\mathrm{ED}}$. 
\end{tablenotes}
\vspace{-0.5cm}
\end{table} 


\subsection{Case 5: Real-World Data Test Case}
\label{sec:empirical_openmeteo_case}
To further evaluate the robustness of ASSE under realistic uncertainty patterns, we add a real-world historical data test, based on the modified IEEE 118-bus system. 
This case adopts the same wind/solar integration settings as Case 2, including the same number of wind farms and solar PV plants as well as the same bus locations. The key difference is that the 12-dimensional uncertainty inputs are constructed from Open-Meteo historical weather profiles~\cite{OpenMeteoHistoricalWeatherAPI}, rather than from prescribed parametric distributions. Specifically, hourly wind-speed profiles at a height of 100 m from 2010 to 2012 are used for the six wind farms, while hourly shortwave-radiation profiles are used for the six PV plants. Before constructing the training and testing datasets, short missing intervals are filled using linear interpolation, and abnormal records are screened using a three-standard-deviation rule. Unlike the previous parametric cases, Case 5 directly uses empirical historical data and does not impose Weibull, Beta, or other prescribed marginal distributions. The full dataset contains 20,000 samples, and the detailed data-extraction settings are provided in the \href{https://github.com/TxiaoWang/ASSE.git}{referenced configuration}.

\noindent{\textbf{ASSE Model Construction:}} 
Similar to the previous cases, we construct the ASSE surrogate for the probabilistic AC-OPF problem~\eqref{eq:DOPF}. 
For this real-world data case, all surrogate models are trained using the same training sample size, $N_{\mathrm{ED}}=840$ and $N_{\mathrm{ref}}=72$. 
The PCE order and truncation norm settings are kept the same as in Case 2. The proposed ASSE method is compared with SPCE, GPR, DNN, and the MC benchmark.

\begin{figure}[htbp!]
\setlength{\abovecaptionskip}{-0.41cm}
\setlength{\belowcaptionskip}{-0.5cm}
\centering
\includegraphics[width=0.5\textwidth]{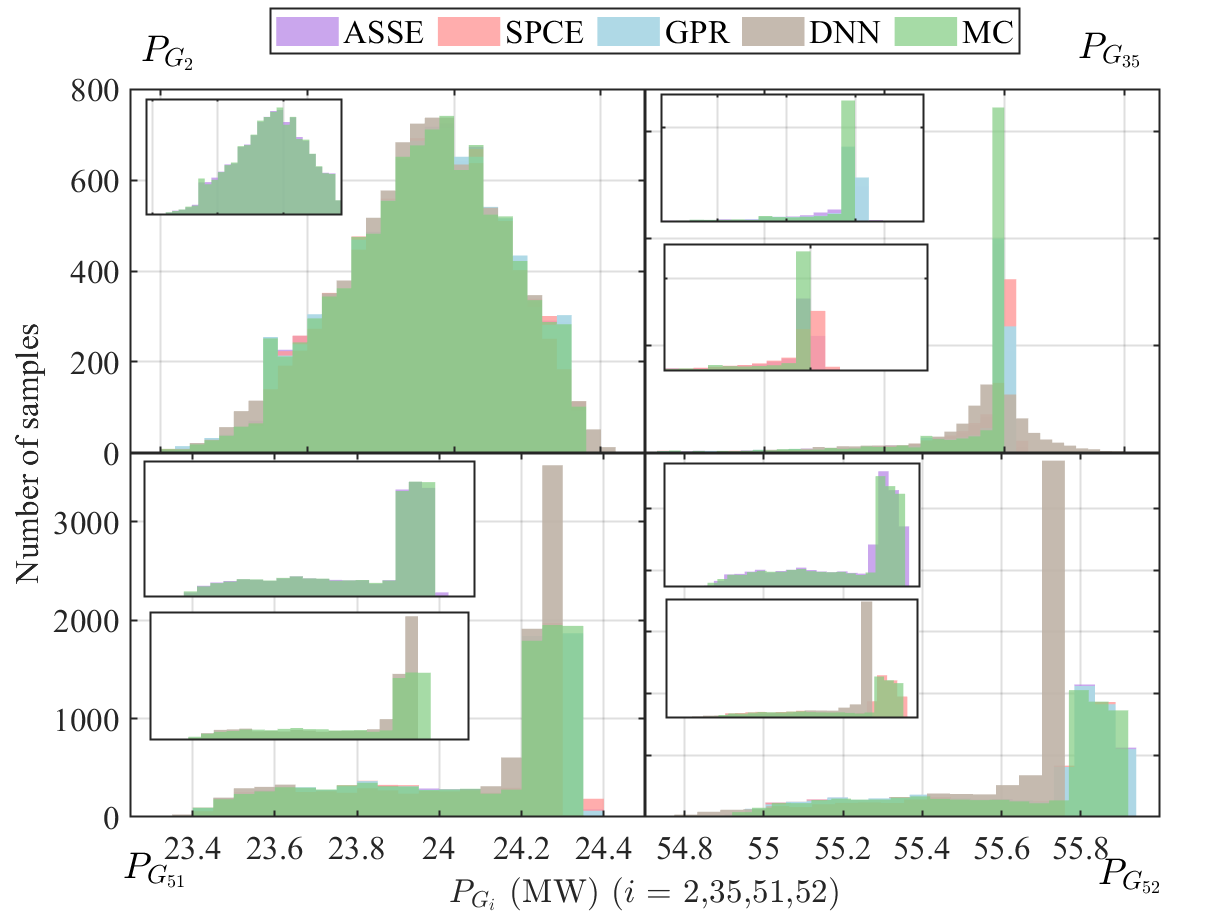}
\caption{ Case 5: Estimated PDFs of the generator output power ($P_{G_{i}}, i=\{2,35,51,52\}$) from the ASSE (purple), SPCE (red), GPR (blue), DNN (brown), and MC simulations (green). For comparison, all surrogate models are constructed using the same training sample size ($N_{\mathrm{ED}} = 840$). ASSE remains good accuracy for both Gaussian-like ($P_{G_{2}}$) and skewed output distributions ($P_{G_{35}}$, $P_{G_{51}}$, $P_{G_{52}}$). }
\label{fig:ASSE_PDFall_case118_real}
\vspace{-10pt}
\end{figure}

\noindent \textbf{Performance Evaluation:} 
A validation set with $N_{\mathrm{val}}=10,000$ samples is directly  acquired from the real-data set to evaluate the constructed surrogate models. 
Fig.~\ref{fig:ASSE_PDFall_case118_real} compares the estimated PDFs of representative generator outputs $P_{G_{i}},i=\{2, 35,51,52\}$, obtained by ASSE, SPCE, GPR, DNN, and MC simulations. The results show that ASSE captures the output distributions well under the limited training sample size. 
GPR also provides comparable distributional accuracy for some outputs, whereas SPCE and DNN show larger deviations in several localized regions. These results indicate that ASSE remains stable when the input uncertainties are derived from empirical historical profiles rather than prescribed Weibull or Beta distributions.

\noindent\textbf{Decisions under Uncertainties:} 
Table~\ref{tab:tail_pg_comp_real_case118} reports the PBL~\eqref{eq:pbl}, KSD~\eqref{eq:KSD}, and WD~\eqref{eq:WD_1} for the estimated generator output distributions. 
The results show that ASSE achieves the lowest KSD and competitive PBL and WD values under empirical wind and solar uncertainty, indicating that the proposed ASSE remains stable under real-data conditions.
Moreover, Table~\ref{tab:IEEE118_real_cost_comp} summarizes the corresponding lower and upper production cost bounds, $C^{\mathrm{cost}}_{\mathrm{low}}$ and $C^{\mathrm{cost}}_{\mathrm{up}}$. 
Compared with the MC benchmark, ASSE yields the smallest errors in the lower and upper bound cost estimates. 
This demonstrates that the proposed ASSE framework can provide reliable decision-oriented cost bounds under empirical renewable uncertainty.

\begin{table}[!t]

\setlength{\abovecaptionskip}{0pt}   
\setlength{\belowcaptionskip}{0pt} 
\renewcommand{\arraystretch}{1.3}
\caption{Case 5: Comparisons of metrics to evaluate the estimated distribution and tail behaviour for $P_{G_{35}}$ from  ASSE, SPCE, GPR, DNN, and MC simulations. }
\label{tab:tail_pg_comp_real_case118}
\centering
\begin{tabular}{c|c|c|c|c}
\hline
Methods & $\operatorname{PBL}_{0.05}$ & $\operatorname{PBL}_{0.95}$ & $\operatorname{KSD}$ & $\operatorname{WD}$ \\
\hline
ASSE & \textbf{0.1239} & 0.0204 & \textbf{0.2021} & {0.0328} \\ \hline
SPCE & 0.1241 & 0.0222 & 0.3481 & 0.0483  \\ \hline
GPR & 0.1239 & 0.0190 & 0.2367 & 0.0062 \\ \hline
DNN & 0.1246 & 0.0446 & 0.3093 & 0.1740 \\ \hline
\end{tabular}
\begin{tablenotes}
\item *All surrogate models are trained with the same sample size  $N_{\mathrm{ED}}=840, N_{\mathrm{ref}}=72$.
\end{tablenotes}
\end{table}

\begin{table}[htbp!]

\setlength{\abovecaptionskip}{0.cm}
\renewcommand{\arraystretch}{1.3}
\caption{Case 5: Comparison of the estimated lower and upper bounds of cost $C^{\mathrm{cost}}_{\mathrm{Index}}$  and their normalized errors in \% by ASSE, SPCE, GPR, DNN, and MC simulations.}
\label{tab:IEEE118_real_cost_comp}
\centering
\begin{tabular}{c|c|c|c|c}
\hline
Methods & $C_{\mathrm{low}}^{\mathrm{cost}}$ & $\Delta_{\mathrm{low}}^{\mathrm{cost}} \%$ & $C_{\mathrm{up}}^{\mathrm{cost}}$ & $\Delta_{\mathrm{up}}^{\mathrm{cost}} \%$ \\ 
\hline
ASSE &  $8.9634\times 10^{5}$ & {$\bf{0.0408}$} & $1.3764\times 10^{6}$ & $\bf{-0.0100}$ \\\hline
SPCE &  $8.8536\times 10^{5}$ & $-1.1839$ & $1.3642\times 10^{6}$ & $-0.8984$ \\\hline
GPR  & $8.6707\times 10^{5}$ & $-3.2254$ & $1.3474\times 10^{6}$ & $-2.1222$ \\\hline
DNN  &  $9.0027\times 10^{5}$ & $0.4798$ & $1.3693\times 10^{6}$ & $-0.5259$ \\\hline
MC   &  $8.9597\times 10^{5}$ & --      & $1.3766\times 10^{6}$ & -- \\
\hline
\end{tabular}
\end{table}

%

\begin{remark} 
During data generation, each operating point is obtained by solving the probabilistic AC-OPF in \eqref{eq:DOPF}. Under the uncertainty settings in Case~1--Case~5, all OPF runs converged and no infeasible samples were observed. Under stressed settings (e.g., enlarged load perturbations or tightened generation limits), a small fraction of OPF runs became infeasible or non-convergent; for example, in the $\mathcal{M}_{\mathrm{in}}=101$ test (Appendix~\ref{re:101_dimension}), the failure rate was $4/10000$ ($0.04\%$). All failed OPF samples were discarded and excluded from both training and evaluation. Constraint activations (e.g., generators/voltages hitting limits) can introduce non-smooth regime changes; however, ASSE maintains accuracy in tail regions without requiring explicit manual partitioning.
\end{remark}

\begin{remark} {\textbf{Discussions:}}
The benchmark methods considered in this paper include SPCE, GPR, and DNN, which represent sparse spectral, probabilistic kernel-based, and neural-network surrogate families, respectively. 
Recent uncertainty-quantification studies have also explored more advanced alternatives, such as normalizing flows \cite{rezende2015variational}, Transformer-based models \cite{vaswani2017attention}, Bayesian optimization \cite{wang2016bayesian}, and ensemble-based methods~\cite{zhou2025ensemble}. 
These methods are promising for high-dimensional uncertainty modeling, but they generally require different modeling formulations or additional design choices. For example, normalizing flows are more naturally formulated as density or conditional-density estimators, Bayesian optimization focuses on sequential sample acquisition, Transformer-based surrogates require architecture-specific attention design, and ensemble methods require choices of base learners, ensemble size, combination rules, and uncertainty calibration. In contrast to these alternatives,  ASSE targets the small-to-medium sample regime common in expensive simulations. 
Its sparse closed-form expansion is sample-efficient and lightweight to train.  A distinguishing feature of ASSE is its ability to analytically compute key statistics (e.g., mean and variance) and global sensitivity measures (e.g., Sobol' indices) directly from the PCE coefficient structure without requiring additional MC sampling. 

In Case~3, ASSE attains distribution and tail-quantile accuracy comparable to a DNN surrogate with substantially fewer samples, highlighting its complementary advantages. 
Extending ASSE with flow-based density models, Transformer-based architectures, adaptive Bayesian-optimization sampling, or ensemble-based uncertainty calibration is a promising direction for future work.
\end{remark}

%
%

\section{Conclusions} \label{sec:cons}
This paper proposed an ASSE method to effectively and accurately address the probabilistic AC-OPF problem in the presence of uncertainties. The proposed method efficiently  and accurately estimated the complete probability distributions (e.g., PDFs, means, variances, and quantile-based  CI bounds) 
of probabilistic AC-OPF solutions while minimizing power losses. These distributions are then mapped to confidence-interval production-cost bands, providing operators with explicit upper and lower cost bounds under uncertainty. 
Specially, an adaptive domain partition strategy was incorporated for domain refinement, and a Bayesian compressive sensing–based coefficient calculation algorithm guaranteed flexible parameter selection and enhanced performance with limited training data.   Numerical studies on modified IEEE 9-bus and IEEE 118-bus systems demonstrated that the proposed method verified accurate probabilistic characteristic estimations of OPF solutions (e.g., generator power outputs and the objective function). The efficacy of the proposed method was validated by comparing it with the MC simulations and   representative surrogate models, 
including SPCE, GPR, and DNN.  
Furthermore, ASSE remained markedly more accurate for skewed, heavy-tailed, or multimodal output distributions, making it a reliable tool for robust operational planning.
\vspace{-10pt}

\appendices
\section{The BCS Algorithm}
\label{sec:append_bcs}
To solve the coefficients in the PCE-based model \eqref{eq:PCE}, the optimization problem \eqref{eq:coe_optmization} is reformulated under a Bayesian framework to facilitate sparsity, probabilistic predictions, and automatic hyperparameter estimations. The observation model assumes independent Gaussian noise:
\small
\begin{equation}
\setlength{\abovedisplayskip}{3pt}
\setlength{\belowdisplayskip}{3pt}
    p(\bm{Z}^{\bm{\alpha}}|\bm{C}^{\bm{\alpha}}, \sigma^2) = \mathcal{N}(\bm{Z}^{\bm{\alpha}}|\bm{\Phi}^{\bm{\alpha}}\bm{C}^{\bm{\alpha}}, \sigma^2\bm{I})
\end{equation}
\normalsize
with variance $\sigma^2$ as a hyperparameter and identity matrix $\bm{I}$.  This Gaussian observation model is used as a working likelihood for sparse coefficient estimation in each local PCE model. It does not imply that the uncertain inputs, OPF outputs, or final estimated output distributions are Gaussian. 
Hierarchical priors are introduced to promote sparsity in $\bm{C}^{\bm{\alpha}}$. Specifically, the coefficient priors are modeled as:
\begin{equation}
\small
\setlength{\abovedisplayskip}{2pt}
\setlength{\belowdisplayskip}{2pt}
    p(\bm{C}^{\bm{\alpha}}|\bm{\gamma}) = \prod_{m=1}^{M}\mathcal{N}(c_{m}^{\bm{\alpha}}|0, \gamma_{m}),
\end{equation}
\normalsize
with hyperparameter $\bm{\gamma}=\{\gamma_{1},\dots,\gamma_{M}\}$, each following an exponential prior:
\begin{equation}
\small
\setlength{\abovedisplayskip}{2pt}
\setlength{\belowdisplayskip}{2pt}
    p(\gamma_{m}|\lambda) = \Gamma\left(\gamma_{m}|1,\frac{\lambda}{2}\right)=\frac{\lambda}{2}\exp\left(-\frac{\lambda \gamma_{m}}{2}\right), \quad \gamma_{m}\geq0.
\end{equation}
\normalsize
where the hyperparameter $\lambda$ itself is drawn from a Gamma hyperprior:
%
$ p(\lambda|\nu) = \Gamma\left(\lambda|\frac{\nu}{2},\frac{\nu}{2}\right), \nu\rightarrow 0$. Then, the coefficients $\bm{C}^{\bm{\alpha}}$ is inferred by maximizing the posterior distribution $p(\bm{C}^{\bm{\alpha}},\bm{\gamma},\lambda|\bm{Z}^{\bm{\alpha}})$ via:
%
\small
\begin{equation}
\setlength{\abovedisplayskip}{2pt}
\setlength{\belowdisplayskip}{2pt}
\label{eq:final_likelihood}
\widehat{\bm{C}}^{\bm{\alpha}}
= \arg\max_{\bm{C}^{\bm{\alpha}},\bm{\gamma},\lambda} \left\{ p(\bm{Z}^{\bm{\alpha}}|\bm{C}^{\bm{\alpha}},\sigma^2)\, p(\bm{C}^{\bm{\alpha}}|\bm{\gamma})\, p(\bm{\gamma}|\lambda)\, p(\lambda|\nu) \right\}
\end{equation}
\normalsize
which is solved via a fast Laplace method as described in \cite{Babacan2010,Tsilifis2019compressive}.  In this Bayesian framework, the hyperparameters $\gamma_m$ and $\lambda$ are updated iteratively by the Fast Laplace algorithm rather than tuned manually, while the noise variance $\sigma^2$ is selected from a logarithmically spaced grid by minimizing the modified cross-validation error $e_{\mathrm{mcv}}$. This results in a fully data-driven choice of $\gamma_m$, $\lambda$, and $\sigma^2$ without case-specific manual tuning. 
Detailed algorithmic implementation is provided in Algorithm 1 of \cite{Babacan2010} and Appendix B of \cite{Tsilifis2019compressive}. 

Although the current BCS formulation uses a Gaussian working likelihood, robust Bayesian sparse regression based on heavier-tailed likelihoods (e.g., Student-$t$ likelihoods), could further improve robustness when local regression residuals exhibit strong heavy-tailed behavior. This extension is left as an important direction for future work.

\vspace{-12pt}
\section{Evaluation Metrics} \label{appen:evaluation}
This appendix defines the evaluation metrics: validation error $e_{\mathrm{val}}$, PBL, KSD, and WD, discussed in Remark~\ref{rem:remark2}, which measure pointwise prediction error, quantile accuracy, worst-case CDF mismatch, and overall distributional discrepancy, respectively.

The validation error adopted to evaluate the performance is computed by:
\begin{equation}
\label{eq:val_error}
e_{\mathrm{Val}}=\frac{N_{\mathrm{Val}}-1}{N_{\mathrm{Val}}}\left[\frac{\sum_{n=1}^{N_{\mathrm{Val}}}\left(Z^{(n)}-F_{\mathrm{SSE}}(\bm{\zeta}^{(n)})\right)^2}{\sum_{n=1}^{N_{\mathrm{Val}}}\left(Z^{(n)}-\hat{\mu}_{Z_{\mathrm{Val}}}\right)^2}\right]
\end{equation}
where $\hat{\mu}_{Z{_\mathrm{Val}}} = \frac{1}{N_{\mathrm{Val}}}\sum_{n=1}^{N_{\mathrm{Val}}}Z^{(n)}$ represents the sample mean of the response, derived from $N_{\mathrm{Val}}$ random input samples.


The PBL  at a certain quantile $\tau_{\mathrm{qtl}} \in (0,1)$ is given by \cite{wang2019probabilistic}:
\small
\begin{equation}
\label{eq:pbl}
\operatorname{PBL}(Z, \hat Z_{\tau_{\mathrm{qtl}}})=
\begin{cases}
(1-\tau_{\mathrm{qtl}})\left(\hat{Z}_{\tau_{\mathrm{qtl}}}-Z\right), & \hat{Z}_{\tau_{\mathrm{qtl}}}\ge Z\\
\tau_{\mathrm{qtl}}\left(Z-\hat{Z}_{\tau_{\mathrm{qtl}}}\right), & \hat{Z}_{\tau_{\mathrm{qtl}}}< Z
\end{cases}
\end{equation}
\normalsize
where $\hat{Z}_{\tau_{\mathrm{qtl}}}$ denotes the estimated $\tau_{\mathrm{qtl}}$-th quantile from the ASSE method. Smaller PBL values indicate a more accurate prediction at the specified quantile level. PBL has the same physical units as ${Z}$ (e.g., MW for $P_{G_i}$).

The empirical KSD can be determined by the maximum vertical difference between two CDFs \cite{massey1951kolmogorov}:
\begin{equation}
\label{eq:KSD}
   \operatorname{KSD}(Z, \hat Z)
  = \\
  \sup_{z \in \mathbb{R}}
    \left|
      \Pr(Z \le z)
      -
       \Pr(\hat{Z}\le z)
    \right|
\end{equation}
where $\Pr(Z \le z)$ and $\Pr(\hat Z \le z)$ denote the empirical CDFs of $Z$ and $\hat{Z}$,
respectively. If KSD is small, the two distributions are uniformly close in terms of their CDF (i.e., the maximum CDF mismatch is small).

The empirical WD metric can be calculated by \cite{panaretos2019}: 
\begin{equation}
\label{eq:WD_1}
 \operatorname{WD}(Z, \hat Z) = \frac{1}{N_{\mathrm{val}}}\sum_{w=1}^{N_{\mathrm{val}}}
    \bigl| Z_{(w)} - \hat Z_{(w)}\bigr|
\end{equation}
where $Z_{(w)}$ and $\hat Z_{(w)}$ denote the $w$-th order statistics of the empirical samples from true and predicted distributions of $Z$, respectively. A smaller $\operatorname{WD}(Z,\hat Z)$ indicates close overall agreement between the two distributions, and it can be interpreted as the average absolute discrepancy between matched (sorted) samples. Moreover, WD has the same physical units as $Z$.

\section{Training-Time Scaling of ASSE Parameters}
\label{app:training_time}

\begin{figure}[htbp!]
\vspace{-8pt}
\centering
\setlength{\abovecaptionskip}{-0.51cm}
\includegraphics[width=0.5\textwidth]{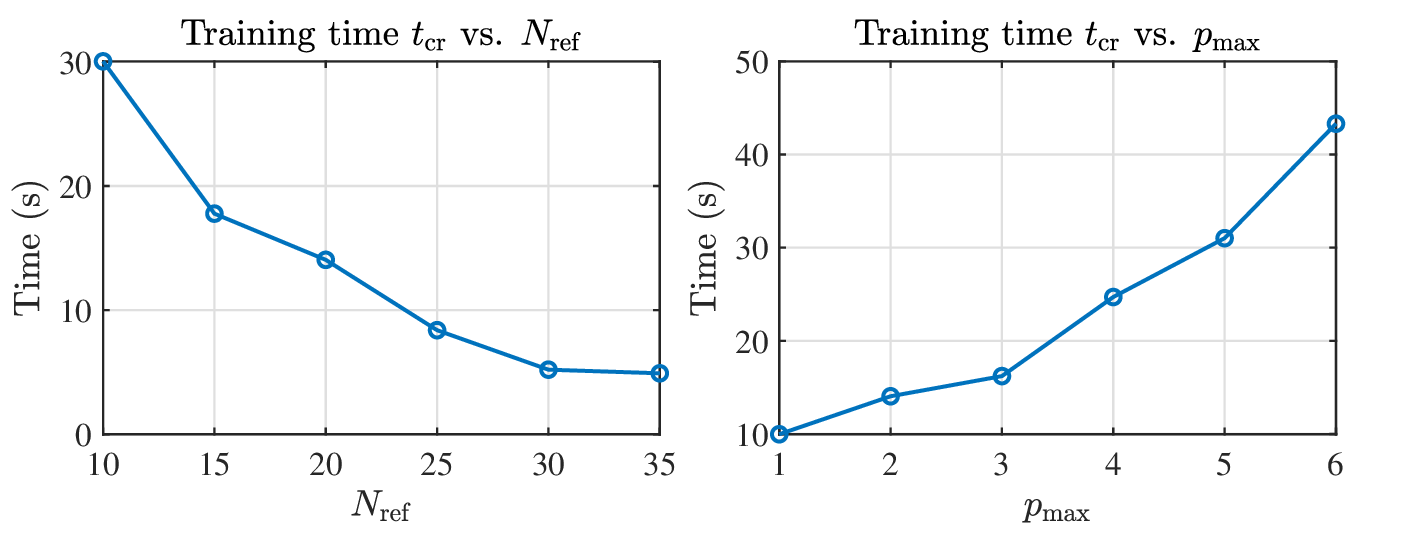}
\caption{Training-time scaling of ASSE parameters. Left: model training time $t_\mathrm{cr}$ versus $N_{\mathrm{ref}}$. Training number of sample size $N_{\mathrm{ED}} =350$,  $p_{\mathrm{max}}=2$ and $q_{\mathrm{max}}=0.8$ are set the same for $N_{\mathrm{ref}}$ variation test; Right:   model training time $t_\mathrm{cr}$ versus max order $p_{\mathrm{max}}$. Training number of sample size $N_{\mathrm{ED}}=350$,   $N_{\mathrm{ref}} =20$ and $q_{\mathrm{max}}=0.8$ are set the same for $p_{\mathrm{max}}$ variation test.}
\label{fig:case_1_ASSE_N_ref_pmax_metrics}
 \end{figure}

Fig.~\ref{fig:case_1_ASSE_N_ref_pmax_metrics} reports the empirical scaling of ASSE training time with respect to key algorithmic parameters.  All results are obtained on the IEEE 9-bus system (Case~1), while keeping all unspecified parameters fixed. 

\vspace{-14pt}\section{$\mathcal{M}_{\mathrm{in}}=101$ Test}
  \label{re:101_dimension}
 To further evaluate the proposed ASSE under high dimensionality, we increase the input dimension to $\mathcal{M}_{\mathrm{in}}=101$ by augmenting Case~4 ($\mathcal{M}_{\mathrm{in}}=51$) with 50 additional random load inputs, each set to a 10\% variation of the base load power. For highly skewed and multimodal output distributions in this 101-dimensional setting, ASSE continues to outperform the baseline methods. In contrast, GPR and SPCE generally require more training samples and a longer total construction time ($t_{\mathrm{ed}}+t_{\mathrm{cr}}$) to achieve accuracy comparable to ASSE. For complex responses (e.g., strongly skewed or nonsmooth),  ASSE attains reasonable good accuracy with $N_{\mathrm{ED}} $ around 1000, whereas SPCE and GPR typically require substantially more samples (e.g., $N_{\mathrm{ED}}\ge 2000$), yet still lag behind.



\bibliographystyle{IEEEtran}
\bibliography{Final_main}

@INPROCEEDINGS{Leeton2010,
  author={Leeton, U. and Uthitsunthorn, D. and Kwannetr, U. and Sinsuphun, N. and Kulworawanichpong, T.},
  booktitle={ECTI-CON2010: The 2010 ECTI International Confernce on Electrical Engineering/Electronics, Computer, Telecommunications and Information Technology}, 
  title={Power loss minimization using optimal power flow based on particle swarm optimization}, 
  year={2010},
  volume={},
  number={},
  pages={440-444},
  doi={}}

@ARTICLE{Wang2024,
  author={Wang, Zhengcheng and Zhou, Yanzhen and Guo, Qinglai and Sun, Hongbin},
  journal={IEEE Transactions on Power Systems}, 
  year={2024},
  volume={39},
  number={1},
  pages={453-464},
  doi={10.1109/TPWRS.2023.3236330}}

@article{Rajabi2019review,
  title={Review and comparison of two meta-model-based uncertainty propagation analysis methods in groundwater applications: polynomial chaos expansion and Gaussian process emulation},
  author={Rajabi, Mohammad Mahdi},
  journal={Stochastic environmental research and risk assessment},
  volume={33},
  pages={607--631},
  year={2019},
  publisher={Springer}
}

@ARTICLE{Pareek2021,
  author={Pareek, Parikshit and Nguyen, Hung D.},
  journal={IEEE Trans. Power Syst.}, 
  title={Gaussian Process Learning-Based Probabilistic Optimal Power Flow}, 
  year={2021},
  volume={36},
  number={1},
  pages={541-544},
  doi={10.1109/TPWRS.2020.3031765}}

@ARTICLE{Mühlpfordt2019,
  author={Mühlpfordt, Tillmann and Roald, Line and Hagenmeyer, Veit and Faulwasser, Timm and Misra, Sidhant},
  journal={IEEE Trans. Power Syst.}, 
  title={Chance-Constrained AC Optimal Power Flow: A Polynomial Chaos Approach}, 
  year={2019},
  volume={34},
  number={6},
  pages={4806-4816},
  doi={10.1109/TPWRS.2019.2918363}}

@ARTICLE{Sheng2018,
  author={Sheng, Hao and Wang, Xiaozhe},
  journal={IEEE Trans. Power Syst.}, 
  title={Applying Polynomial Chaos Expansion to Assess Probabilistic Available Delivery Capability for Distribution Networks With Renewables}, 
  year={2018},
  volume={33},
  number={6},
  pages={6726-6735},
  doi={10.1109/TPWRS.2018.2825657}}

@ARTICLE{Ly2023,
  author={Ly, Sel and Pareek, Parikshit and Nguyen, Hung D.},
  journal={IEEE Systems Journal}, 
  title={Scalable Probabilistic Optimal Power Flow for High Renewables Using Lite Polynomial Chaos Expansion}, 
  year={2023},
  volume={17},
  number={2},
  pages={2282-2293},
  doi={10.1109/JSYST.2022.3198726}}

@ARTICLE{Pan2023,
  author={Pan, Xiang and Chen, Minghua and Zhao, Tianyu and Low, Steven H.},
  journal={IEEE Systems Journal}, 
  title={DeepOPF: A Feasibility-Optimized Deep Neural Network Approach for AC Optimal Power Flow Problems}, 
  year={2023},
  volume={17},
  number={1},
  pages={673-683},
  doi={10.1109/JSYST.2022.3201041}}

@ARTICLE{Gao2023,
  author={Gao, Maosheng and Yu, Juan and Yang, Zhifang and Zhao, Junbo},
  journal={IEEE Trans. Power Syst.}, 
  title={A Physics-Guided Graph Convolution Neural Network for Optimal Power Flow}, 
  year={2023},
  volume={},
  number={},
  pages={1-11},
  doi={10.1109/TPWRS.2023.3238377}}

@ARTICLE{Falconer2022,
  author={Falconer, Thomas and Mones, Letif},
  journal={IEEE Transactions on Power Systems}, 
  title={Leveraging Power Grid Topology in Machine Learning Assisted Optimal Power Flow}, 
  year={2023},
  volume={38},
  number={3},
  pages={2234-2246},
  doi={10.1109/TPWRS.2022.3187218}}

@ARTICLE{Park2024,
  author={Park, Seonho and Chen, Wenbo and Mak, Terrence W.K. and Van Hentenryck, Pascal},
  journal={IEEE Transactions on Power Systems}, 
  title={Compact Optimization Learning for AC Optimal Power Flow}, 
  year={2024},
  volume={39},
  number={2},
  pages={4350-4359},
  doi={10.1109/TPWRS.2023.3313438}}

@ARTICLE{Hasan2021,
  author={Hasan, Fouad and Kargarian, Amin and Mohammadi, Javad},
  journal={IEEE Transactions on Industry Applications}, 
  title={Hybrid Learning Aided Inactive Constraints Filtering Algorithm to Enhance AC OPF Solution Time}, 
  year={2021},
  volume={57},
  number={2},
  pages={1325-1334},
  doi={10.1109/TIA.2021.3053516}}

@ARTICLE{Liu2023OPF,
  author={Liu, Shudi and Guo, Ye and Tang, Wenjun and Sun, Hongbin and Huang, Wenqi and Hou, Jiaxuan},
  journal={IEEE Transactions on Power Systems}, 
  title={Varying Condition SCOPF Based on Deep Learning and Knowledge Graph}, 
  year={2023},
  volume={38},
  number={4},
  pages={3189-3200},
  doi={10.1109/TPWRS.2022.3199238}}

@INPROCEEDINGS{Dong2020,
  author={Dong, Wenqian and Xie, Zhen and Kestor, Gokcen and Li, Dong},
  booktitle={SC20: International Conference for High Performance Computing, Networking, Storage and Analysis}, 
  title={Smart-PGSim: Using Neural Network to Accelerate AC-OPF Power Grid Simulation}, 
  year={2020},
  volume={},
  number={},
  pages={1-15},
  doi={10.1109/SC41405.2020.00067}}

@ARTICLE{Zou2014,
  author={Zou, Bin and Xiao, Qing},
  journal={IEEE Trans. Power Syst.}, 
  title={Solving Probabilistic Optimal Power Flow Problem Using Quasi Monte Carlo Method and Ninth-Order Polynomial Normal Transformation}, 
  year={2014},
  volume={29},
  number={1},
  pages={300-306},
  doi={10.1109/TPWRS.2013.2278986}}

@ARTICLE{Schellenberg2005,
  author={Schellenberg, A. and Rosehart, W. and Aguado, J.},
  journal={IEEE Trans. Power Syst.}, 
  title={Cumulant-based probabilistic optimal power flow (P-OPF) with Gaussian and gamma distributions}, 
  year={2005},
  volume={20},
  number={2},
  pages={773-781},
  doi={10.1109/TPWRS.2005.846184}}

@ARTICLE{Tamtum2009,
  author={Tamtum, Ali and Schellenberg, Antony and Rosehart, William D.},
  journal={IEEE Trans. Power Syst.}, 
  title={Enhancements to the Cumulant Method for Probabilistic Optimal Power Flow Studies}, 
  year={2009},
  volume={24},
  number={4},
  pages={1739-1746},
  doi={10.1109/TPWRS.2009.2030409}}

@ARTICLE{Yang2018,
  author={Yang, Zhifang and Zhong, Haiwang and Bose, Anjan and Zheng, Tongxin and Xia, Qing and Kang, Chongqing},
  journal={IEEE Transactions on Power Systems}, 
  title={A Linearized OPF Model With Reactive Power and Voltage Magnitude: A Pathway to Improve the MW-Only DC OPF}, 
  year={2018},
  volume={33},
  number={2},
  pages={1734-1745},
  doi={10.1109/TPWRS.2017.2718551}}

@ARTICLE{Niu2014,
  author={Niu, Ming and Wan, Can and Xu, Zhao},
  journal={Journal of Modern Power Systems and Clean Energy}, 
  title={A review on applications of heuristic optimization algorithms for optimal power flow in modern power systems}, 
  year={2014},
  volume={2},
  number={4},
  pages={289-297},
  doi={10.1007/s40565-014-0089-4}}

@ARTICLE{Onate2008,
  author={Onate Yumbla, Pablo E. and Ramirez, Juan M. and Coello Coello, Carlos A.},
  journal={IEEE Transactions on Power Systems}, 
  title={Optimal Power Flow Subject to Security Constraints Solved With a Particle Swarm Optimizer}, 
  year={2008},
  volume={23},
  number={1},
  pages={33-40},
  doi={10.1109/TPWRS.2007.913196}}

@article{Gao2023physics,
  title={A Physics-Guided Graph Convolution Neural Network for Optimal Power Flow},
  author={Gao, Maosheng and Yu, Juan and Yang, Zhifang and Zhao, Junbo},
  journal={IEEE Trans. Power Syst.},
  year={2023},
  publisher={IEEE}
}

@ARTICLE{wang2022,
  author={Wang, Xiaoting and Liu, Rong-Peng and Wang, Xiaozhe and Hou, Yunhe and Bouffard, François},
  journal={IEEE Transactions on Power Systems}, 
  title={A Data-Driven Uncertainty Quantification Method for Stochastic Economic Dispatch}, 
  year={2022},
  volume={37},
  number={1},
  pages={812-815},
  doi={10.1109/TPWRS.2021.3114083}}

@ARTICLE{Liu2020,
  author={Liu, Rong-Peng and Lei, Shunbo and Peng, Chaoyi and Sun, Wei and Hou, Yunhe},
  journal={IEEE Transactions on Smart Grid}, 
  title={Data-Based Resilience Enhancement Strategies for Electric-Gas Systems Against Sequential Extreme Weather Events}, 
  year={2020},
  volume={11},
  number={6},
  pages={5383-5395},
  doi={10.1109/TSG.2020.3007479}}

@article{Marelli2021stochastic,
  title={Stochastic spectral embedding},
  author={Marelli, Stefano and Wagner, Paul-Remo and Lataniotis, Christos and Sudret, Bruno},
  journal={International Journal for Uncertainty Quantification},
  volume={11},
  number={2},
  year={2021},
  publisher={Begel House Inc.}
}

@article{Xiu2002wiener,
  title={The Wiener--Askey polynomial chaos for stochastic differential equations},
  author={Xiu, Dongbin and Karniadakis, George Em},
  journal={SIAM journal on scientific computing},
  volume={24},
  number={2},
  pages={619--644},
  year={2002},
  publisher={SIAM}
}

@ARTICLE{Wang2021,
  author={Wang, Xiaoting and Wang, Xiaozhe and Sheng, Hao and Lin, Xi},
  journal={IEEE Trans. Power Syst.}, 
  title={A Data-Driven Sparse Polynomial Chaos Expansion Method to Assess Probabilistic Total Transfer Capability for Power Systems With Renewables}, 
  year={2021},
  volume={36},
  number={3},
  pages={2573-2583}}

@ARTICLE{Xu2019PPF,
  author={Xu, Yijun and Mili, Lamine and Zhao, Junbo},
  journal={IEEE Trans. Power Syst.}, 
  title={Probabilistic Power Flow Calculation and Variance Analysis Based on Hierarchical Adaptive Polynomial Chaos-ANOVA Method}, 
  year={2019},
  volume={34},
  number={5},
  pages={3316-3325},
  doi={10.1109/TPWRS.2019.2903164}}

@ARTICLE{Babacan2010,
  author={Babacan, S. Derin and Molina, Rafael and Katsaggelos, Aggelos K.},
  journal={IEEE Transactions on Image Processing}, 
  title={Bayesian Compressive Sensing Using Laplace Priors}, 
  year={2010},
  volume={19},
  number={1},
  pages={53-63},
  doi={10.1109/TIP.2009.2032894}}

@article{Tsilifis2019compressive,
  title={Compressive sensing adaptation for polynomial chaos expansions},
  author={Tsilifis, Panagiotis and Huan, Xun and Safta, Cosmin and Sargsyan, Khachik and Lacaze, Guilhem and Oefelein, Joseph C and Najm, Habib N and Ghanem, Roger G},
  journal={Journal of Computational Physics},
  volume={380},
  pages={29--47},
  year={2019},
  publisher={Elsevier}
}

@TechReport{UQdoc_20_104,
author = {Marelli, S. and L\"uthen, N. and Sudret, B.},
title = {{UQLab user manual -- Polynomial chaos expansions}},
institution = {Chair of Risk, Safety and Uncertainty Quantification, ETH Zurich,
Switzerland},
year = {2022},
note = {Report UQLab-V2.0-104}
}

@article{Sudret2008global,
  title={Global sensitivity analysis using polynomial chaos expansions},
  author={Sudret, Bruno},
  journal={Reliability engineering \& system safety},
  volume={93},
  number={7},
  pages={964--979},
  year={2008},
  publisher={Elsevier}
}

@ARTICLE{Wang2024gsa,
  author={Wang, Xiaoting and Liu, Rong-Peng and Wang, Xiaozhe and Bouffard, François},
  journal={IEEE Transactions on Circuits and Systems II: Express Briefs}, 
  title={A Comparative Study of Polynomial Chaos Expansion-Based Methods for Global Sensitivity Analysis in Power System Uncertainty Control}, 
  year={2024},
  volume={71},
  number={1},
  pages={216-220},
  doi={10.1109/TCSII.2023.3295805}}

@article{Zimmerman2011,
  author = {Zimmerman, R. D. and Murillo-S\'{a}nchez, C. E. and Thomas, R. J.},
  title = {Matpower: Steady-State Operations, Planning and Analysis Tools for Power Systems Research and Education},
  journal = {IEEE Trans. Power Syst.},
  volume = {26},
  number = {1},
  pages = {12--19},
  month = {Feb},
  year = {2011}
}

@TechReport{UQdoc_20_118,
author = {Wagner, P.-R. and Marelli, S. and Sudret, B.},
title = {{UQLab user manual -- Stochastic Spectral Embedding}},
institution = {Chair of Risk, Safety and Uncertainty Quantification, ETH Zurich,
Switzerland},
year = {2022},
note = {Report UQLab-V2.0-118}
}

@article{Karki2006,
  title={A simplified wind power generation model for reliability evaluation},
  author={Karki, Rajesh and Hu, Po and Billinton, Roy},
  journal={IEEE transactions on Energy conversion},
  volume={21},
  number={2},
  pages={533--540},
  year={2006},
  publisher={IEEE}
}

@article{Salameh1995,
  title={Photovoltaic module-site matching based on the capacity factors},
  author={Salameh, Ziyad M and Borowy, Bogdan S and Amin, Atia RA},
  journal={IEEE transactions on Energy conversion},
  volume={10},
  number={2},
  pages={326--332},
  year={1995},
  publisher={IEEE}
}

@ARTICLE{Billinton2008,
  author={Billinton, Roy and Huang, Dange},
  journal={IEEE Trans. Power Syst.}, 
  title={Effects of Load Forecast Uncertainty on Bulk Electric System Reliability Evaluation}, 
  year={2008},
  volume={23},
  number={2},
  pages={418-425},
  doi={10.1109/TPWRS.2008.920078}}

@article{blatman2009adaptive,
  title={Adaptive sparse polynomial chaos expansions for uncertainty propagation and sensitivity analysis},
  author={Blatman, G{\'e}raud},
  year={2009}
}

@ARTICLE{Lorca2015,
  author={Lorca, Alvaro and Sun, Xu Andy},
  journal={IEEE Trans. Power Syst.}, 
  title={Adaptive Robust Optimization With Dynamic Uncertainty Sets for Multi-Period Economic Dispatch Under Significant Wind}, 
  year={2015},
  volume={30},
  number={4},
  pages={1702-1713},
  doi={10.1109/TPWRS.2014.2357714}}

@ARTICLE{Ruiz2009,
  author={Ruiz, Pablo A. and Philbrick, C. Russ and Zak, Eugene and Cheung, Kwok W. and Sauer, Peter W.},
  journal={IEEE Trans. Power Syst.}, 
  title={Uncertainty Management in the Unit Commitment Problem}, 
  year={2009},
  volume={24},
  number={2},
  pages={642-651},
  doi={10.1109/TPWRS.2008.2012180}}

@article{wagner2021bayesian,
  title={Bayesian model inversion using stochastic spectral embedding},
  author={Wagner, Paul-Remo and Marelli, Stefano and Sudret, Bruno},
  journal={Journal of Computational Physics},
  volume={436},
  pages={110141},
  year={2021},
  publisher={Elsevier}
}

@article{luthen2021sparse,
  title={Sparse polynomial chaos expansions: Literature survey and benchmark},
  author={Luthen, Nora and Marelli, Stefano and Sudret, Bruno},
  journal={SIAM/ASA Journal on Uncertainty Quantification},
  volume={9},
  number={2},
  pages={593--649},
  year={2021},
  publisher={SIAM}
}

@article{chi2024hybrid,
  title={Hybrid model-data-driven dynamic VAR planning for wind-penetrated power system using spectral surrogate techniques},
  author={Chi, Yuan and Zou, Yao and Zheng, Xinying and Wang, Qianggang},
  journal={International Journal of Electrical Power \& Energy Systems},
  volume={159},
  pages={109998},
  year={2024},
  publisher={Elsevier}
}

@ARTICLE{Hu2021,
  author={Hu, Zhixiong and Xu, Yijun and Korkali, Mert and Chen, Xiao and Mili, Lamine and Valinejad, Jaber},
  journal={IEEE Transactions on Sustainable Energy}, 
  title={A Bayesian Approach for Estimating Uncertainty in Stochastic Economic Dispatch Considering Wind Power Penetration}, 
  year={2021},
  volume={12},
  number={1},
  pages={671-681},
  doi={10.1109/TSTE.2020.3015353}}

@article{panaretos2019,
  title={Statistical aspects of Wasserstein distances},
  author={Panaretos, Victor M and Zemel, Yoav},
  journal={Annual review of statistics and its application},
  volume={6},
  number={1},
  pages={405--431},
  year={2019},
  publisher={Annual Reviews}
}

@article{wang2019probabilistic,
  title={Probabilistic individual load forecasting using pinball loss guided LSTM},
  author={Wang, Yi and Gan, Dahua and Sun, Mingyang and Zhang, Ning and Lu, Zongxiang and Kang, Chongqing},
  journal={Applied Energy},
  volume={235},
  pages={10--20},
  year={2019},
  publisher={Elsevier}
}

@article{massey1951kolmogorov,
  title={The Kolmogorov-Smirnov test for goodness of fit},
  author={Massey Jr, Frank J},
  journal={Journal of the American statistical Association},
  volume={46},
  number={253},
  pages={68--78},
  year={1951},
  publisher={Taylor \& Francis}
}

@ARTICLE{Xu2020coula,
  author={Xu, Yijun and Mili, Lamine and Korkali, Mert and Karra, Kiran and Zheng, Zongsheng and Chen, Xiao},
  journal={IEEE Transactions on Power Systems}, 
  title={A Data-Driven Nonparametric Approach for Probabilistic Load-Margin Assessment Considering Wind Power Penetration}, 
  year={2020},
  volume={35},
  number={6},
  pages={4756-4768},
  doi={10.1109/TPWRS.2020.2987900}}

@article{guha2019machine,
  title={Machine learning for AC optimal power flow},
  author={Guha, Neel and Wang, Zhecheng and Wytock, Matt and Majumdar, Arun},
  journal={arXiv preprint arXiv:1910.08842},
  year={2019}
}

@misc{OpenMeteoHistoricalWeatherAPI,
  author       = {{Open-Meteo}},
  title        = {{Historical Weather API Documentation}},
  year         = {2026},
  howpublished = {\url{https://open-meteo.com/en/docs/historical-weather-api}},
  note         = {Accessed: 2026-04-26} 
}

@article{wang2016bayesian,
  title={Bayesian optimization in a billion dimensions via random embeddings},
  author={Wang, Ziyu and Hutter, Frank and Zoghi, Masrour and Matheson, David and De Feitas, Nando},
  journal={Journal of Artificial Intelligence Research},
  volume={55},
  pages={361--387},
  year={2016}
}

@inproceedings{rezende2015variational,
  author    = {Rezende, Danilo J. and Mohamed, Shakir},
  title     = {{Variational Inference with Normalizing Flows}},
  booktitle = {{Proceedings of the 32nd International Conference on Machine Learning}},
  pages     = {{1530--1538}},
  year      = {{2015}},
  address   = {{Lille, France}}
}

@inproceedings{vaswani2017attention,
  author    = {Vaswani, Ashish and Shazeer, Noam and Parmar, Niki and Uszkoreit, Jakob and Jones, Llion and Gomez, Aidan N. and Kaiser, Lukasz and Polosukhin, Illia},
  title     = {{Attention Is All You Need}},
  booktitle = {{Proceedings of the 31st International Conference on Neural Information Processing Systems}},
  pages     = {{5998--6008}},
  year      = {{2017}},
  address   = {{Long Beach, CA, USA}}
}

@book{zhou2025ensemble,
  author    = {Zhou, Zhi-Hua},
  title     = {{Ensemble Methods: Foundations and Algorithms}},
  edition   = {{2}},
  publisher = {CRC Press},
  address   = {Boca Raton, FL, USA},
  year      = {2025}
}
\begin{IEEEbiography}[{\includegraphics[width=1in,height=1.25in,clip,keepaspectratio]{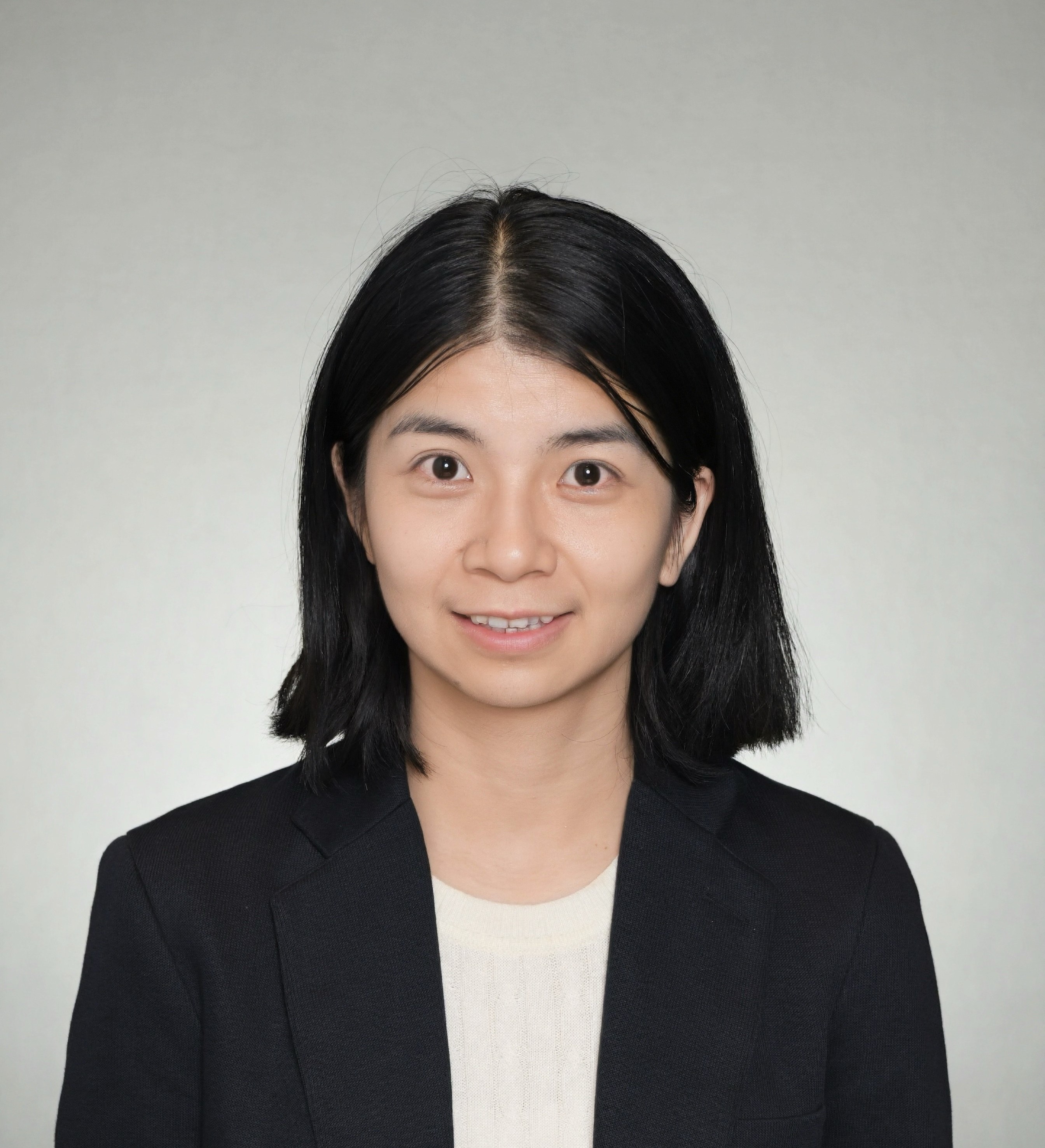}}]{Xiaoting Wang}
(Member, IEEE) is currently a Postdoctoral Fellow with the Department of Electrical and Computer Engineering, University of Alberta, Edmonton, AB, Canada.  She received the B.S. degree in Electrical Engineering and Automation from Fuzhou University, Fuzhou, China, in 2016, the M.S. degree in Control Science and Engineering from Harbin Institute of Technology, Shenzhen, China, in 2019 and the Ph.D. degree in Electrical Engineering from the Department of Electrical and Computer Engineering, McGill University, Montreal, QC, Canada, in 2024. Her research interests include uncertainty quantification, power system security, resilience, and distributed energy management.
\end{IEEEbiography} 

\begin{IEEEbiography}[{\includegraphics[width=1in,height=1.25in,clip,keepaspectratio]{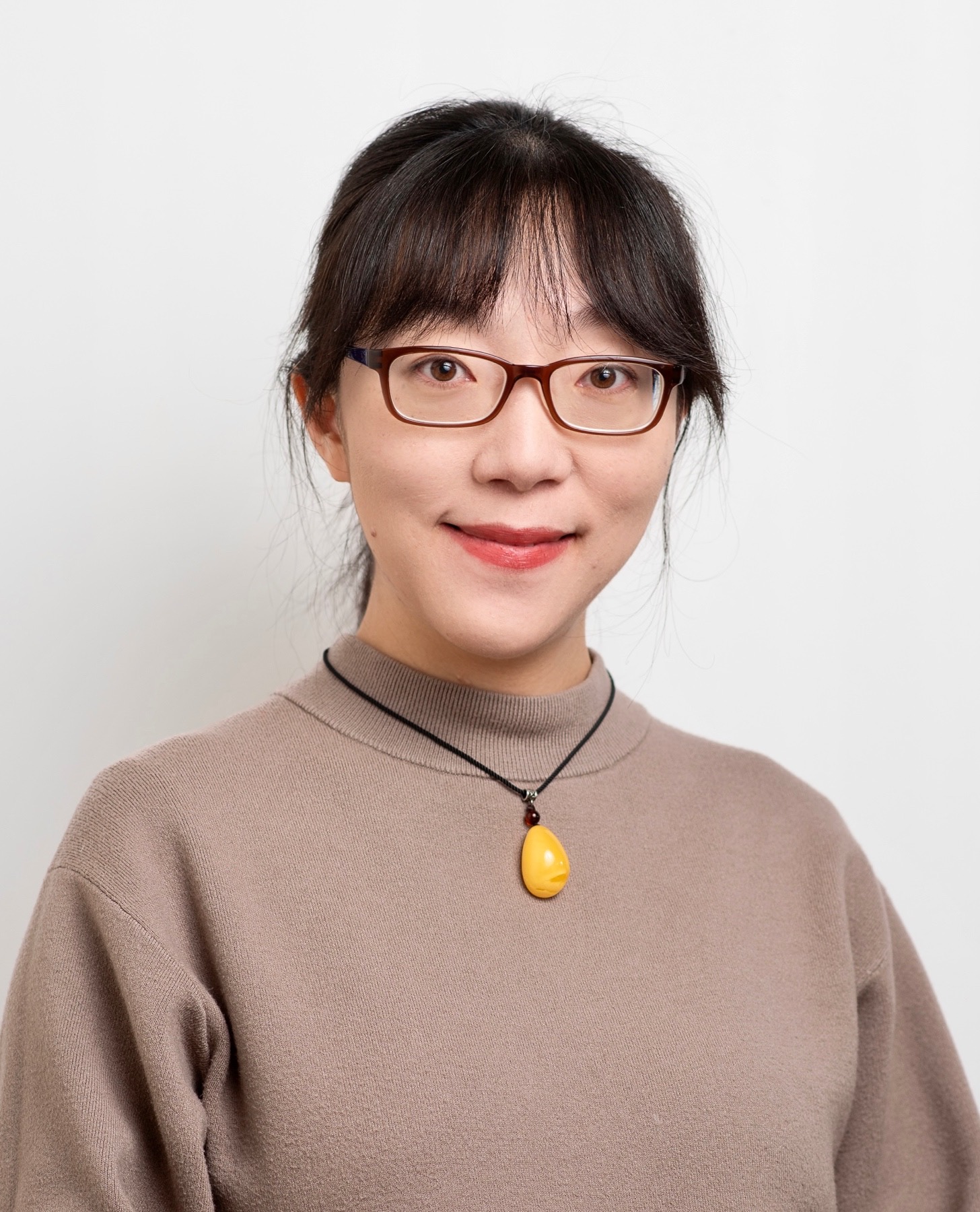}}]{Xiaozhe Wang}
(Senior Member, IEEE) received the B.S. degree in information science and electronic engineering from Zhejiang University, Zhejiang, China, in 2010, and the Ph.D. degree from the School of Electrical and Computer Engineering, Cornell University, Ithaca, NY, USA, in 2015. She is currently an Associate Professor and Canada Research Chair with the Department of Electrical and Computer Engineering, McGill University, Montreal, QC, Canada. She was a recipient of the Alexander von Humboldt Research Fellowship. Her research interests include data-driven power system stability monitoring and control, uncertainty quantification in power system security, stability and resilience, and cybersecurity in power systems.

\end{IEEEbiography}

\begin{IEEEbiography}[{\includegraphics[width=1in,height=1.25in,clip,keepaspectratio]{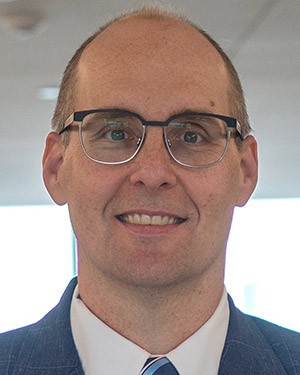}}]{Gregory Kish} (Senior Member, IEEE) received the B.E.Sc. degree from the University of Western Ontario, London, ON, Canada, in 2009, and the M.A.Sc. and Ph.D. degrees from the University of Toronto, Toronto, ON, Canada, in 2011 and 2016, respectively, all in electrical engineering. 
He is currently an Associate Professor with the University of Alberta, Edmonton, AB, Canada. His research interests include the development and application of power electronic converter systems in electric grids.

\end{IEEEbiography}

\begin{IEEEbiography}[{\includegraphics[width=1in,height=1.25in,clip,keepaspectratio]{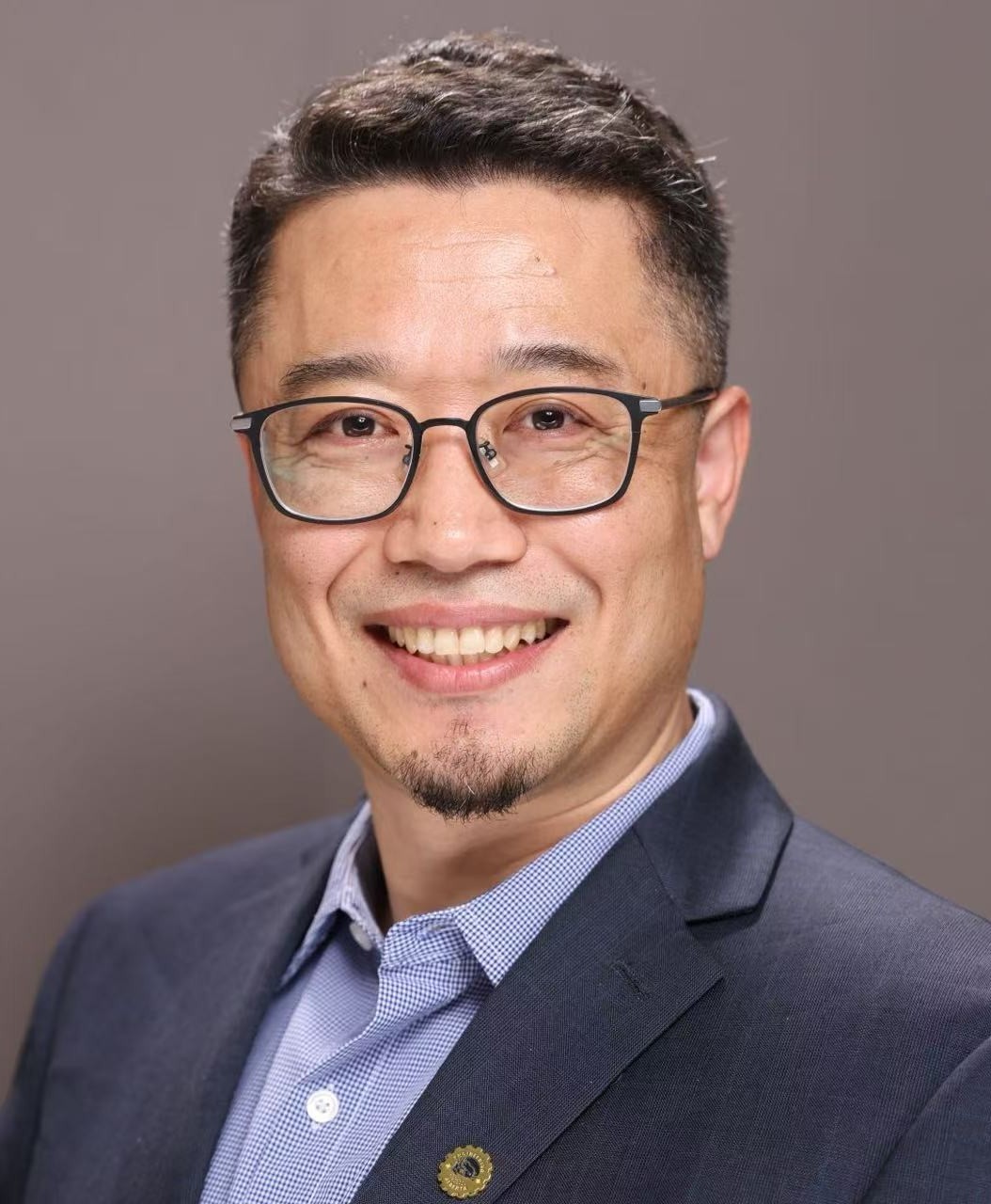}}]{Yunwei (Ryan) Li}
(S'04-M'05-SM'11-F'20)  is a professor, University of Alberta Senior Engineering Research Chair, and Chair of the Department of Electrical and Computer Engineering.

He received the B.Sc. in Engineering degree in electrical engineering from Tianjin University, Tianjin, China, in 2002, and the Ph.D. degree from Nanyang Technological University, Singapore, in 2006. In 2005, Dr. Li was a Visiting Scholar with Aalborg University, Denmark. From 2006 to 2007, he was a Postdoctoral Research Fellow at the Toronto Metropolitan University, Canada. In 2007, he also worked at Rockwell Automation Canada before he joined University of Alberta, Canada in the same year. His research interests include distributed generation, microgrid, renewable energy, high power converters and electric motor drives. 

Dr. Li is the Vice President for Products of IEEE Power Electronics Society (PELS) 2022-2026. He was the Editor-in-Chief for IEEE Transactions on Power Electronics Letters 2019-2023. Dr. Li served as the general chair of IEEE Energy Conversion Congress of Exposition (ECCE) in 2020 for the first ever virtual version during the pandemic. Dr. Li received the Research Excellence Summit Award by the Association of Professional Engineers and Geoscientists of Alberta in 2025, Nagamori Foundation Award in 2022, and the Richard M. Bass Outstanding Young Power Electronics Engineer Award from IEEE PELS in 2013. He is a Fellow of the Canadian Academy of Engineering, and recognized as the Clarivate Highly Cited Researcher.

\end{IEEEbiography} 
\end{document}